\documentclass[prd,preprint,floatfix,nofootinbib,preprintnumbers,superscriptaddress,showkeys]{revtex4} %,twocolumn
\pdfoutput=1
\usepackage[caption=false]{subfig}
\usepackage{amsmath}
\usepackage{amsfonts}
\usepackage{amssymb}
\usepackage{float}
\usepackage{color}
\usepackage{graphicx}
\usepackage{graphics}
\usepackage{hyperref}
\usepackage{adjustbox,array}
\usepackage{enumitem} 
\hypersetup{}
\usepackage[utf8x]{inputenc} 
\usepackage{multirow}
\usepackage[normalem]{ulem}
\usepackage{booktabs}
\usepackage{mathrsfs}
\usepackage{diagbox}
\usepackage[english]{babel}
\usepackage[autostyle]{csquotes}
\usepackage[normalem]{ulem}
\usepackage{xspace}
\usepackage{bm}
\usepackage{xcolor}

\graphicspath{{./figures/}}

\definecolor{rosso}{cmyk}{0,1,1,0.4}
\definecolor{rossos}{cmyk}{0,1,1,0.55}
\definecolor{rossoc}{cmyk}{0,1,1,0.2}
\definecolor{blu}{cmyk}{1,1,0,0.3}
\definecolor{blus}{cmyk}{1,1,0,0.6}
\definecolor{bluc}{cmyk}{1,1,0,0.1}
\definecolor{verde}{cmyk}{0.92,0,0.59,0.25}
\definecolor{verdec}{cmyk}{0.92,0,0.59,0.15}
\definecolor{verdes}{cmyk}{0.92,0,0.59,0.4}

\AtBeginDocument{
\heavyrulewidth=.08em
\lightrulewidth=.05em
\cmidrulewidth=.03em
\belowrulesep=.65ex
\belowbottomsep=0pt
\aboverulesep=.4ex
\abovetopsep=0pt
\cmidrulesep=\doublerulesep
\cmidrulekern=.5em
\defaultaddspace=.5em
}

\hypersetup{
 colorlinks=true,
 linkcolor=verdec,
 urlcolor=verdec,
 citecolor=verdec%gray
}

\newcommand{\lf}{\left(}
\newcommand{\ri}{\right)}

\newcommand{\nn}{\nonumber}

\renewcommand{\lg}{\mathscr{L}} % Amplitude
\newcommand{\br}{\text{Br}}
\newcommand{\hc}{{\rm H.c.}}

\newcommand{\iab}{{\;{\rm ab}^{-1}}}

\newcommand{\tev}{{\;{\rm TeV}}}

\newcommand{\beq}{\begin{equation}}
\newcommand{\eeq}{\end{equation}}
\newcommand{\bea}{\begin{eqnarray}}
\newcommand{\eea}{\end{eqnarray}}
\newcommand{\barr}{\begin{array}}
\newcommand{\earr}{\end{array}}
\newcommand{\bc}{\begin{center}}
\newcommand{\ec}{\end{center}}
\newcommand{\bit}{\begin{itemize}}
\newcommand{\eit}{\end{itemize}}
\newcommand{\ben}{\begin{enumerate}}
\newcommand{\een}{\end{enumerate}}

\newcommand{\al}{\alpha}

\newcommand{\kp}{\kappa}

\newcommand{\gm}{\gamma}

\newcommand{\hsm}{{h_{125}}}
\newcommand{\mch}{M_{H^\pm}}

\newcommand{\tb}{t_\beta}

\newcommand{\cb}{c_\beta}
\renewcommand{\sb}{s_\beta}

\newcommand{\ttau}      {{\tau^+\tau^-}} %{{\tau\tau}} %

\newcommand{\ttop}      {{t\bar{t}}}

\newcommand{\bb}      {{b \bar{b}}}
\newcommand{\ww}      {{W^+ W^-}}

\newcommand{\cw}      {c_{\rm W}}

\newcommand{\xiv}      {\xi_V}
\newcommand{\ztt}      {\zeta_t}
\newcommand{\ztb}      {\zeta_{b}}

\newcommand{\ztta}      {\zeta_\tau}

\definecolor{mint}{rgb}{0.24, 0.71, 0.54}

\newcommand{\code}[1]{\textsc{\small #1}}

\begin{document}

\title{\color{verdes} Intrinsic Properties of Large CP Violation\\ in the  
Complex Two-Higgs-Doublet Model}

\author{Soojin Lee}
\email{soojin.lee@gapp.nthu.edu.tw}
\affiliation{Department of Physics, National Tsing Hua University, Hsinchu 30013, Taiwan} 
\affiliation{Center for Theory and Computation, National Tsing Hua University, Hsinchu 30013, Taiwan}

\author{A. Hammad}
\email{hamed@post.kek.jp}
\affiliation{Theory Center, IPNS, KEK, 1-1 Oho, Tsukuba, Ibaraki 305-0801, Japan} 

\author{Dongjoo Kim}
\email{dongjookim.phys@gmail.com}
\affiliation{Department of Physics, Konkuk University, 120 Neungdong-ro, Gwangjin-gu, Seoul 05029, Republic of Korea}

\author{Jeonghyeon Song}
\email{jhsong@konkuk.ac.kr}
\affiliation{Department of Physics, Konkuk University, 120 Neungdong-ro, Gwangjin-gu, Seoul 05029, Republic of Korea} 

\begin{abstract} 
We investigate the parameter space supporting large CP violation (CPV) in the complex two-Higgs-doublet model with softly broken $Z_{2}$ symmetry, where the 125~GeV Higgs boson is identified as the lightest neutral Higgs boson $H_1$. Through a comprehensive global scan of Type-I and Type-II models under theoretical, collider, and eEDM constraints, we identify distinct structures that facilitate large CPV. In Type-I, gauge-sector CPV is maximized when the 125~GeV Higgs boson is nearly degenerate with a second neutral scalar. For the ensemble of physically viable points, the predicted eEDM values typically exceed $10^{-31}\,e\cdot\mathrm{cm}$, placing the model largely within the sensitivity of next-generation experiments. Conversely, Type-II models strongly suppress gauge-sector CPV while allowing for nearly maximal CPV in the Yukawa sector. Destructive interference among various contributions allows for $|d_e|$ values as low as $\mathcal{O}(10^{-35})\,e\cdot\mathrm{cm}$, resulting in no phenomenologically relevant lower bound. Finally, we uncover the phenomenon of ``hidden CPV'' in the near-alignment limit, characterized by CP-violating mixing between the heavy neutral Higgs bosons governed by the angle $\alpha_3$. We demonstrate that this hidden CPV can be experimentally probed at future colliders via CP-violating Yukawa interactions of $H_2$ and $H_3$, as well as the robust $H_2$-$H_3$-$Z$ coupling.
\end{abstract}

\vspace{1cm}
\keywords{CP violation, Beyond the Standard Model, Higgs Precision Data}

\maketitle
\tableofcontents
\flushbottom 

\section{Introduction}

The discovery of the 125~GeV Higgs boson~\cite{ATLAS:2012yve,CMS:2012qbp} has firmly established the mechanism of electroweak symmetry breaking within the Standard Model (SM), yet it has simultaneously intensified the search for new scalar dynamics. A compelling motivation for this search arises from the observed matter--antimatter asymmetry of the Universe~\cite{Planck:2015fie}, whose dynamical generation requires the three Sakharov conditions: baryon number violation, C and CP violation (CPV), and a departure from thermal equilibrium~\cite{Sakharov:1967dj}.

While the SM satisfies these conditions in principle, it fails in two critical aspects. First, for a 125~GeV Higgs boson, the electroweak phase transition is a smooth crossover~\cite{Huet:1994jb,Kajantie:1996mn,Csikor:1998eu}, failing to provide the required departure from thermal equilibrium. Second, the CPV arising from the Cabibbo--Kobayashi--Maskawa (CKM) matrix is far too small---by roughly ten orders of magnitude---to account for the observed baryon-to-photon ratio~\cite{Gavela:1993ts,Gavela:1994ds,Gavela:1994dt}. These shortcomings point to the necessity of new sources of CPV and additional scalar degrees of freedom near the electroweak scale, making extended Higgs sectors with intrinsic CPV a particularly attractive and well-motivated framework for electroweak baryogenesis.

Among such extensions, the complex two-Higgs-doublet model (C2HDM) with a softly broken $Z_2$ symmetry stands out as a crucial benchmark~\cite{Gunion:1989we,Branco:2011iw,Basler:2021kgq,Goncalves:2023svb}. In this framework, a single independent CP-violating phase in the scalar potential mixes the three neutral Higgs mass eigenstates $H_1$, $H_2$, and $H_3$, ordered as $M_{H_1}<M_{H_2}<M_{H_3}$. Depending on the $Z_2$ charge assignments of the right-handed fermions, four Yukawa types emerge---Type-I, Type-II, Type-X, and Type-Y. This minimal extension provides all the necessary ingredients for electroweak baryogenesis~\cite{Basler:2021kgq} while remaining predictive and testable through correlated signatures in collider searches~\cite{Basso:2012st,Keus:2015hva,Aoki:2018zgq,Arhrib:2018qmw,deBlas:2018mhx,Wang:2019pet,Biekotter:2019kde,Basler:2019iuu,Chen:2020soj,Han:2020lta,Abouabid:2021yvw,Biekotter:2022jyr,Azevedo:2023zkg,Biekotter:2024ykp,Biekotter:2025fjx}, Higgs precision measurements~\cite{Coy:2019rfr,Frank:2021pkc,Azevedo:2023zkg,Biekotter:2024ykp}, and electric dipole moment (EDM) experiments~\cite{Jung:2013hka,Yamanaka:2018uud,Fuyuto:2019svr,Altmannshofer:2020shb,Cheung:2020ugr,Frank:2021pkc,Biekotter:2025fjx,Davila:2025goc}.

Yet, the very predictivity of the C2HDM leads to strong phenomenological tensions. In particular, the non-observation of the electron EDM (eEDM)~\cite{Inoue:2014nva,Cheung:2014oaa,Chen:2015gaa,Biekotter:2025fjx} imposes exceptionally stringent constraints on CP violation. The most recent JILA measurement, $|d_e| < 4.1\times10^{-30}\,e\cdot\mathrm{cm}$ (90\%~C.L.)~\cite{Roussy:2022cmp}, improves upon the previous ACME bound, $|d_e| < 1.1\times10^{-29}\,e\cdot\mathrm{cm}$ (90\%~C.L.)~\cite{ACME:2018yjb}. This dramatic tightening of the eEDM limit severely restricts the amount of CP violation in the C2HDM, confining the allowed parameter space to highly specific, nontrivial regions.

%comment 1

A primary candidate for such a restricted parameter space is the Higgs alignment regime, which is strongly motivated by the observed SM-like couplings of the 125~GeV Higgs boson. Such alignment may arise naturally from global symmetries of the scalar potential---a framework known as Natural Higgs Alignment (NHAL)~\cite{BhupalDev:2014bir}---where the exact symmetric limit typically precludes explicit CP violation. In these scenarios, sizable CP violation can only be attained through the soft or explicit breaking of the underlying NHAL symmetries~\cite{Darvishi:2023fjh,Pilaftsis:2025wpz}. In addition, the Type-I and Type-II Yukawa structures are also motivated by symmetry arguments~\cite{Glashow:1976nt,Khater:2003ym}.

Since the exact Higgs alignment limit is associated with CP conservation, and our focus is on the phenomenology of sizable CP violation, we do not assume such a symmetry-enforced setup.
By identifying the 125~GeV state as the lightest neutral Higgs boson and stochastically sampling the remaining degrees of freedom, we determine the viable parameter regions that support sizable CP violation. This random-scan methodology provides direct access to the Higgs-sector interactions responsible for CP violation and their associated collider phenomenology. 

While a substantial body of research has already explored collider observables across various benchmark channels~\cite{Fontes:2017zfn,Frank:2021pkc,Abouabid:2021yvw,Fontes:2022izp,Capucha:2025qgr}, these works often focus on specific final states. What remains missing is a systematic characterization of the \emph{intrinsic structure}---namely, the specific mass hierarchies and theoretical patterns---that allows the C2HDM to support large or even maximal CP violation while remaining consistent with the full suite of theoretical and experimental constraints, including the stringent eEDM bounds.

To rigorously analyze this structure, we quantify the magnitude of CP violation using two normalized measures: $\xi_V$ for the gauge sector~\cite{Mendez:1991gp} and $\zeta_t$ for the fermionic sector~\cite{Khater:2003ym}. These metrics allow us to map the landscape of CP violation across the viable parameter space. After isolating all points with sizable $\xi_V$ and $\zeta_t$, we address the following fundamental questions:
\begin{enumerate}[label=(\roman*)]
    \item Does the requirement of large CP violation favor specific mass hierarchies, and how do the resulting mass spectra distinguish Type-I from Type-II scenarios?
    \item How are the gauge- and Yukawa-sector CPV measures correlated with the eEDM, and what characteristic ranges of $|d_e|$ are realized within the viable parameter space?
    \item Can sizable CP violation manifest in the near-alignment regime, where it decouples from the 125~GeV Higgs boson?    
\end{enumerate}

Our results reveal that Type-I and Type-II follow two qualitatively distinct pathways to large CP violation when $H_1$ is identified as the 125~GeV Higgs boson. In Type-I, both the gauge- and Yukawa-sector CPV measures can reach significant magnitudes---with $\xiv$ rising to $\sim 0.4$ and $\ztt$ to $\sim 0.8$---but only in regions where the 125~GeV Higgs boson is nearly degenerate with $H_2$. In contrast, Type-II yields a strongly suppressed gauge-sector CPV measure ($\xi_V \lesssim 10^{-3}$), while still allowing for sizable Yukawa CPV with $\ztt \sim 0.7$ and $\ztb \sim 0.8$. In this case, the maximal values of $\ztt$ and $\ztb$ lead to nearly maximal CP violation within the heavy scalar sector involving $H_2$ and $H_3$.

Finally, addressing the last question leads us to uncover a notable phenomenon emerging near the Higgs alignment limit: \emph{hidden CP violation}. We demonstrate that CP-violating mixing between $H_2$ and $H_3$ becomes increasingly \emph{maximal} as the Yukawa-sector CPV measures grow.

The remainder of this paper is organized as follows. In Sec.~\ref{sec:model}, we briefly review the structure of the C2HDM and define the CPV measures $\xiv$ and $\ztt $. Section~\ref{sec:scan} describes our parameter scan and discusses the characteristics of the viable parameter space with large CPV measures. In Sec.~\ref{sec:results}, we present the future prospects and observability of the large CPV parameter space, including eEDM prospects, Higgs signal strengths in the near-degenerate Type-I scenario, and the CP mixing in the top and $\tau$ couplings in Type-II. Additionally, we study the hidden CP violation in the near-alignment limit.
We conclude in Sec.~\ref{sec:conclusion}.

\section{Brief review of C2HDM}
\label{sec:model}

\subsection{Scalar potential, vacuum structure and neutral scalar mixing}

The 2HDM extends the SM by introducing two complex $SU(2)_L$ scalar doublet fields, $\Phi_1$ and $\Phi_2$, both with hypercharge $Y=1/2$ under the $Q = T_3 + Y$ convention~\cite{Branco:2011iw}.
These fields are expressed as:
\beq
\Phi_1 = \left(
\begin{array}{c}
\phi_1^+ \\
\dfrac{v_1 + \rho_1 + i \eta_1}{\sqrt{2}}
\end{array}
\right) \qquad \mbox{and} \qquad
\Phi_2 = \left(
\begin{array}{c}
\phi_2^+ \\
\dfrac{v_2 + \rho_2 + i \eta_2}{\sqrt{2}}
\end{array}
\right) \;, \label{eq-phi12}
\eeq
where $v_1$ and $v_2$ are the \emph{real} vacuum expectation values (VEVs) of $\Phi_1$ and $\Phi_2$, respectively, assuming no spontaneous CP violation.
Their ratio defines the key model parameter $\tb = v_2/v_1$. For notational simplicity, we use the abbreviations $s_x = \sin x$, $c_x = \cos x$, and $t_x = \tan x$.
The combined VEV, $v = \sqrt{v_1^2+v_2^2} \approx 246\,\text{GeV}$, induces spontaneous electroweak symmetry breaking.

To eliminate tree-level flavor-changing neutral currents (FCNCs), a discrete $Z_2$ symmetry is imposed, under which the fields transform as $\Phi_1 \to \Phi_1$ and $\Phi_2 \to -\Phi_2$~\cite{Glashow:1976nt,Paschos:1976ay}.
The scalar potential with a softly broken $Z_2$ symmetry is given by:
\begin{equation}
\label{eq:VPhi}
\begin{split}
V_\Phi (\Phi_1,\Phi_2) &=  m^2_{11} \Phi^\dagger_1 \Phi_1 + m^2_{22} \Phi^\dagger_2 \Phi_2 - ( m^2_{12}  \Phi^\dagger_1 \Phi_2 + \hc) \\[3pt]
&\quad + \tfrac{1}{2}\lambda_1 (\Phi^\dagger_1 \Phi_1)^2 + \tfrac{1}{2}\lambda_2 (\Phi^\dagger_2 \Phi_2)^2  + \lambda_3 (\Phi^\dagger_1 \Phi_1) (\Phi^\dagger_2 \Phi_2)  \\[3pt]
&\quad + \lambda_4 (\Phi^\dagger_1 \Phi_2)(\Phi^\dagger_2 \Phi_1) + \tfrac{1}{2}  \bigl[ \lambda_5 (\Phi^\dagger_1 \Phi_2)^2 + \hc \bigr],
\end{split}
\end{equation}
where the $m_{12}^2$ term acts as the soft breaking parameter.
Due to the hermiticity of the Lagrangian, only the parameters $m_{12}^2$ and $\lambda_5$ can be complex.

The minimization conditions for the potential are:
\bea
m_{11}^2 v_1 + \frac{\lambda_1}{2} v_1^3 + \frac{\lambda_{345}}{2} v_1
v_2^2 &=& \text{Re} \left(m_{12}^2\right) v_2 \;, \label{eq:mincond1} \\
m_{22}^2 v_2 + \frac{\lambda_2}{2} v_2^3 + \frac{\lambda_{345}}{2} v_1^2
v_2 &=& \text{Re} \left(m_{12}^2\right) v_1 \;, \label{eq:mincond2} \\
2\, \text{Im} (m_{12}^2) &=& v_1 v_2 \text{Im} (\lambda_5) 
\;, \label{eq:mincond3}
\eea
where $\lambda_{345} \equiv \lambda_3 + \lambda_4 + \text{Re} (\lambda_5)$.
The condition in \autoref{eq:mincond3} relates two imaginary parts, leaving a single independent CP phase.
To accommodate explicit CP violation, we further assume 
$\phi(\lambda_5) \neq 2\, \phi(m_{12}^2)$~\cite{Ginzburg:2002wt},
ensuring that the two phases cannot be simultaneously removed by rephasing $\Phi_1$ and $\Phi_2$.
This framework constitutes the C2HDM.

In the C2HDM, the three neutral Higgs bosons do not possess definite CP quantum numbers.
The mass eigenstates are denoted as $H_1$, $H_2$, and $H_3$, with the mass hierarchy $M_{H_1}<M_{H_2}<M_{H_3}$.
While the observed $125\,\text{GeV}$ Higgs boson could correspond to any of these states,
we focus on the ``normal scenario'' where the lightest state is identified with the observed particle:
\beq
\text{Normal scenario: } M_{H_1}=125\,\text{GeV}.
\eeq

The Higgs basis
$\{ {\cal H}_1, {\cal H}_2 \}$ is particularly advantageous 
as it aligns the electroweak symmetry-breaking VEV with a single doublet (defined here as ${\cal H}_1$)~\cite{Lavoura:1994fv,Botella:1994cs,Davidson:2005cw,Haber:2010bw,Ginzburg:2002wt}.
The relation between  $\{ {\cal H}_1, {\cal H}_2 \}$ and  $\{ \Phi_1, \Phi_2 \}$ is given by the rotation:
\beq\label{eq-H-Phi-basis}
\left( \begin{array}{c} {\cal H}_1 \\ {\cal H}_2 \end{array} \right) 
= 
\left( \begin{array}{cc} c_\beta & s_\beta \\ - s_\beta &
    c_\beta \end{array} \right) \left( \begin{array}{c} \Phi_1 \\
    \Phi_2 \end{array} \right) \;,
\eeq
which leads to the component decomposition:
\beq
\label{eq-H12}
{\cal H}_1 = \left( \begin{array}{c} G^\pm \\ \frac{1}{\sqrt{2}} (v + H^0
    + i G^0) \end{array} \right) \quad \mbox{and} \qquad
{\cal H}_2 = \left( \begin{array}{c} H^\pm \\ \frac{1}{\sqrt{2}} (R_2
    + i I_2) \end{array} \right) \;.
\eeq
Here, $H^0$ represents the SM Higgs interaction eigenstate before any mixing with the additional neutral scalar components $R_2$ and $I_2$.
The Goldstone bosons $G^\pm$ and $G^0$ reside entirely in ${\cal H}_1$ and are absorbed by the $W^\pm$ and $Z$ gauge bosons, respectively, whereas the physical charged Higgs eigenstates $H^\pm$ originate from ${\cal H}_2$.

The three neutral mass eigenstates $H_i$ ($i=1,2,3$) arise from the mixing of the two neutral CP-even components ($\rho_{1,2}$ in \autoref{eq-phi12}) and the neutral CP-odd component ($I_2$ in \autoref{eq-H12}). 
The mass-squared matrix of the neutral Higgs bosons, $\bm{\mathcal{M}}^2$, is defined through
\beq
V_\Phi \supset  \frac{1}{2} \begin{pmatrix} \rho_1 & \rho_2 & I_2 \end{pmatrix} \bm{\mathcal{M}}^2
\begin{pmatrix} \rho_1 \\ \rho_2 \\ I_2 \end{pmatrix},
\eeq
where 
\beq
\label{eq-Msq-def}
\bm{\mathcal{M}}^2 = 
\begin{pmatrix} 
\bar{m}^2 s_\beta^2 + \lambda_1 v^2 c_\beta^2 & -(\bar{m}^2 - \lambda_{345} v^2) s_\beta c_\beta & -\frac{1}{2} \text{Im}(\lambda_5) v^2 s_\beta \\ 
-(\bar{m}^2 - \lambda_{345} v^2) s_\beta c_\beta & \bar{m}^2 c_\beta^2 + \lambda_2 v^2 s_\beta^2 & -\frac{1}{2} \text{Im}(\lambda_5) v^2 c_\beta \\ 
-\frac{1}{2} \text{Im}(\lambda_5) v^2 s_\beta & -\frac{1}{2} \text{Im}(\lambda_5) v^2 c_\beta & \bar{m}^2 - \text{Re}(\lambda_5) v^2 
\end{pmatrix},
\eeq
with $\bar{m}^2 = \text{Re}(m_{12}^2)/(\sb\cb)$.

The orthogonal mixing matrix $\mathbf{R}$ diagonalizes $\bm{\mathcal{M}}^2$, yielding mass eigenstates $\{H_a, H_b, H_c\}$
\beq
\label{eq-R-def}
\bm{\mathcal{M}}^2 = \mathbf{R}^T 
\begin{pmatrix}
m_{H_a}^2 & 0 & 0 \\
0 & m_{H_b}^2 & 0\\
0 & 0 & m_{H_c}^2
\end{pmatrix}
 \mathbf{R}.
\eeq 
The matrix $\mathbf{R}$ is parameterized by three mixing angles $\al_{1,2,3}$ as:
\begin{align}
\label{eq-R}
\mathbf{R} &= \begin{pmatrix}
1 & 0 & 0 \\
0 & c_{\alpha_3} & s_{\alpha_3} \\
0 & -s_{\alpha_3} & c_{\alpha_3}
\end{pmatrix}
\begin{pmatrix}
c_{\alpha_2} & 0 & s_{\alpha_2} \\
0 & 1 & 0 \\
- s_{\alpha_2} & 0 & c_{\alpha_2}
\end{pmatrix}
\begin{pmatrix}
c_{\alpha_1} & s_{\alpha_1} & 0 \\
- s_{\alpha_1} & c_{\alpha_1} & 0 \\
0 & 0 & 1
\end{pmatrix} \\ \nn
&=
\addtolength{\arraycolsep}{7pt}\begin{pmatrix}
c_{\alpha_1} c_{\alpha_2} & s_{\alpha_1} c_{\alpha_2} & s_{\alpha_2} \\[6pt]
- c_{\alpha_1} s_{\alpha_2} s_{\alpha_3} -\,s_{\alpha_1} c_{\alpha_3}  &
c_{\alpha_1} c_{\alpha_3} - s_{\alpha_1} s_{\alpha_2} s_{\alpha_3} &
c_{\alpha_2} s_{\alpha_3} \\[6pt]
- c_{\alpha_1} s_{\alpha_2} c_{\alpha_3} + s_{\alpha_1} s_{\alpha_3}  &
-\,c_{\alpha_1} s_{\alpha_3} - s_{\alpha_1} s_{\alpha_2} c_{\alpha_3} &
c_{\alpha_2} c_{\alpha_3}
\end{pmatrix}.
\end{align}
The mixing angles lie in the range $\alpha_{1,2,3} \in [-\pi/2, \pi/2]$.
Since the hierarchy among $m_{H_a}$, $m_{H_b}$, and $m_{H_c}$ is not imposed a priori, the states must be reordered according to their masses to define the mass eigenstates $\{H_1, H_2, H_3\}$, satisfying the standard hierarchy $M_{H_1} < M_{H_2} < M_{H_3}$.
In summary, the mass eigenstates $(H_1,H_2,H_3)$ are related to the states $(\rho_1, \rho_2 , I_2)$ via
\beq
\label{eq-matrix-R}
\left( \begin{array}{c} H_1 \\ H_2 \\ H_3 \end{array} \right) = \mathbf{R}
\left( \begin{array}{c} \rho_1 \\ \rho_2 \\ I_2 \end{array} \right)
\;.
\eeq

An interesting relation emerges between the mass parameters and the mixing angles. The explicit form of the mass matrix $\bm{\mathcal{M}}^2$ in \autoref{eq-Msq-def} leads to the following identity:
\beq
\label{eq-M13sq/M23sq}
\mathcal{M}_{13}^2 = \tb \, \mathcal{M}_{23}^2 .
\eeq
Using the relation between $\bm{\mathcal{M}}^2$ and the mixing matrix $\mathbf{R}$ in \autoref{eq-R-def}, the mass of the third neutral scalar, $m_{H_c}$, can be expressed as~\cite{Fontes:2017zfn}
\begin{equation}
\label{eq-mHc}
m_{H_c}^2 =
\dfrac{
m_{H_a}^2 R_{a3}(R_{a2}\tb - R_{a1}) + m_{H_b}^2 R_{b3}(R_{b2}\tb - R_{b1})
}{
R_{c3}(R_{c1} - R_{c2}\tb )
}\;.
\end{equation}
As a derived quantity, $m_{H_c}$ may emerge as the lightest, intermediate, or heaviest state, depending on the choice of input parameters. The states $\{H_a, H_b, H_c\}$ are then reordered accordingly to define the mass eigenstates $\{H_1, H_2, H_3\}$.

Furthermore, the physical mass basis is related to the Higgs basis via the mixing matrix $\mathbf{T}$~\cite{Branco:2011iw}:
\begin{align}
\left( \begin{array}{c} H_1 \\ H_2 \\ H_3 \end{array} \right) 
&\equiv \mathbf{T}^T \left( \begin{array}{c} H^0 \\ R_2 \\ I_2 \end{array} \right).
\end{align}
Combining \autoref{eq-H-Phi-basis} and \autoref{eq-matrix-R}, the explicit form of $\mathbf{T}$ is:
\beq
\label{eq-fullT}
\mathbf{T}=
\begin{pmatrix}
c_{\beta-\alpha_1} c_{\alpha_2} &
~~-\,c_{\beta-\alpha_1} s_{\alpha_2} s_{\alpha_3} + s_{\beta-\alpha_1} c_{\alpha_3}~~ & 
-\,c_{\beta-\alpha_1} s_{\alpha_2} c_{\alpha_3} - s_{\beta-\alpha_1} s_{\alpha_3} \\[6pt]
-\,s_{\beta-\alpha_1} c_{\alpha_2} & 
c_{\beta-\alpha_1} c_{\alpha_3} + s_{\beta-\alpha_1} s_{\alpha_2} s_{\alpha_3} &
-\,c_{\beta-\alpha_1} s_{\alpha_3} + s_{\beta-\alpha_1} s_{\alpha_2} c_{\alpha_3} \\[6pt]
s_{\alpha_2} & c_{\alpha_2} s_{\alpha_3} & 
c_{\alpha_2} c_{\alpha_3}
\end{pmatrix}
\eeq

Finally, we identify CP-conserving limits corresponding to scenarios in which  one state becomes pure CP-odd. Given the identification $M_{H_1} = 125~\mathrm{GeV}$, this decoupling occurs in two distinct cases~\cite{ElKaffas:2007rq}:
    \ben[label=(\roman*)]
    \item $H_2$ is CP-odd: This is realized by setting $\al_2=0$ and $\al_3=\pm\pi/2$, yielding
    \beq
\label{eq-T-H2-CPodd}
\mathbf{T}^T \big|_{H_2:\text{CP-odd}} = 
\begin{pmatrix}
c_{\beta-\alpha_1} &  -s_{\beta-\alpha_1} & 0 \\
0 &0 & \pm 1  \\
\mp s_{\beta-\alpha_1} & \mp c_{\beta-\alpha_1}& 0 \\
\end{pmatrix}.
\eeq

     \item $H_3$ is CP-odd: This is realized by setting $\alpha_2=\alpha_3=0$, yielding
\beq
\label{eq-T-H3-CPodd}
\mathbf{T}^T\big|_{H_3:\text{CP-odd}}  = 
\begin{pmatrix}
c_{\beta-\alpha_1} & -s_{\beta-\alpha_1} & 0\\
s_{\beta-\alpha_1} & c_{\beta-\alpha_1} & 0 \\
0 & 0& 1
\end{pmatrix}.
\eeq
    \een

\subsection{Neutral Higgs couplings and global measures of CP violation}

The interactions of the neutral Higgs bosons with the gauge and fermionic sectors are central to the phenomenology of the C2HDM.\footnote{For the complete set of C2HDM couplings, we refer the reader to Appendix A of Ref.~\cite{Fontes:2017zfn}.}
The interaction vertex $H_i$-$V_\mu$-$V_\nu$ (where $V=W^\pm,Z$) is parameterized as
\beq\label{eq-ci-def}
i \, c(H_i VV) \, g_{\mu\nu} \, g_V m_V \;, 
\eeq
where $g_W= g$ and $g_Z= g /\cw$, with $\cw=\cos\theta_W$ denoting the cosine of the Weinberg angle.
The gauge coupling modifiers are given in terms of the mixing matrix elements by
\beq
\label{eq-ci-T}
c(H_i VV) = T_{1i} \;.
\eeq
Since current Higgs precision data favor a positive sign for the gauge coupling modifier $\kappa_V$~\cite{ATLAS:2022vkf,CMS:2022dwd,Sopczak:2025ptc}, we restrict our analysis to the case where
\beq
\label{eq-CH1VV-sign}
c(H_1 VV)>0.
\eeq

The Yukawa Lagrangian for a fermion field $\psi_f$ with mass $m_f$ is parameterized by
\beq\label{eq-Yukawa-Lg}
{\cal L}_Y = - \sum_{i=1}^3 \frac{m_f}{v} \bar{\psi}_f \big[ c^e(H_i
  ff) + i c^o(H_i ff) \gamma_5 \big] \psi_f H_i \;,
\eeq
where $c^e(H_i ff)$ and $c^o(H_i ff)$ denote the CP-even and CP-odd Yukawa coupling modifiers, respectively.
For Type-I and Type-II models, the modifiers for the top quark, bottom quark, and $\tau$ lepton take the following forms:
\begin{align}
\label{eq-cf-typeI}
\text{Type-I:} & & 
c^e(H_i \ttop) &= c^e(H_i bb) = c^e(H_i \tau\tau) = \frac{R_{i2}}{\sb}, \\ \nn 
& & 
c^o(H_i \ttop) &= -c^o(H_i bb) = -c^o(H_i \tau\tau) = -\frac{R_{i3}}{\tb}, \\[7pt] 
\label{eq-cf-typeII}
\text{Type-II:} & & 
c^e(H_i \ttop) &= \frac{R_{i2}}{\sb}, \quad c^e(H_i bb) = c^e(H_i \tau\tau) =\frac{R_{i1}}{\cb}, 
\\ \nn 
& & 
c^o(H_i \ttop) &= - \frac{R_{i3}}{\tb}, \quad c^o(H_i bb) = c^o(H_i \tau\tau) = -\tb R_{i3}.
\end{align}

To quantify the extent of CP violation in the scalar sector, various measures have been proposed and studied extensively~\cite{Mendez:1991gp,Branco:1999fs,Khater:2003ym,Fontes:2015xva}.
In this work, we adopt two normalized global CPV measures: $\xiv$ for the gauge sector and $\zeta_{t,b}$ for the Yukawa sector.
By definition, both measures satisfy $\xiv \in [0,1]$ and $\zeta_{t,b}  \in [0,1]$.

The gauge-sector measure $\xiv$ is defined as~\cite{Mendez:1991gp}
\begin{equation} \label{eq-xiv}
\xiv=27\left[c(H_1 VV) c(H_2 VV) c(H_3 VV) \right]^2 = 27\prod_{i=1}^3T_{1i}^2 \, .
\end{equation}
For the Yukawa sector, we introduce two normalized sum variables~\cite{Khater:2003ym}:
\begin{equation}\label{eq-zeta2313}
\zeta_{23}=2\sum_{i=1}^3 \lf R_{i2}\,R_{i3} \ri^2, \quad \zeta_{13}=2\sum_{i=1}^3 \lf R_{i1}\,R_{i3}\ri^2 \, .
\end{equation}
Given the distinct structures of the CP-even and CP-odd Yukawa couplings in Type-I and Type-II models (see \autoref{eq-cf-typeI} and \autoref{eq-cf-typeII}), the normalized CPV measures for the top quark, $b$ quark, and $\tau$ lepton are identified as follows:
\begin{equation}
\label{eq-zeta-tbtau}
\begin{cases} 
\text{Type-I:}  & \ztt = \ztb = \ztta = \zeta_{23} \\[6pt]
\text{Type-II:} & \ztt = \zeta_{23}, \quad \ztb = \ztta = \zeta_{13}
\end{cases}
\end{equation}

\subsection{Higgs Alignment Limit}

Before presenting our numerical results, we investigate the Higgs alignment limit, a critical regime where the lightest neutral scalar, $H_1$, exhibits gauge and Yukawa couplings identical to those of the SM Higgs boson. In the C2HDM, the exact Higgs alignment limit is realized when $\alpha_1 = \beta$ and $\alpha_2 = 0$. Under these conditions, the mixing matrix $\mathbf{T}$ simplifies to:
\begin{equation}
\label{eq-T-H-alignment}
\mathbf{T} \big|_{\alpha_1=\beta,\,\alpha_2=0} = 
\begin{pmatrix}
1 & 0 & 0\\
0 & c_{\alpha_3} & -s_{\alpha_3} \\
0 & s_{\alpha_3} & c_{\alpha_3}
\end{pmatrix}.
\end{equation}

A primary implication of the Higgs alignment limit is the recovery of CP conservation in the scalar potential. This is manifest in the behavior of the imaginary part of the quartic coupling $\lambda_5$:~\cite{ElKaffas:2007rq}
\begin{equation}
\begin{aligned}
\text{Im} \, \lambda_5 = & -\frac{1}{c_\beta s_\beta v^2} \Big\{ c_\beta \left[ c_{\alpha_1} c_{\alpha_2} s_{\alpha_2} M_{H_1}^2 - c_{\alpha_2} s_{\alpha_3} (c_{\alpha_1} s_{\alpha_2} s_{\alpha_3} + s_{\alpha_1} c_{\alpha_3}) M_{H_2}^2 \right. \\
& + c_{\alpha_2} c_{\alpha_3} (s_{\alpha_1} s_{\alpha_3} - c_{\alpha_1} s_{\alpha_2} c_{\alpha_3}) M_{H_3}^2 \left. \right] + s_\beta \left[ s_{\alpha_1} c_{\alpha_2} s_{\alpha_2} M_{H_1}^2 \right. \\
& + \left. c_{\alpha_2} s_{\alpha_3} (c_{\alpha_1} c_{\alpha_3} - s_{\alpha_1} s_{\alpha_2} s_{\alpha_3}) M_{H_2}^2 - c_{\alpha_2} c_{\alpha_3} (c_{\alpha_1} s_{\alpha_3} + s_{\alpha_1} s_{\alpha_2} c_{\alpha_3}) M_{H_3}^2 \right] \Big\},
\end{aligned}
\end{equation}
which vanishes identically upon substituting the alignment conditions $\alpha_1 = \beta$ and $\alpha_2 = 0$. This symmetry requirement forces the CP-violating off-diagonal elements of the mass matrix to vanish—specifically $\mathcal{M}_{13}^2 = \mathcal{M}_{23}^2 = 0$ in \autoref{eq-Msq-def}—thereby decoupling the CP-odd state from the CP-even sector. Furthermore, the condition $\text{Im} \, \lambda_5 = 0$ 
leads to the indeterminacy of $M_{H_c}$ in \autoref{eq-mHc} as the exact alignment limit is approached.

In the exact Higgs alignment limit, the mixing angle $\alpha_3$ is further constrained by the requirement of CP conservation. Applying the alignment conditions $\alpha_1 = \beta$ and $\alpha_2 = 0$ to the mixing matrix in \autoref{eq-matrix-R}, the heavy neutral states can be expressed as:
\begin{align} \label{eq-H23-HAL}
H_2 &= c_{\alpha_3} \left( -s_\beta \rho_1 + c_\beta \rho_2 \right) + s_{\alpha_3} I_2, \\
H_3 &= -s_{\alpha_3} \left( -s_\beta \rho_1 + c_\beta \rho_2 \right) + c_{\alpha_3} I_2,
\end{align}
where $\rho_{1,2}$ are CP-even components and $I_2$ is CP-odd. Since CP conservation requires $H_2$ and $H_3$ to be pure CP eigenstates, $\alpha_3$ must take the values $0$ or $\pm \pi/2$.

In summary, the exact Higgs alignment limit within the C2HDM with softly broken $Z_2$ symmetry is characterized by the following simultaneous conditions:
\begin{equation}
\label{eq-H-align-condition}
\text{Higgs alignment: } \alpha_1 = \beta, \quad \alpha_2 = 0, \quad \text{Im} \, \lambda_5 = 0, \quad \alpha_3 \in \{0, \pm \pi/2\}.
\end{equation}
To quantify small but finite deviations from this limit, we introduce the normalized alignment measure:
\begin{equation}
\label{eq-dha}
\delta_{H\text{-align}} = \sqrt{ \frac{s_{\beta-\alpha_1}^2 + s_{\alpha_2}^2}{2} } \;,
\end{equation}
where $\delta_{H\text{-align}} = 0$ corresponds to the case of exact alignment.

In the exact Higgs alignment limit, the coupling modifiers for $H_1$ recover their SM values: $c(H_1 VV)=1$, $c^e(H_1 ff)=1$, and $c^o(H_1 ff)=0$. For the heavy scalars $H_{2,3}$, the gauge coupling modifiers vanish, $c(H_{2,3} VV)=0$, while the Yukawa coupling modifiers are given by:
\begin{align}
\label{eq-H23-modifiers-HAL}
&\text{\underline{Higgs alignment limit}}  \ (\alpha_1 = \beta, \, \alpha_2 = 0): \nonumber \\[10pt]
& \text{Type-I:} 
\begin{cases}
c^e(H_2 ff) = \frac{c_{\alpha_3}}{t_\beta}, \quad & c^o(H_2 tt) = -c^o(H_2 bb) = -c^o(H_2 \tau\tau) = -\frac{s_{\alpha_3}}{t_\beta}, \\
c^e(H_3 ff) = -\frac{s_{\alpha_3}}{t_\beta}, \quad & c^o(H_3 tt) = -c^o(H_3 bb) = -c^o(H_3 \tau\tau) = -\frac{c_{\alpha_3}}{t_\beta},
\end{cases} \\[15pt]
& \text{Type-II:} 
\begin{cases}
c^e(H_2 tt) = \frac{c_{\alpha_3}}{t_\beta}, \quad & c^e(H_2 bb) = c^e(H_2 \tau\tau) = -t_\beta c_{\alpha_3}, \\
c^o(H_2 tt) = -\frac{s_{\alpha_3}}{t_\beta}, \quad & c^o(H_2 bb) = c^o(H_2 \tau\tau) = -t_\beta s_{\alpha_3}, \\
c^e(H_3 tt) = -\frac{s_{\alpha_3}}{t_\beta}, \quad & c^e(H_3 bb) = c^e(H_3 \tau\tau) = t_\beta s_{\alpha_3}, \\
c^o(H_3 tt) = -\frac{c_{\alpha_3}}{t_\beta}, \quad & c^o(H_3 bb) = c^o(H_3 \tau\tau) = -t_\beta c_{\alpha_3}.
\end{cases}
\end{align}
We explicitly retain the $\alpha_3$ dependence in these expressions, as it plays a central role in the emergence of hidden CP violation discussed in Sec.~\ref{subsec-hidden}.

\section{Comprehensive Scan}
\label{sec:scan}

In this section, we describe the comprehensive global scan performed to identify C2HDM parameter regions capable of supporting large CP violation.
We begin by detailing our scanning methodology, including the choice of input parameters and the rigorous set of theoretical and experimental constraints---most notably the eEDM---applied to ensure phenomenological viability.
We then characterize the intrinsic properties of the surviving parameter space, revealing that Type-I and Type-II models realize large CP violation through fundamentally different phenomenological patterns.
For Type-I, we identify a unique near-degenerate mass hierarchy that is essential for generating large CP violation in the gauge sector.
For Type-II, we show that large CP violation is restricted to the Yukawa sector of heavy scalars. 
\subsection{Parameter setup and scanning methodology}
\label{subsec-scan}

We investigate the parameter space of the C2HDM with a softly broken $Z_2$ symmetry.
The scalar potential in \autoref{eq:VPhi} depends on the parameters
$\{m_{11}^2,\, m_{22}^2,\, m_{12}^2,\, \lambda_{1,\cdots,5}\}$,
where $m_{12}^2$ and $\lambda_5$ can, in general, be complex.
A global $U(1)$ rephasing of the Higgs doublets allows us to remove one of these two phases,
leaving a single physical CP-violating phase $\arg(\lambda_5 (m_{12}^2)^{-2})$.
Consequently, the model is characterized by nine independent real parameters.

A physical basis for these nine input parameters consists of:
\beq
\label{eq-9param}
\{\, m_{H_a},~ m_{H_b},~ m_{H^\pm},~ \alpha_1,~ \alpha_2,~ \alpha_3,~
\tb,~ \text{Re}(m_{12}^2),~ v\,\},
\eeq
where $m_{H_a}$ and $m_{H_b}$ denote the masses of two neutral scalars.

To assess the theoretical consistency and phenomenological viability of the parameter points, we employ the public code \code{ScannerS~v2.0.0}~\cite{Coimbra:2013qq,Muhlleitner:2020wwk}, which fully implements the C2HDM. For efficient sampling, particularly to ensure consistency with Higgs precision data, \code{ScannerS} utilizes an alternative input parameterization focused on the properties of one neutral Higgs boson, identified here as the observed 125~GeV state ($H_a \equiv H_1$)~\cite{Muhlleitner:2020wwk}:
\begin{equation} \label{eq-param-C2HDM-ScannerS}
\{\, m_{H_a}=125\,\text{GeV},~ m_{H_b},~ m_{H^\pm},~ \tb,~ \text{Re}(m_{12}^2),~ c^2(H_a VV),~ |c(H_a t\bar{t})|^2,~ \mathrm{sign}(R_{a3}),~ R_{b3}\,\}.
\end{equation}
Here, $c(H_a VV)$ is the effective $H_aVV$ coupling normalized to the SM value (see \autoref{eq-ci-T}), and $|c(H_a t\bar{t})|^2 = (c^e(H_a tt))^2 + (c^o(H_a tt))^2$ parameterizes the coupling strength to top quarks.

We perform a random scan over the following parameter ranges:
\begin{equation} 
\label{eq:scan_ranges}
\begin{alignedat}{3} 
m_{H_b} &\in [130, 2000] \,\text{GeV}, &\quad
\mch &\in [90, 2000]\,\text{GeV}, &\quad \tb &\in [0.5, 50] ,\\
\text{Re}(m_{12}^2) &\in [0, 2\times 10^6]\,\text{GeV}^2, 
 &\quad \mathrm{sign}(R_{a3}) &\in \{\pm 1\}, &\quad
R_{b3} &\in [-1,1], \\
c^2(H_a VV) &\in [0.7,1.0], &\quad
 |c(H_a t\bar{t})|^2 &\in [0.6,1.6].
 \end{alignedat}
\end{equation}
%comment 4
We restrict our scan to the mass range below 2 TeV to focus on the parameter space most relevant for current and near-future collider searches, where the additional Higgs bosons remain within the direct reach of the High-Luminosity LHC and potential future Higgs factories.

Each generated point is subjected to a comprehensive set of constraints implemented within \code{ScannerS}, as outlined below:
\begin{description}
\item[(i) Theoretical requirements:]
We require the scalar potential to be bounded from below~\cite{Ivanov:2006yq}, ensure perturbative unitarity by constraining the eigenvalues of the $2\to2$ scalar scattering matrix to be less than $8\pi$, and enforce the stability of the electroweak vacuum against decay to deeper minima~\cite{Ivanov:2008cxa,Barroso:2012mj,Barroso:2013awa}.

\item[(ii) Electroweak and flavor constraints:] 
Parameter points must be consistent with electroweak precision data, specifically the oblique parameters ($S, T, U$)~\cite{Peskin:1991sw}, compatible with the global fit values from the Particle Data Group at 95\% C.L.\cite{ParticleDataGroup:2024cfk}. Constraints from flavor physics are applied at 95\% C.L., including bounds from $B \to X_s \gamma$, neutral meson mixing, and various rare $B$-meson decays ($B_{d,s}\to \mu^+\mu^-$, $B \to \tau\nu$, $B_s \to \phi \gamma$)~\cite{Haller:2018nnx,Arbey:2017gmh,Sanyal:2019xcp,Misiak:2017bgg,Belle:2017hum,Belle:2014sac}.

\item[(iii) Higgs collider data:] 
Agreement with Higgs data is evaluated using the \code{HiggsTools} framework (version~1.1.3)~\cite{Bahl:2022igd}, which comprises three sub-libraries: \code{HiggsPredictions}, \code{HiggsSignals}~\cite{Bechtle:2020uwn}, and \code{HiggsBounds}~\cite{Bechtle:2020pkv}.
\code{HiggsSignals} tests the compatibility of the 125~GeV Higgs boson ($H_1$) with LHC signal strength measurements, requiring the deviation $\Delta\chi^2 = \chi^2_{\text{C2HDM}} - \chi^2_{\text{SM}}$ to lie within the $2\sigma$ region of the SM best-fit value.
\code{HiggsBounds} checks that the predicted signal rates for additional Higgs bosons ($H_2$, $H_3$, $H^\pm$) are not excluded by direct searches at the LEP, Tevatron, or LHC.
Since the constraint on the CP-mixing angle $\alpha_{\tau}^{H_1}$, defined as $\alpha_{\tau}^{H_1} = \arctan(c^o (H_1\tau\tau) / c^e(H_1\tau\tau) )$, is not natively implemented in \code{ScannerS}, we explicitly impose the ATLAS bound derived from $\hsm  \to \tau^+\tau^-$ decays. We require consistency with the ATLAS measurement $\alpha_{\tau}^{H_1} = 9^\circ \pm 34^\circ$ (95\%~C.L.)~\cite{ATLAS:2022akr}.

\item[(iv) eEDM:] 
The predicted eEDM, which primarily arises from two-loop Barr--Zee diagrams~\cite{Barr:1990vd} and has been extensively studied in two-Higgs-doublet models~\cite{Pospelov:2005pr,Abe:2013qla}, must satisfy the stringent 90\% C.L. upper limit $|d_e| < 4.1\times10^{-30}\,e\cdot\mathrm{cm}$ set by the JILA collaboration~\cite{Roussy:2022cmp}.
\end{description}

In evaluating consistency with Higgs precision data, we also include near-degenerate scenarios where the $H_2$ mass is close to $125\,\text{GeV}$. To avoid numerical singularities associated with exact mass degeneracy, we impose the following mass conditions:
\begin{equation} \label{eq-mass-condition}
M_{H_1}=125\,\text{GeV},\quad M_{H_2} \geq 126\,\text{GeV}, \quad M_{H_3} \geq 130\,\text{GeV}.
\end{equation}
Through extensive random scans, we collected $1.1\times10^5$ viable points for Type-I and an equal number for Type-II, all of which satisfy the full set of constraints including the mass conditions in \autoref{eq-mass-condition}. The eEDM constraint proves to be the most restrictive factor; only about 1\% of the points satisfying all other requirements survive this bound in both Type-I and Type-II, thereby significantly shaping the allowed parameter space.

We now clarify the treatment of near-degenerate scenarios involving the observed state and the second neutral scalar ($M_{H_1} \approx M_{H_2}$) in \texttt{HiggsSignals}~\cite{Bechtle:2013xfa}, which requires careful consideration in the signal-strength evaluation.
\code{HiggsSignals} compares theoretical predictions to experimental measurements by summing the signal strengths ($\mu_i$) of all neutral scalars whose masses fall within the effective mass window of a given channel, $m_h \pm \Delta m$. 
However, the widths of these mass windows differ substantially between two categories of analysis channels.

The first category comprises the \emph{mass-sensitive} channels, namely the high-resolution $H\!\to\!\gamma\gamma$ and $H\!\to\!ZZ$ analyses based on Refs.~\cite{ParticleDataGroup:2020ssz,ATLAS:2016neq}. 
These channels employ exceptionally narrow mass windows, approximately $[124.74,125.76]\,\text{GeV}$ and $[124.37,125.81]\,\text{GeV}$, respectively.
\code{HiggsSignals} includes only the contributions from neutral Higgs bosons whose masses fall strictly within these windows.\footnote{%
This restriction is necessary because the experimental inputs for these channels~\cite{ParticleDataGroup:2020ssz,ATLAS:2016neq} are provided as signal-strength fits fixed at the 125~GeV peak, rather than as model-independent invariant-mass spectra. Consequently, a simultaneous two-state fit to overlapping $H_1$ and $H_2$ contributions is not feasible, making the use of predefined narrow windows the only statistically consistent approach.}
Our mass condition of $M_{H_2} \ge 126\,\text{GeV}$ ensures that $H_2$ lies \textbf{outside} these high-resolution windows. 
Consequently, $H_2$ does not contribute to the predicted signal strengths of these dominant precision channels, which drive the global $\chi^2$ fit.

In contrast, the remaining \emph{mass-insensitive} channels---covering almost all other Higgs decay modes such as $H\!\to\!WW$, $H\!\to\!\tau\tau$, $H\!\to\!b\bar{b}$, and $H\!\to\!\mu\mu$ (and lower-resolution measurements of $\gamma\gamma$ and $ZZ$)---employ substantially wider mass windows. %, typically $\Delta m \simeq \pm(5$--$20)\,\text{GeV}$. 
In these analyses, the second Higgs state $H_2$ often falls within the corresponding mass window, and \code{HiggsSignals} appropriately accounts for its presence by summing its contribution to the total predicted signal strength:
\beq
\mu_i^{\rm pred} = \mu_i(H_1) + \mu_i(H_2)\;.
\eeq

This distinction is central to the viability of the near-degenerate scenario. 
The overall $\chi^2$ is dominated by the high-resolution channels, which possess much smaller experimental uncertainties $\delta\mu_i^{\text{obs}}$ and thus carry the greatest statistical weight in the global fit~\cite{Bechtle:2020uwn}.\footnote{%
Note that the full $\chi^2$ evaluation in \code{HiggsSignals} uses the covariance matrix of correlated experimental and theoretical uncertainties~\cite{Bechtle:2020uwn}.} 
By construction, $H_2$ with $M_{H_2}>126\,\text{GeV}$ is excluded from the dominant, mass-sensitive channels~\cite{ParticleDataGroup:2020ssz,ATLAS:2016neq}. Consequently, although it contributes to the less precise, mass-insensitive channels, its impact on the global $\chi^2$ remains marginal.
As shown in the following analysis, this treatment allows near-degenerate scenarios to remain phenomenologically viable, particularly within the Type-I C2HDM.

\subsection{Characteristics of the large CP violation parameter space in Type-I}
\label{subsec-characteristics-Type-I}

Having detailed our random scanning methodology in the previous subsection, we now analyze the intrinsic characteristics of the viable parameter points for Type-I. Our analysis places particular focus on regions exhibiting sizable CP violation, as quantified by the gauge and Yukawa CPV measures, $\xiv$ and $\ztt $.

 %------------------------
\begin{figure}[t]
\centering
\includegraphics[width=\textwidth]{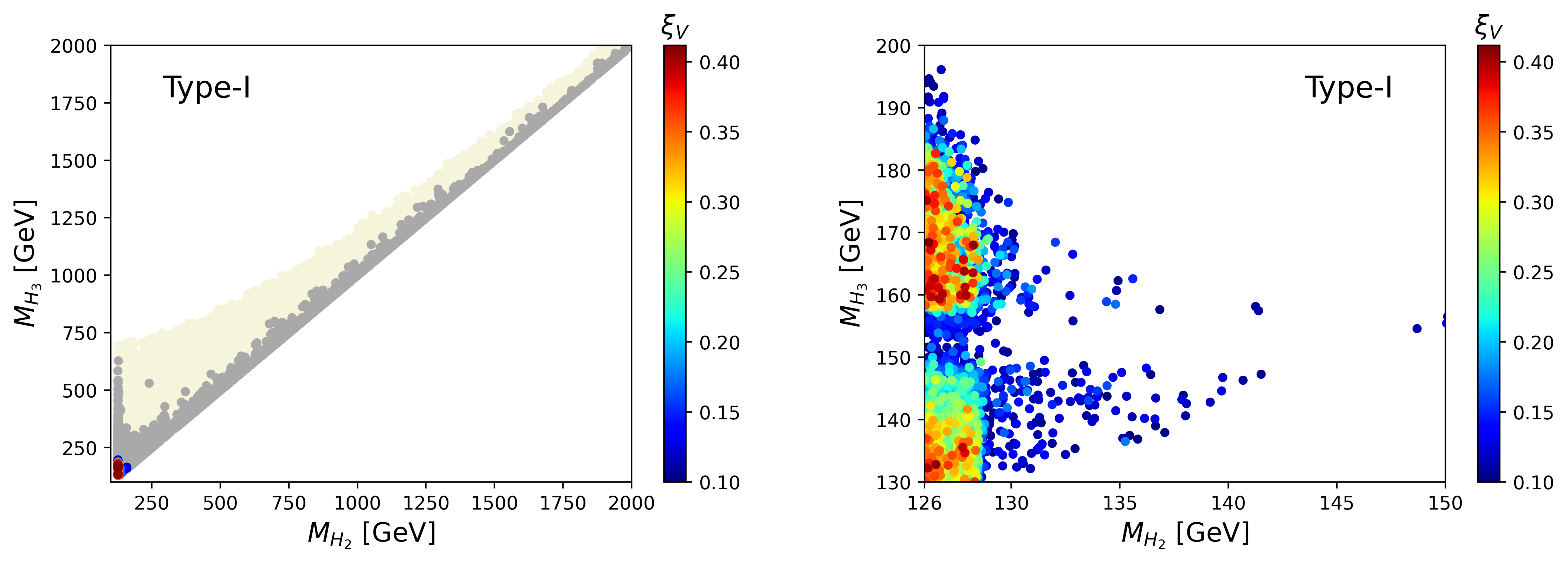}
\caption{%
Viable parameter points in the $(M_{H_2}, M_{H_3})$ plane for the Type-I C2HDM,
with the color code of $\xiv$. Beige points satisfy theoretical requirements and constraints from electroweak precision data, flavor physics, and Higgs collider searches. Gray points represent the subset additionally satisfying the eEDM bound. Colored points highlight the region with large CP violation ($\xiv > 0.1$), layered by the value of $\xiv$, with higher values plotted on top. The right panel provides a magnified view focused on the large $\xiv$ region.
}
\label{fig-type1-xiv-mh3-mh2}
\end{figure}
%------------------------

We begin by mapping the viable Type-I parameter space in the $(M_{H_2}, M_{H_3})$ plane, as illustrated in \autoref{fig-type1-xiv-mh3-mh2}.
The color coding reflects the hierarchy of constraints: beige points satisfy standard theoretical and experimental bounds (constraints (i)--(iii)), while gray points represent the subset that additionally satisfies the eEDM bound (constraints (i)--(iv)). The profound impact of the eEDM constraint is immediately apparent; the EDM bound excludes approximately 99\% of the parameter points that were otherwise allowed by collider and flavor physics. Interestingly, the surviving (gray) points exhibit a preference for near-degenerate $H_2$ and $H_3$ masses when $M_{H_2} \gtrsim 250\,\text{GeV}$, and permit the decoupling limit where the BSM scalar masses can reach up to the 2~TeV scan limit.

To systematically explore the phenomenology of large CP violation, we introduce the benchmark condition $\xiv > 0.1$, which is highlighted by the colored points in \autoref{fig-type1-xiv-mh3-mh2}.
While numerically arbitrary, this threshold serves as a practical tool to isolate regions with non-negligible CP-violating effects.
Furthermore, this criterion underscores a fundamental difference between Type-I and Type-II: $\xiv$ can reach values up to $\sim 0.4$ in Type-I but remains suppressed below $\mathcal{O}(10^{-3})$ in Type-II.
Quantitatively, $\xiv=0.1$ corresponds to a geometric mean of the gauge couplings $\left[ \prod_i c(H_i VV)^2 \right]^{1/3} \approx 0.155$, representing a physically meaningful level of CP mixing in the gauge sector.

Remarkably, requiring large gauge-sector CP violation ($\xiv > 0.1$) imposes severe restrictions on the masses of the heavy neutral scalars.
Imposing the large CPV condition $\xiv > 0.1$ selects approximately 8.8\% of the viable (gray) parameter points.
However, these surviving points are not distributed uniformly; rather, they are strictly confined to the low-mass region, as evident from the left panel of \autoref{fig-type1-xiv-mh3-mh2}.
The magnified view in the right panel reveals that $\xiv > 0.1$ requires $M_{H_2} \lesssim 159 \,\text{GeV}$ and $M_{H_3} \lesssim 197\,\text{GeV}$.

%------------------------
\begin{figure}[t]
\centering
\includegraphics[width=0.6\textwidth]{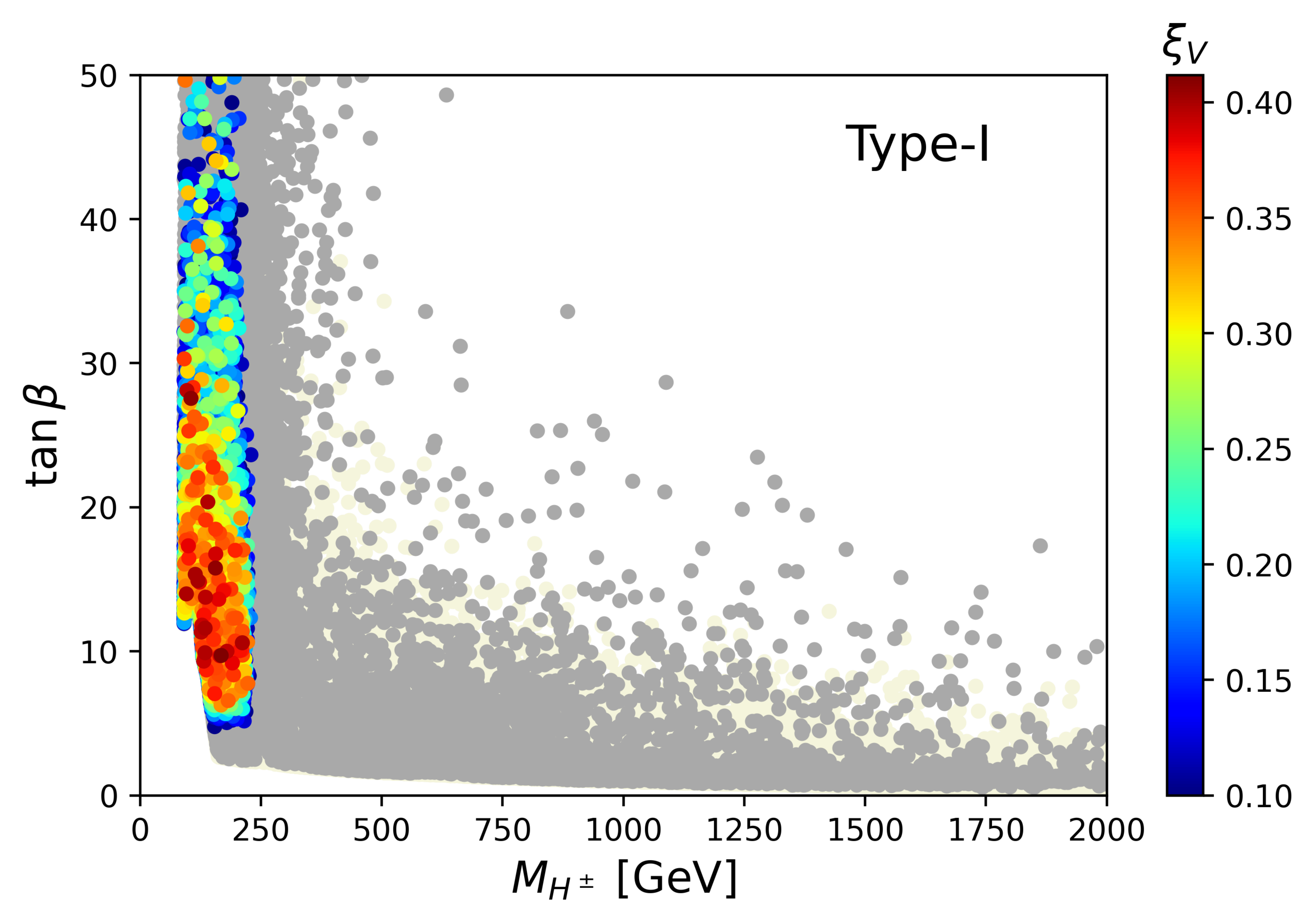}
\caption{%
Viable parameter points in the $(\mch,\,\tb)$ plane for the Type-I C2HDM.
The color bar indicates the CPV measure $\xiv$.
The beige, gray, and colored points follow the same convention as in \autoref{fig-type1-xiv-mh3-mh2}.
}
\label{fig-type1-xiv-tb-mch}
\end{figure}
%------------------------

Turning to the charged Higgs sector, \autoref{fig-type1-xiv-tb-mch} maps the viable parameter points in the $(\mch, \tb)$ plane.
Unlike the neutral scalar masses, $\mch$ and $\tb$ are not significantly restricted by EDM measurements alone, as evidenced by the considerable overlap between the beige and gray points. However, imposing the large CPV condition $\xiv > 0.1$ significantly narrows the allowed space to $\tb > 5.1$ and $\mch < 229.4 \,\text{GeV}$. 
It is encouraging that large CPV measures systematically favor low $\mch$, that is readily accessible to direct searches at high-energy colliders.

%------------------------
\begin{figure}[h]
\centering
\includegraphics[width=\textwidth]{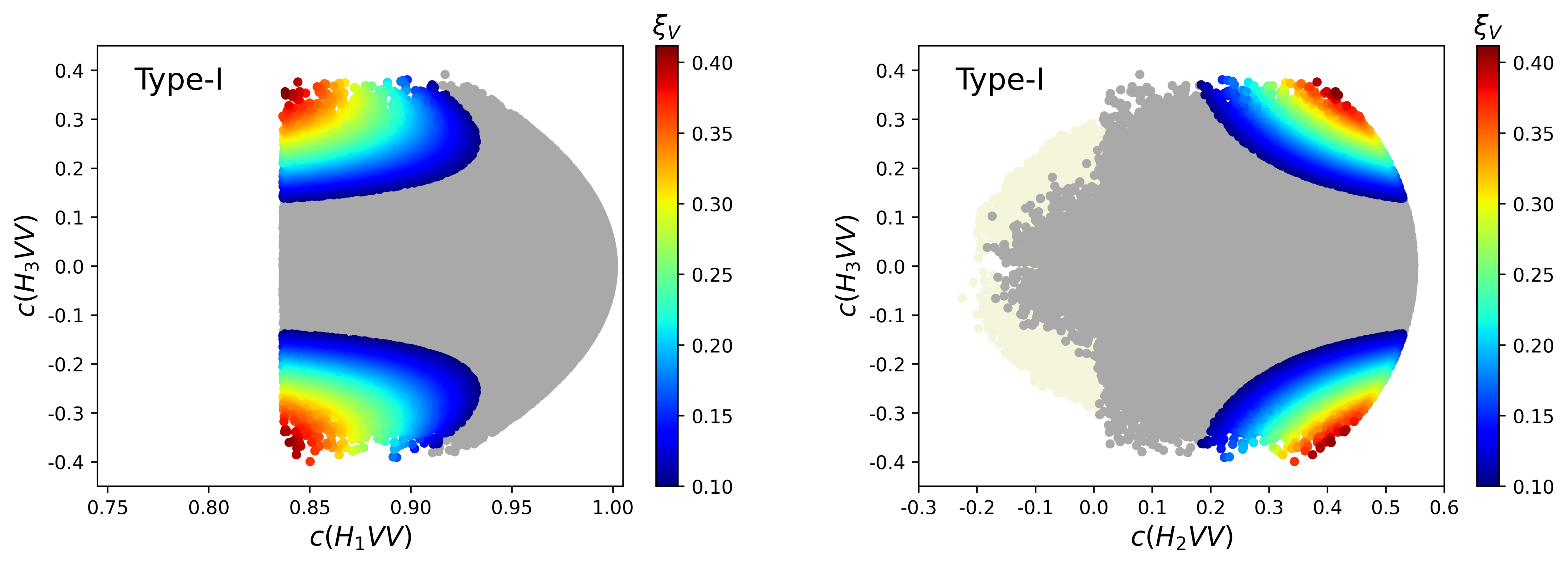}
\caption{%
Viable parameter points in the $(c(H_1VV), c(H_3 VV))$ plane (left)
and in the $(c(H_2VV), c(H_3 VV))$ plane (right) for the Type-I C2HDM. 
The beige, gray, and colored points follow the convention from \autoref{fig-type1-xiv-mh3-mh2}, 
where the color bar indicates the value of the CP-violating measure $\xiv$.
}
\label{fig-type1-xiv-CHVV}
\end{figure}
%------------------------

Having identified the specific mass regions favored by large CP violation, we now examine the underlying correlations among the individual Higgs-vector-boson couplings, $c(H_iVV)$, which directly dictate the collider phenomenology.
In \autoref{fig-type1-xiv-CHVV}, we present the correlations of $c(H_3 VV)$ with $c(H_1 VV)$ (left panel) and with $c(H_2 VV)$ (right panel).

In the left panel, we observe that the viable (gray) points for $c(H_3 VV)$ are distributed broadly and symmetrically around zero, while $c(H_1 VV)$ ranges between approximately 0.84 and 1, consistent with an SM-like 125 GeV Higgs. 
The Higgs alignment limit, where $c(H_1 VV)=1$ and $c(H_3 VV)=0$, remains viable, corresponding to the decoupling of $H_3$ from the gauge sector.
However, the large $\xiv$ region (colored points) necessitates a sizable deviation from alignment: specifically, $|c(H_3 VV)|\gtrsim 0.14$ and $c(H_1VV) \lesssim 0.935$. Furthermore, as $\xiv$ increases, both $|c(H_3 VV)|$ and $|1- c(H_1VV)|$ increase. Consequently, future precision measurements of the Higgs couplings will critically test this scenario; for example, if $c(H_1VV)$ is constrained to be $> 0.95$, the CPV measure $\xiv$ cannot exceed 0.1.

The right panel of \autoref{fig-type1-xiv-CHVV} illustrates the correlation between $c(H_3 VV)$ and $c(H_2 VV)$.
In contrast to the symmetric distribution of $c(H_3 VV)$, the distribution of $c(H_2 VV)$ is notably asymmetric, exhibiting a clear preference for positive values.
Crucially, in the parameter space with large $\xiv$, only $c(H_2 VV) \gtrsim 0.2$ is permitted.
We also observe that $\xiv$ increases concurrently with both $c(H_2 VV)$ and $|c(H_3 VV)|$, confirming that $\xiv$ serves as a robust representative measure for CP violation in the gauge sector.

%------------------------
\begin{figure}[t]
\centering
\includegraphics[width=\textwidth]{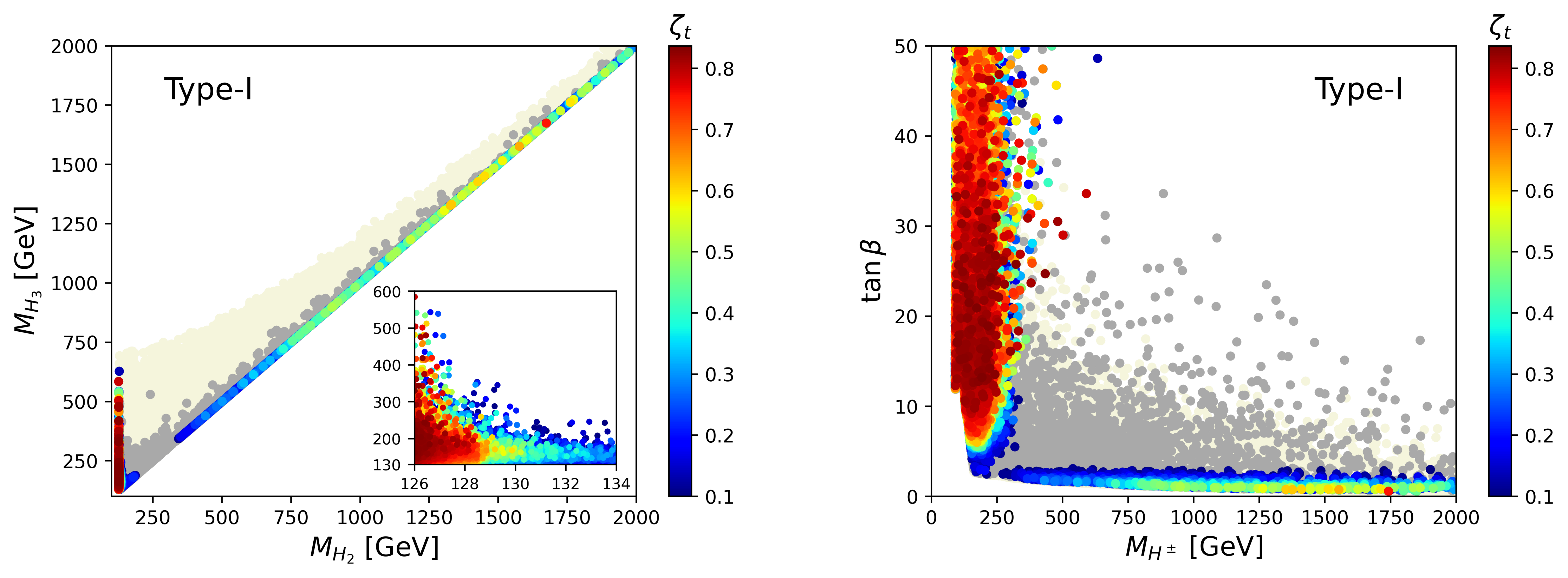}
\caption{%
Viable parameter points in the $(M_{H_2},\,M_{H_3})$ plane (left) and the $(\mch,\,\tb)$ plane (right) for the Type-I C2HDM.
The color bar indicates the value of the CP-violating measure $\ztt $.
The beige and gray points follow the convention from \autoref{fig-type1-xiv-mh3-mh2},
while the colored points represent the subset with $\ztt  > 0.1$.
The inset in the left panel provides a magnified view of the low-mass region.
}
\label{fig-type1-zetat}
\end{figure}

We next shift our focus to CP violation in the Yukawa sector, quantified by $\ztt $.
\autoref{fig-type1-zetat} displays the distribution of viable points in the $(M_{H_2},\,M_{H_3})$ and $(\mch,\,\tb)$ planes, color-coded by $\ztt $.
Similar to the gauge CPV case, the subset with $\ztt  > 0.1$ constitutes roughly 10\% of the viable parameter space. However, a key distinction is that $\ztt $ can reach significantly higher values with $\max (\ztt ) \approx 0.836$, compared to $\xiv$. This difference arises because $\ztt $ is a sum variable, whereas $\xiv$ is multiplicative; for instance, if $H_3$ is purely CP-odd, $\xiv$ vanishes, yet $\ztt $ remains non-zero provided $\beta \neq \alpha_1$.

Furthermore, $\ztt  > 0.1$ allows for the decoupling limit where $H_2$ and $H_3$ are heavy and nearly degenerate (up to the 2~TeV scan limit)---a region strictly excluded by the $\xiv > 0.1$ condition. Nevertheless, the highest $\ztt $ values still correlate with lighter BSM masses. The inset in the left panel of \autoref{fig-type1-zetat} clearly shows that $\ztt  > 0.6$ requires $M_{H_2} < 129\,\text{GeV}$, again confining the model to a scenario where $H_2$ is nearly degenerate with $H_1$.

The right panel of \autoref{fig-type1-zetat} reveals two distinct clusters in the $(\mch,\,\tb)$ plane for large $\ztt $.
The first cluster features a lighter charged Higgs ($\mch \lesssim 250\,\text{GeV}$) and spans a broad range of $\tb \approx 2$--$50$, whereas the second cluster extends to the heavy $\mch$ limit (up to 2~TeV) but is realized only at low $\tb$ ($\tb \lesssim 3$).
This latter restriction arises because the Type-I Yukawa couplings scale approximately as $1/\tb$; consequently, small $\tb$ values are necessary to prevent strong suppression of the couplings in the heavy mass regime.
Notably, the most extreme CPV values ($\ztt  > 0.8$) are predominantly found within the lighter charged Higgs cluster.

%------------------------
\begin{figure}[t]
\centering
\includegraphics[width=\textwidth]{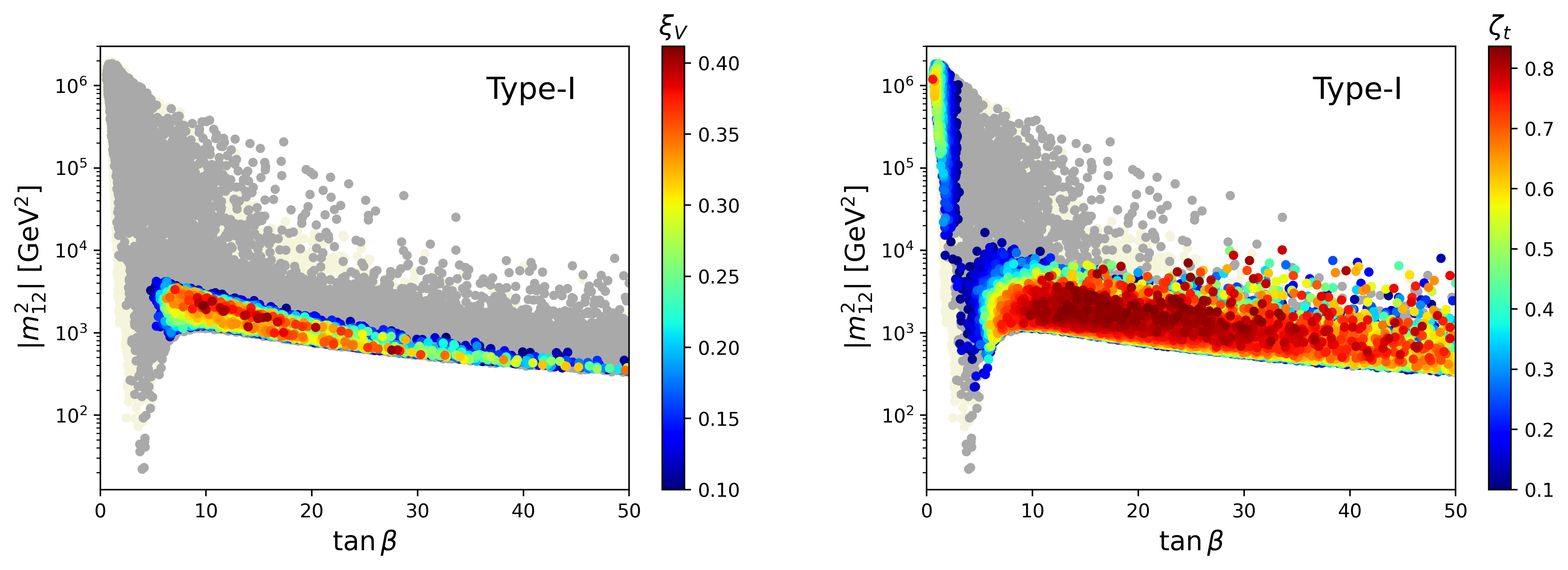}
\caption{%
Viable parameter points in the $(\tb,\,|m_{12}^2|)$ plane for the Type-I C2HDM.
The color bar indicates the value of the CPV measure $\xiv$ (left panel)
and $\ztt $ (right panel).
The beige and gray points follow the convention from \autoref{fig-type1-xiv-mh3-mh2},
while the colored points represent the subset with $\xiv > 0.1$ (left) and $\ztt  > 0.1$ (right).
}
\label{fig-type1-m12sq-tb}
\end{figure}
%------------------------

Finally, we address the implications of large CP violation for the soft $Z_2$ breaking scale.
\autoref{fig-type1-m12sq-tb} illustrates the viable points in the $(\tb,\,|m_{12}^2|)$ plane. Even within the general viable space (gray points), $|m_{12}^2|$ deviates significantly from zero, confirming that soft $Z_2$ breaking is essential. The lower bound is approximately $|m_{12}^2|_{\min} \approx 53\,\text{GeV}^2$ at $\tb \simeq 4$, rising as $\tb$ deviates from this value. Conversely, the upper bound decreases with $\tb$, reaching $\mathcal{O}(10^6)\,\text{GeV}^2$ for $\tb \lesssim 1$.

Remarkably, requiring large CP violation imposes much tighter and distinct restrictions on $|m_{12}^2|$.
As shown in the left panel, the condition $\xiv > 0.1$ favors a narrow band characterized by $|m_{12}^2| \sim \mathcal{O}(10^3)\,\text{GeV}^2$.
In contrast, the condition $\ztt  > 0.1$ (right panel) permits two separate regions: a vertical band at low $\tb$ with $|m_{12}^2| \sim \mathcal{O}(10^4\text{--}10^6)\,\text{GeV}^2$ (corresponding to heavier $H_2$), and a horizontal band at larger $\tb$ with $|m_{12}^2| \sim \mathcal{O}(10^3)\,\text{GeV}^2$ (corresponding to lighter $H_2$).
This distinction demonstrates that the preferred scale of $|m_{12}^2|$ depends critically on whether the CP violation originates in the gauge or Yukawa sector.

In conclusion, we summarize the parameter ranges identifying large CPV regions in Type-I.
For large gauge CPV ($\xiv > 0.1$), the parameters are confined to:
\begin{align}
 \text{Type-I with } \xiv > 0.1: \quad
 &M_{H_2} \in [126.0, 148.7] \,\text{GeV}, \quad M_{H_3} \in [130.0, 194.7]\,\text{GeV}, \nonumber \\
 & \mch \in [90.9, 229.4]\,\text{GeV}, \nonumber \\
 & |m_{12}^2| \in [363.2, 4202.4] \,\text{GeV}^2, \quad \tb \in [5.1, 49.7].
\end{align}
For large Yukawa CPV ($\ztt  > 0.1$), the allowed ranges bifurcate depending on the mass of $H_2$:
\begin{align}
 \text{Type-I with }& \ztt  > 0.1: \quad
 \\
 & \text{Case 1 (Light $H_2$):} \nonumber \\
 & \quad M_{H_2} \in [126.0, 182.9] \,\text{GeV}, \quad M_{H_3} \in [130.0, 543.1]\,\text{GeV}, \nonumber \\
 & \quad \mch \in [90.9, 482.4] \,\text{GeV}, \quad |m_{12}^2| \in [219.0, 10991.1] \,\text{GeV}^2, \nonumber \\
 & \quad \tb \in [2.89, 50.0]. \nonumber \\[5pt]
 & \text{Case 2 (Heavy $H_2$):} \nonumber \\
 & \quad M_{H_2}, M_{H_3} \in [346.0, 1992.9]\,\text{GeV}, \nonumber \\
 & \quad \mch \in [285.4, 1991.9]\,\text{GeV}, \quad |m_{12}^2| \in [1.3\times 10^4, 1.8 \times 10^6]\,\text{GeV}^2, \nonumber \\
 & \quad \tb \in [0.58, 3.07].
\end{align}
We emphasize that parameter points are not uniformly distributed across these ranges, as evident from the scatter plots presented in this subsection.

\subsection{Characteristics of the large CP violation parameter space in Type-II}
\label{subsec-characteristics-Type-II}

We now turn our attention to the characteristics of the large CPV parameter space in Type-II.
In sharp contrast to Type-I, the gauge-sector CPV measure $\xiv$ in Type-II is extremely suppressed:
\beq
\label{eq-type2-xiv}
\text{Type-II: } \quad \xiv < 3.28\times 10^{-3}.
\eeq
However, the Yukawa sector in Type-II is richer, featuring two distinct CPV measures: $\ztt$ for the top quark and $\ztb$ for the bottom quark and $\tau$ lepton (see \autoref{eq-zeta-tbtau}).  Consequently, we focus our analysis on the viable parameter space satisfying $\ztt > 0.1$ and $\ztb > 0.1$.

%------------------------
\begin{figure}[h]
\centering
\includegraphics[width=\textwidth]{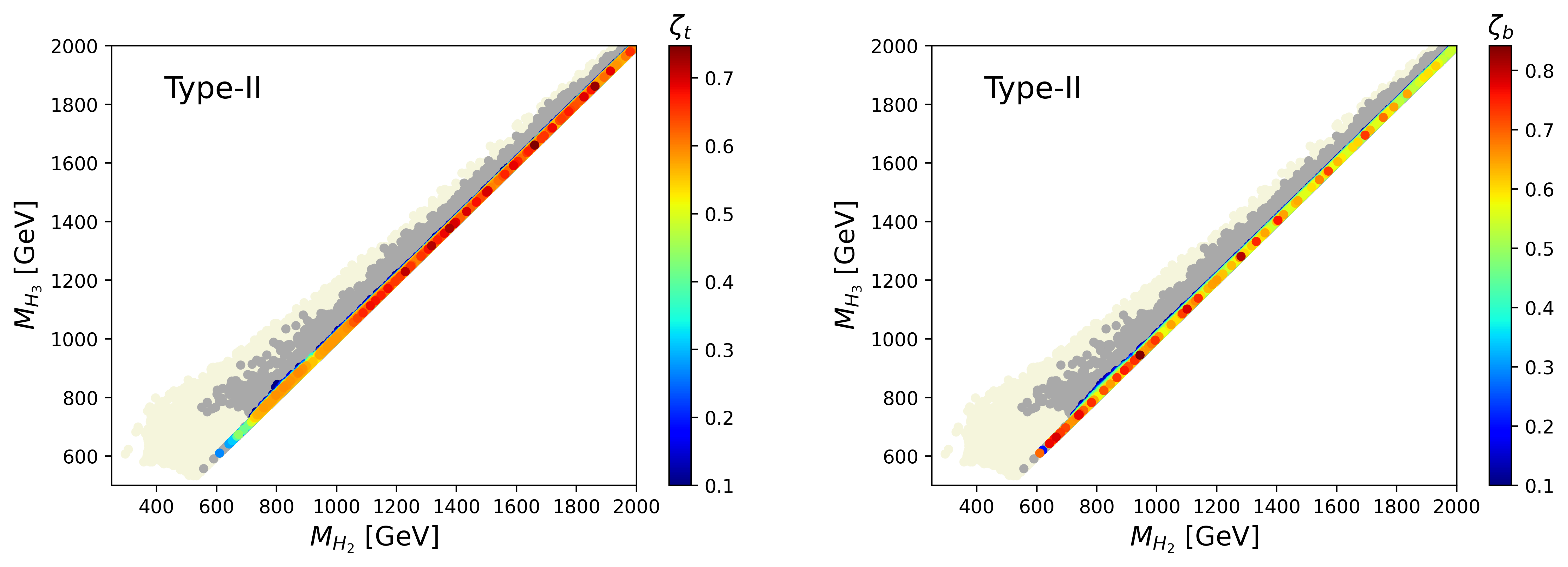}
\caption{%
Viable parameter points in the $(M_{H_2},M_{H_3})$ plane for the Type-II C2HDM.
The color bar indicates the value of the CP-violating measure $\ztt$ (left panel)
and $\ztb$ (right panel).
The beige and gray points follow the convention from \autoref{fig-type1-xiv-mh3-mh2},
while the colored points represent the subset with $\ztt > 0.1$ (left) and $\ztb > 0.1$ (right).
}
\label{fig-type2-zeta-mh3-mh2}
\end{figure}
%------------------------

\autoref{fig-type2-zeta-mh3-mh2} displays the viable parameter points in the $(M_{H_2},M_{H_3})$ plane, color-coded by $\ztt$ (left) and $\ztb$ (right).
The beige and gray points adhere to the convention established in \autoref{fig-type1-xiv-mh3-mh2}.
Notably, unlike the suppressed gauge measure, the Yukawa CPV measures in Type-II can be quite large, reaching maximum values of:
\beq
\text{Type-II: } \quad \max (\ztt) \approx 0.747, \quad
\max (\ztb) \approx 0.842.
\eeq

A striking difference from Type-I is immediately visible in the distribution of the beige points:
Type-II requires both $H_2$ and $H_3$ to be significantly heavier than 125~GeV.
This restriction stems primarily from the charged Higgs sector. Strong constraints from flavor physics, particularly the $B \to X_s \gamma$ measurement, impose a lower bound on the charged Higgs mass of $\mch \gtrsim 580\,\text{GeV}$~\cite{Haller:2018nnx,Misiak:2020vlo,Biekotter:2024ykp,Biekotter:2025fjx}.
To satisfy electroweak precision constraints (specifically the oblique parameters $S, T, U$~\cite{Peskin:1991sw}), the neutral BSM scalars must effectively track the mass of this heavy charged Higgs, forcing them to be heavy as well.
While the eEDM constraint (gray points) excludes a significant portion of the beige points, the excluded region is noticeably smaller than in Type-I (compare with \autoref{fig-type1-xiv-mh3-mh2}).

The requirement of large CP violation ($\ztt > 0.1$ or $\ztb > 0.1$) further constrains the mass spectrum.
As shown by the colored points in \autoref{fig-type2-zeta-mh3-mh2}, large Yukawa CP violation enforces a strong mass degeneracy between $H_2$ and $H_3$, in contrast to the broader distribution of the gray points.
However, the absolute mass bounds remain largely unchanged compared to the general viable parameter space: both $M_{H_2}$ and $M_{H_3}$ extend from approximately $580\,\text{GeV}$ up to our 2~TeV scanning limit.

%------------------------
\begin{figure}[h]
\centering
\includegraphics[width=\textwidth]{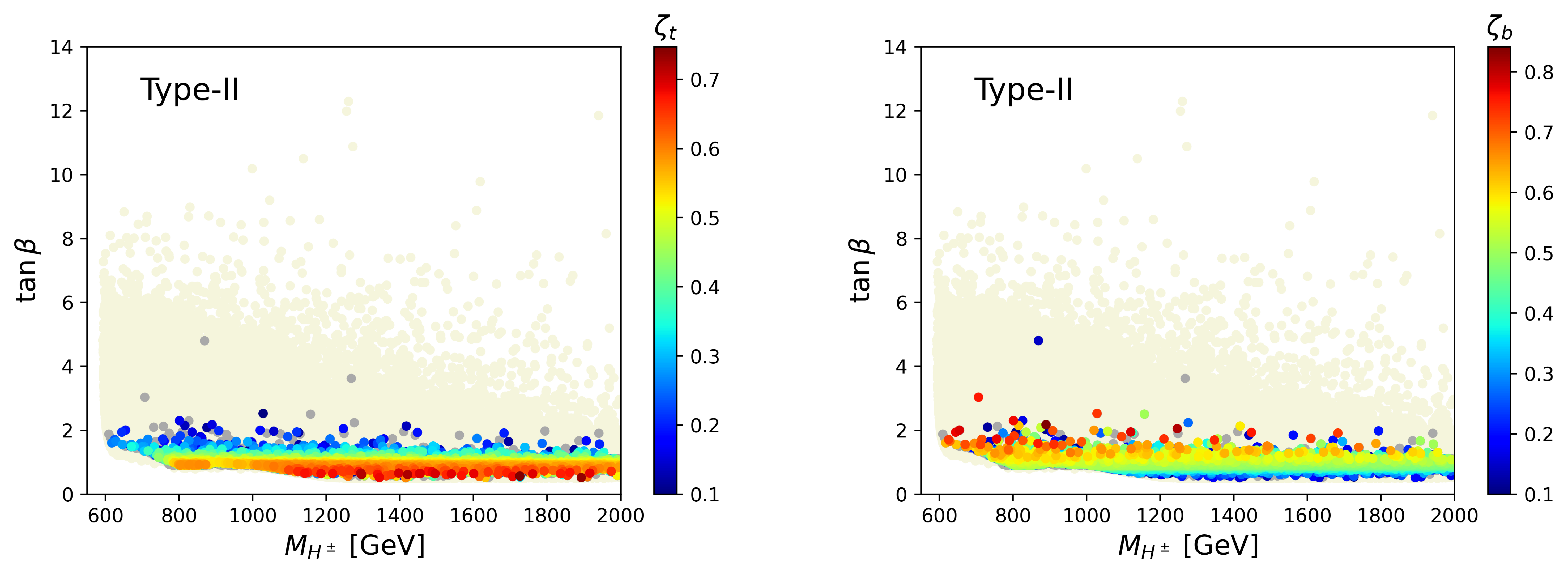}
\caption{%
Viable parameter points in the $(\mch,\tb)$ plane for the Type-II C2HDM.
The color bar indicates the value of the CP-violating measure $\ztt$ (left panel)
and $\ztb$ (right panel).
The beige and gray points follow the convention from \autoref{fig-type1-xiv-mh3-mh2},
while the colored points represent the subset with $\ztt > 0.1$ (left) and $\ztb > 0.1$ (right).
}
\label{fig-type2-zeta-tb-mch}
\end{figure}
%------------------------

We next examine the parameter space in the $(\mch,\tb)$ plane, presented in \autoref{fig-type2-zeta-tb-mch}.
A comparison between the beige and gray points reveals a crucial feature: the eEDM measurements impose a stringent upper limit of $\tb
\lesssim 4$.
This constitutes a dramatic tightening of the parameter space relative to earlier constraints.
For example, under the previous ACME~I limit of $|d_e| < 9.3 \times10^{-29}\,e\cdot\mathrm{cm}$ (95\% C.L.)~\cite{ACME:2013pal}, $\tb$ values as high as $\sim 30$ were permitted~\cite{Fontes:2017zfn}.

The impact of large CP violation on $\tb$ depends on the specific flavor sector involved.
The condition $\ztt > 0.1$ (left panel) moderately restricts the allowed range to $\tb \in [0.52, 2.18]$ and exhibits a clear correlation where larger $\ztt$ values favor smaller $\tb$.
Crucially, this range ensures that the Yukawa couplings of the BSM Higgs bosons to top quarks are either enhanced or, at minimum, remain unsuppressed.
This has significant implications for processes such as four-top-quark production~\cite{ATLAS:2020hpj,Aoude:2022deh,vanBeekveld:2022hty,CMS:2023ftu,ATLAS:2023ajo}.
Conversely, the condition $\ztb > 0.1$ (right panel) allows for a $\tb$ range similar to the general allowed (gray) points but exhibits an opposing correlation, where larger $\ztb$ values prefer larger $\tb$.
Notably, since the parameter space satisfying either $\ztt > 0.1$ or $\ztb > 0.1$ excludes the high $\tb$ regime, the heavy BSM Higgs bosons in Type-II do not benefit from the $\tb$-enhancement of their Yukawa couplings to bottom quarks and $\tau$ leptons.

Regarding the charged Higgs mass, \autoref{fig-type2-zeta-tb-mch} confirms that demanding large CP violation does not significantly alter the lower bound, which increases only mildly to $\mch \approx 610\,\text{GeV}$ compared to the flavor-physics floor of $\sim 580\,\text{GeV}$.
Both the general allowed region and the large CPV subsets extend to the 2~TeV limit, confirming that the decoupling limit remains fully compatible with all constraints in Type-II.

%------------------------
\begin{figure}[t]
\centering
\includegraphics[width=\textwidth]{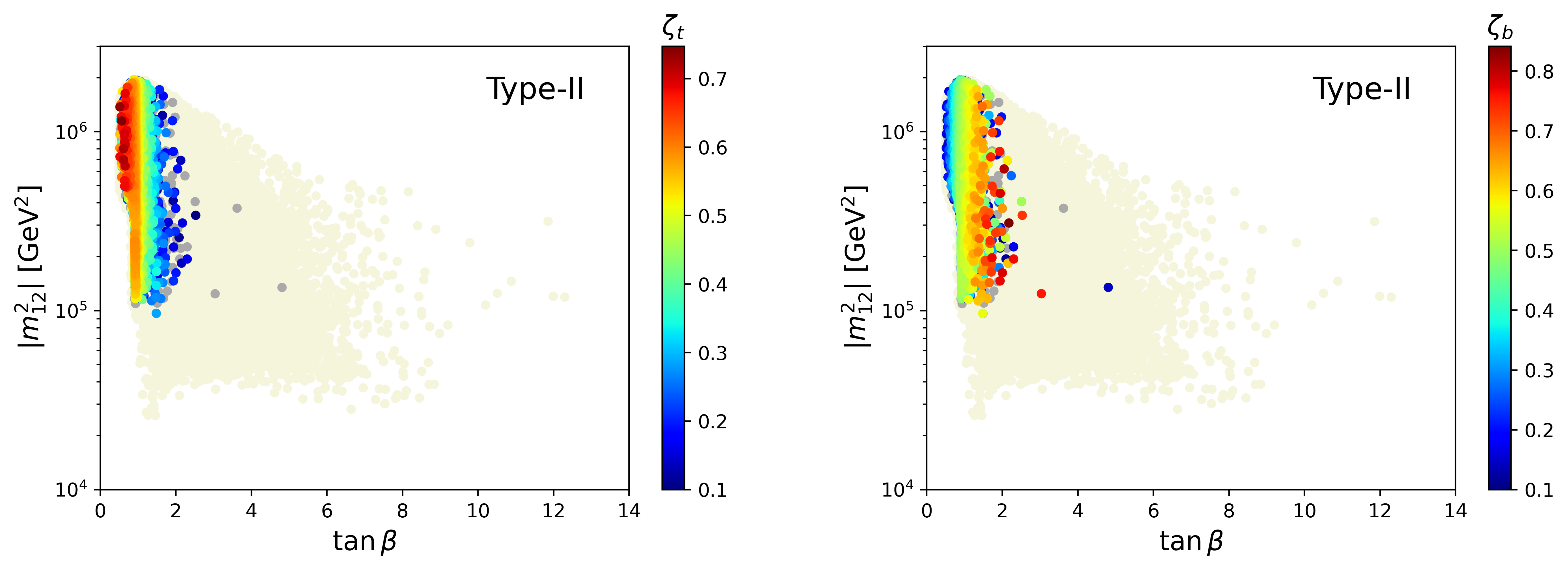}
\caption{%
Viable parameter points in the $(\tb,\,|m_{12}^2|)$ plane for the Type-II C2HDM.
The color bar indicates the value of the CP-violating measure $\ztt$ (left panel)
and $\ztb$ (right panel).
The beige and gray points follow the convention from \autoref{fig-type1-xiv-mh3-mh2},
while the colored points represent the subset with $\ztt > 0.1$ (left) and $\ztb > 0.1$ (right).
}
\label{fig-type2-m12sq-tb}
\end{figure}

%------------------------

Finally, we investigate the soft $Z_2$ breaking scale in the Type-II C2HDM by presenting the viable parameter points in the $(\tb,\,|m_{12}^2|)$ plane, shown in \autoref{fig-type2-m12sq-tb}.
The color bars indicate the values of the CPV measures $\ztt$ (left panel) and $\ztb$ (right panel).
The beige and gray points follow the convention established in \autoref{fig-type1-xiv-mh3-mh2}, while the colored points represent the subsets satisfying $\ztt > 0.1$ (left) and $\ztb > 0.1$ (right).

Similar to the Type-I case (see \autoref{fig-type1-m12sq-tb}), the viable parameter points in Type-II necessitate a non-zero soft breaking term $m_{12}^2$, thereby prohibiting an exact $Z_2$ symmetry.
However, we observe two distinct differences compared to Type-I.
First, the allowed values of $|m_{12}^2|$ are considerably larger in Type-II. This enhancement is driven by the requirement for heavier BSM Higgs bosons in this scenario ($M_{H_2}, M_{H_3} \gtrsim 610\,\text{GeV}$), which naturally scales up the soft breaking term.
Second, the eEDM constraint (gray points) significantly raises the lower bound on $|m_{12}^2|$ compared to the theoretical and flavor constraints alone (beige points).
This elevated lower bound remains robust even under the large CP violation conditions of $\ztt > 0.1$ or $\ztb > 0.1$.
Furthermore, unlike in Type-I, we observe no strong correlation between the magnitude of the CPV measures ($\ztt$ or $\ztb$) and the scale of $|m_{12}^2|$.

In summary, the parameter regions supporting large Yukawa CP violation in Type-II are defined as follows.
For large top-quark Yukawa CPV ($\ztt > 0.1$):
\begin{align}
\label{eq:typeII-ztt-summary}
 \text{Type-II with } \ztt > 0.1: \quad
 &M_{H_2} \in [610.0, 2000]\,\text{GeV}, \quad M_{H_3} \in [610.3, 2000]\,\text{GeV}, \nonumber \\
 & \mch \in [617.3, 2000]\,\text{GeV}, \nonumber \\
 & |m_{12}^2| \in [9.7\times 10^4, 2.0\times 10^6]\,\text{GeV}^2, \quad \tb \in [0.52, 2.53].
\end{align}
For large bottom/tau Yukawa CPV ($\ztb > 0.1$):
\begin{align}
\label{eq:typeII-ztbt-summary}
 \text{Type-II with } \ztb > 0.1: \quad
 &M_{H_2} \in [610.3, 2000]\,\text{GeV}, \quad M_{H_3} \in [610.3, 2000]\,\text{GeV}, \nonumber \\
 & \mch \in [617.3, 2000]\,\text{GeV}, \nonumber \\
 & |m_{12}^2| \in [9.7\times 10^4, 2.0\times 10^6]\,\text{GeV}^2, \quad \tb \in [0.52, 4.8].
\end{align}
As with Type-I, the parameter distribution across these ranges is highly non-uniform, as evident from the corresponding figures.

\section{Future prospects and observability of large CPV}
\label{sec:results}

In the previous section, we delineated the intrinsic characteristics of the C2HDM parameter space capable of supporting large CP violation.
We now turn to the phenomenological consequences and future prospects of these regions.
We begin by evaluating the predicted eEDM values for both Type-I and Type-II, comparing them against the projected sensitivities of next-generation experiments.
Having established these indirect constraints, we then examine the distinct collider signatures associated with the large CP-violation regimes in each model.
For Type-I, our analysis centers on the gauge sector, where large $\xiv$ necessitates a near-degenerate scenario between $H_2$ and the 125~GeV $H_1$; here, we demonstrate how the simultaneous contributions of these two states reproduce the observed Higgs signal strengths.
Conversely, for Type-II, we focus on the Yukawa sector, investigating the magnitude and observability of the CP-violating phases in the top-quark and $\tau$-lepton couplings.
Finally, we investigate the phenomenon of \textit{hidden} CP violation near the Higgs alignment limit, where CP mixing persists in the heavy sector despite vanishing signatures in standard gauge interactions.

\subsection{Prospect of the eEDM in the large CP violation parameter space}

The search for the eEDM has reached an unprecedented level of sensitivity, providing a powerful indirect probe of new physics. 
Currently, the most stringent upper limit on the eEDM is set by the JILA experiment using trapped $\mathrm{HfF}^+$ ions, $|d_e| < 4.1 \times 10^{-30}\,e\cdot\mathrm{cm}$ at 90\% confidence level~\cite{Roussy:2022cmp}. 
This bound is notably more restrictive than the previous benchmark set by the ACME~II collaboration ($|d_e| < 1.1 \times 10^{-29}\,e\cdot\mathrm{cm}$~\cite{ACME:2018yjb}), placing severe pressure on the parameter space of CP-violating extensions of the Standard Model like the C2HDM.

Looking ahead, the experimental frontier is poised to improve sensitivity by another one to two orders of magnitude. 
The ACME~III collaboration targets the $10^{-31}\,e\cdot\mathrm{cm}$ regime using a cryogenic beam of Thorium Monoxide (ThO)~\cite{ACME:2018yjb}, while the next-generation JILA experiment aims for a similar precision of a few times $10^{-31}\,e\cdot\mathrm{cm}$~\cite{Roussy:2022cmp}. 
Complementing these efforts, the NL-eEDM experiment utilizes Barium Fluoride (BaF) to provide a systematic cross-check with a projected sensitivity between $10^{-31}$ and $10^{-32}\,e\cdot\mathrm{cm}$~\cite{NL-eEDM:2018lno}. 
Achieving these goals will extend the reach of eEDM measurements far beyond the direct search capabilities of the LHC~\cite{Chupp:2014gka,Chupp:2017rkp,EuropwanEDMprojects:2025okn,Pospelov:2025vzj}.

Given the current experimental constraints and future prospects, the following questions arise: (i) what underlying mechanisms allow a sizable parameter space to persist despite the stringent JILA eEDM constraint; (ii) what values of $|d_e|$ are predicted across the viable C2HDM parameter space, particularly in regions of maximal CP violation; and (iii) how does $|d_e|$ correlate with the sensitivity of future Higgs precision measurements?

%------------------------
\begin{figure}[t]
\centering
\includegraphics[width=\textwidth]{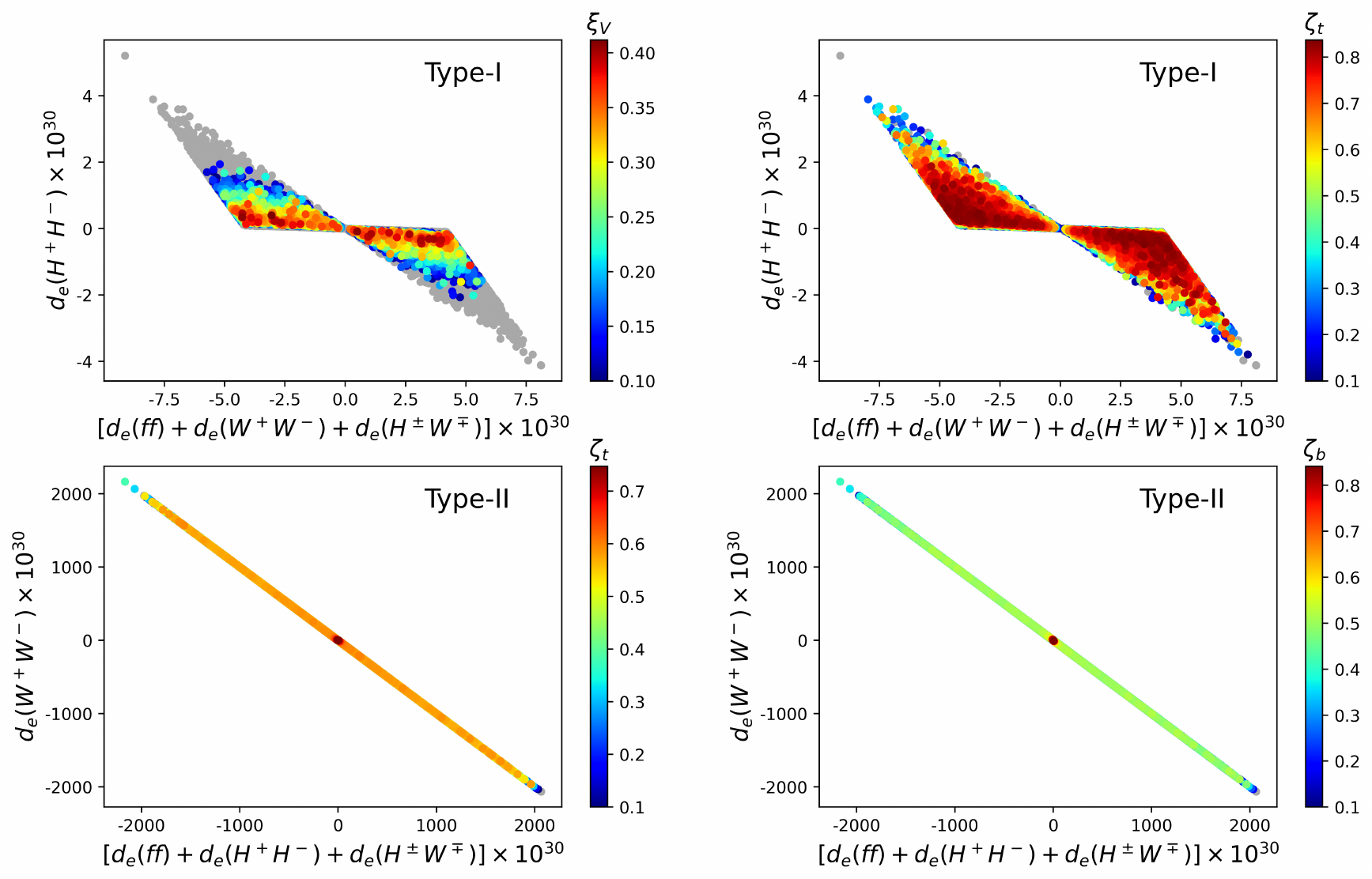}
\caption{Anatomy of individual two-loop Barr--Zee contributions to the electron EDM for Type-I (upper) and Type-II (lower) parameter space under the JILA bound. Here, $d_e(X_i X_j)$ denotes the contribution from the $X_i X_j$ loop class to the total eEDM. Gray and colored points satisfy the same theoretical and experimental constraints as in \autoref{fig-type1-xiv-mh3-mh2}. Color bars indicate the CP-violating measures $\xi_V$, $\zeta_t$, and $\zeta_b$. }
\label{fig-edm-anatomy}
\end{figure}
%------------------------

The resilience of the C2HDM parameter space against eEDM bounds is rooted in the internal cancellation structure of the two-loop Barr--Zee contributions~\cite{Barr:1990vd}. In \autoref{fig-edm-anatomy}, we decompose these contributions for Type-I and Type-II scenarios. The cancellation mechanism differs markedly between the two Yukawa types. 

In Type-I, individual contributions to $|d_e|$ for viable parameter points (gray) are typically of $\mathcal{O}(10^{-30})\,e\cdot\mathrm{cm}$, comparable to the current JILA bound. Nevertheless, the opposite sign of the $H^+H^-$ loop contribution relative to the other contributions leads to a partial cancellation in the total $d_e$. 

In Type-II, individual contributions can exceed the JILA bound by up to three orders of magnitude. Satisfying the experimental constraints therefore requires significant mutual cancellations, as reflected in the strong anti-correlation between the $W^+W^-$ contribution and the remaining terms. Intriguingly, in regions with maximally allowed Yukawa CP violation ($\zeta_{t,b} \gtrsim 0.7$), the individual contributions themselves become suppressed, with each remaining near or below the JILA bound. This behavior is consistent with the findings of Ref.~\cite{Darvishi:2023fjh}, where large CP violation was shown to be compatible with experimental limits without requiring excessive fine-tuning.

%------------------------
\begin{figure}[t]
\centering
\includegraphics[width=\textwidth]{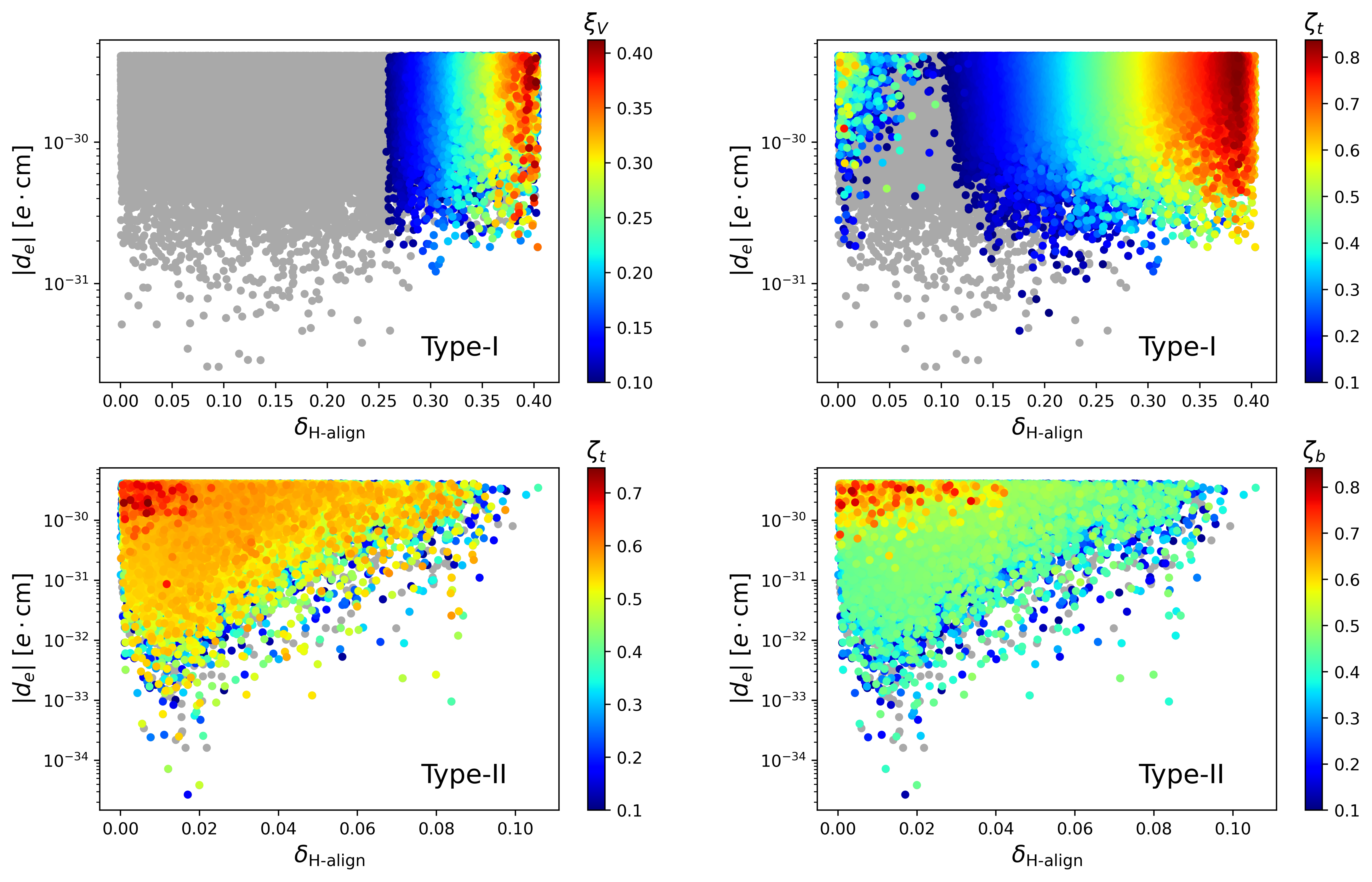}
\caption{Predicted $|d_e|$ as a function of the alignment measure $\delta_\text{H-align} $ for Type-I (upper) and Type-II (lower). Beige, gray, and colored points satisfy the same conditions as in \autoref{fig-type1-xiv-mh3-mh2}. The color scales represent the CP-violating measures $\xi_V$ (upper left), $\zeta_t$ (upper right), $\zeta_t$ (lower left), and $\zeta_b$ (lower right).}
\label{fig-type12-de-dHalign}
\end{figure}
%------------------------

%comment 5
Building on this understanding of the eEDM constraints, \autoref{fig-type12-de-dHalign} illustrates the resulting distribution of $|d_e|$ as a function of the Higgs alignment measure $\delta_{H\text{-align}}$ for Type-I (upper) and Type-II (lower). We first note that the apparent absence of both CP-conserving configurations and exact Higgs alignment is a natural consequence of the numerical random scan over the multidimensional parameter space. While neither the CP-conserving limit ($\alpha_2 = 0$, $\alpha_3 \in \{0, \pm \pi/2\}$) nor the exact Higgs alignment limit ($\beta = \alpha_1$, $\alpha_2 = 0$, $\alpha_3 \in \{0, \pm \pi/2\}$) is theoretically excluded, these limits correspond to measure-zero regions that are statistically inaccessible in a stochastic sampling.

Since our primary objective is to investigate the phenomenology of sizable CP violation—and noting that CP is conserved in the exact alignment limit—we do not fine-tune our scan to populate these regions. As a result, the ensemble of viable points naturally clusters away from these limits, leading to an apparent lower threshold for $|d_e|$, which we interpret as the characteristic scale of CP-violating signals within the considered parameter space.

It is also important to comment on the role of the chosen mass range ($M_{H_{2,3}}, M_{H^\pm} < 2$~TeV). In principle, pushing the model toward the decoupling limit with much heavier Higgs masses would suppress $|d_e|$. However, this limit is not phenomenologically viable in the present framework, as the quartic couplings $|\lambda_{1,\dots,5}|$ scale with the square of the Higgs masses~\cite{ElKaffas:2007rq}, leading to violations of perturbative unitarity at large mass scales.

To demonstrate this, we performed additional scans with fixed heavy scalar masses $M = 3, 4, 5$~TeV, with $m_{H_{b}}$ and $M_{H^\pm}$ fixed to $M$. We find that no parameter points survive theoretical constraints for $M = 4$ and $5$~TeV, while only a negligible number of points remain viable at $M = 3$~TeV, and these are incompatible with current eEDM bounds. Therefore, the distributions shown in \autoref{fig-type12-de-dHalign} represent the physically consistent range of $|d_e|$ predictions within the model at realistic mass scales.

Beyond these constraints, \autoref{fig-type12-de-dHalign} reveals distinct phenomenological features between the two Yukawa types. In Type-I, we observe a pronounced correlation between the eEDM magnitude and the alignment measure $\delta_\text{H-align} $: as the model approaches the alignment limit (smaller $\delta_\text{H-align} $), the minimum values of $|d_e|$ realized in our scan—interpreted as the characteristic eEDM scale within the considered parameter space—shift to larger values. Notably, no comparable correlation is observed between $|d_e|$ and the gauge-sector CPV measure $\xi_V$. 

In contrast, Type-II predictions correlate strongly with the magnitudes of the Yukawa CPV measures $\zeta_t$ and $\zeta_b$, where increased CP violation directly drives larger EDM values. This carries a critical implication: should future eEDM experiments continue to yield null results, they will strictly constrain the possibility of large CP violation within the Type-II Yukawa sector.

In summary, eEDM measurements offer a powerful, type-dependent test of the C2HDM. 
In Type-I, the viable points are concentrated in a range of $|d_e|$ well-matched to the reach of next-generation experiments. 
While the Type-II model can accommodate much smaller eEDM values, it exhibits strong correlations with both the CPV measures and the Higgs alignment parameter. 
Consequently, future EDM bounds will serve as a potent indirect probe of the Type-II C2HDM, tightly constraining both the allowed CP violation and deviations from the Higgs alignment limit.

\subsection{Higgs signal strengths of the near degenerate scenario in Type-I}

A key finding of our comprehensive scan in the Type-I C2HDM is that large CP violation in the gauge sector strongly favors a configuration where $H_2$ is nearly degenerate with the 125~GeV state $H_1$.
As discussed in Sec.~\text{Re}f{subsec-scan}, our \code{HiggsSignals} analysis confirms that such near-degeneracy is fully compatible with current Higgs precision data.
To understand the underlying mechanism, we must examine how the individual contributions from $H_1$ and $H_2$ combine to reproduce the observed Higgs signals.
For the purpose of this analysis, we designate the subset of parameter points satisfying $M_{H_2} \in [126,128]\,\text{GeV}$ and $\xiv > 0.1$ as the \textbf{Near-Degenerate CPV Scenario}.

The first distinctive feature of this scenario is the suppression of the total width of the 125~GeV state $H_1$. We find:
\beq
\label{eq-total-width-H1}
\Gamma_{\rm tot}(H_1) \in [2.65, 3.49]~\text{MeV},
\eeq
which corresponds to a reduction of approximately $15\text{--}35\%$ relative to the SM prediction $\Gamma_{\rm tot}^{\rm SM}(\hsm)\simeq4.1~\text{MeV}$.
The corresponding branching ratios are:
\begin{align}
\br(H_1 \to \bb) &\in [57.25,59.05]\%, & 
\br(H_1 \to \ww) &\in [20.83,22.93]\%, 
\\ \nn
\br(H_1 \to gg) &\in [7.53,7.77]\%, & 
\br(H_1 \to \ttau) &\in [6.16,6.35]\%, 
\\ \nn
\br(H_1 \to c\bar{c}) &\in [2.80,2.89]\%, & 
\br(H_1 \to ZZ) &\in [2.60,2.86]\%, 
\\ \nn
\br(H_1 \to \gamma\gamma) &\in [0.215,0.323]\%, & 
\br(H_1 \to Z\gamma) &\in [0.152,0.181]\%.
\end{align}
These branching ratios are obtained using \code{C2HDM\_HDECAY}~\cite{Fontes:2017zfn}, which evaluates Higgs decay rates via the \code{anyHdecay} interface based on the \code{HDECAY} package~\cite{Djouadi:1997yw,Djouadi:2018xqq}.
This setup ensures state-of-the-art accuracy, including NLO/NNLO QCD and electroweak corrections where available, as well as off-shell effects.

%------------------------
\begin{figure}[t]
\centering
\includegraphics[width=\textwidth]{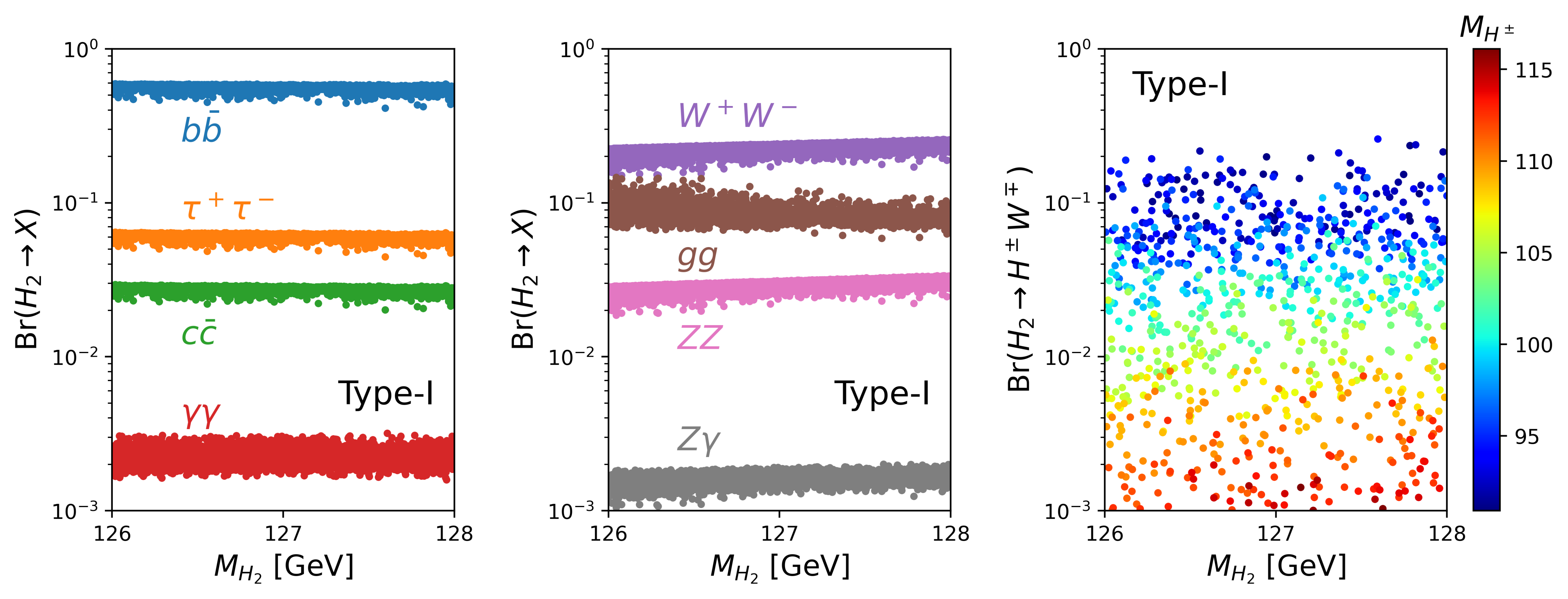}
\caption{%
Branching ratios of $H_2$ in the Near-Degenerate CPV Scenario where $M_{H_2} \in [126,128]\,\text{GeV}$ and $\xiv > 0.1$ in Type-I C2HDM.
The decay modes into $\bb,\ttau,c\bar{c},\gamma\gamma$ are shown in the left panel; $\ww,gg,ZZ,Z\gamma$ in the middle; and $H^\pm W^\mp$ in the right panel.
In the $H^\pm W^\mp$ panel, the color bar indicates $\mch$.
}
\label{fig-type1-BRH2}
\end{figure}
%------------------------

A second notable feature concerns the heavier state $H_2$.  
In this scenario, its total width is even more suppressed:
\beq
\label{eq-total-width-H2}
\Gamma_{\rm tot}(H_2) \in [0.163,1.37]~\text{MeV},
\eeq
which is significantly smaller than the SM prediction for a 125\,GeV Higgs boson.  
The corresponding branching ratios for $M_{H_2} \in [126,128]\,\text{GeV}$ are shown in \autoref{fig-type1-BRH2}.  
They remain nearly constant across this narrow mass window and closely resemble those of the SM Higgs boson.  
The only qualitative difference is the opening  of the decay mode $H_2 \to H^\pm W^\mp$ when it is kinematically accessible. 
As indicated by the color bar, a lighter $\mch$ enhances this mode, although the branching fraction stays below the percent level for $\mch \gtrsim 100\,\text{GeV}$.

These characteristics of $H_2$---a suppressed total width combined with SM-like branching ratios---have two critical consequences for Higgs precision observables.
First, they imply an approximately universal rescaling of the partial widths of $H_2$ into SM final states $f_{\rm SM}$:
\beq
\Gamma(H_2 \to f_{\rm SM}) \simeq \mu(H_2) \, \Gamma^{\rm SM}(\hsm \to f_{\rm SM} ) \quad \text{for } \mu(H_2) < 1 \,.
\eeq
Because the non-SM decay mode $H_2 \to H^\pm W^\mp$ remains subdominant, the total width is well-approximated by the relation $\Gamma_{\rm tot}(H_2) \simeq \mu(H_2) \, \Gamma_{\rm tot}(\hsm)$, which explains the significant suppression of the total width observed in \autoref{eq-total-width-H2}.
Crucially, this factor naturally suppresses the gluon-fusion production rate of $H_2$, thereby ensuring that this additional state remains compatible with current Higgs precision data.

Second, the SM-like nature of the observed $H_1$ (corresponding to a small deviation from the exact Higgs alignment limit, $\alpha_1 = \beta$ and $\alpha_2 = 0$) leads to an approximate sum rule: $c^2(H_1VV) + c^2(H_2VV) \simeq 1$.
This relation is derived by introducing a small misalignment parameter $\epsilon = \beta - \alpha_1 = \alpha_2 \ll 1$, which yields
$c^2(H_1VV) + c^2(H_2VV) \simeq 1 - \epsilon^2(1 + s_{2\alpha_3})$.
This sum rule ensures that the two neutral states, $H_1$ and $H_2$, collectively saturate the SM-like coupling strength to vector bosons.
Consequently, for any channel driven by the Higgs-gauge coupling---whether it is electroweak production via vector-boson fusion (VBF) and associated production (VH), or decay into weak bosons ($WW, ZZ$)---the combined signal strength reproduces the SM expectation:
\beq
\mu_{\rm tot}^{(V)} = \mu^{(V)}(H_1) + \mu^{(V)}(H_2) \simeq 1 \,,
\eeq
where the superscript $(V)$ collectively denotes any process involving vector bosons in either the production or decay.

We now present the individual signal strengths $\mu^{\text{decay}}_{\text{production}}$ for $H_1$ and $H_2$, defined as
\begin{equation}
\mu^{\text{decay}}_{\text{production}}(H_i)
\;\equiv\;
\frac{\big[\sigma(pp\!\to\!H_i)_{\text{production}} \times 
\mathrm{BR}(H_i\!\to\!\text{decay})\big]_{\text{C2HDM}}}
{\big[\sigma(pp\!\to\!H)_{\text{production}} \times 
\mathrm{BR}(H\!\to\!\text{decay})\big]_{\text{SM}}}\;.
\end{equation}
The branching ratios are computed as described above.
The production cross sections are obtained using state-of-the-art tools: gluon fusion (ggF) at NNLO QCD via \code{SusHi}~\cite{Harlander:2012pb,Harlander:2016vzb}, VH production at NNLO QCD via \code{VH@NNLO}~\cite{Brein:2012ne,Harlander:2018zpi}, and VBF via \code{HiggsBounds}~\cite{Bechtle:2020pkv} (which rescales the SM cross sections from the LHCHXSWG to NNLO QCD + NLO EW accuracy).

%------------------------
\begin{figure}[t]
\centering
\includegraphics[width=\textwidth]{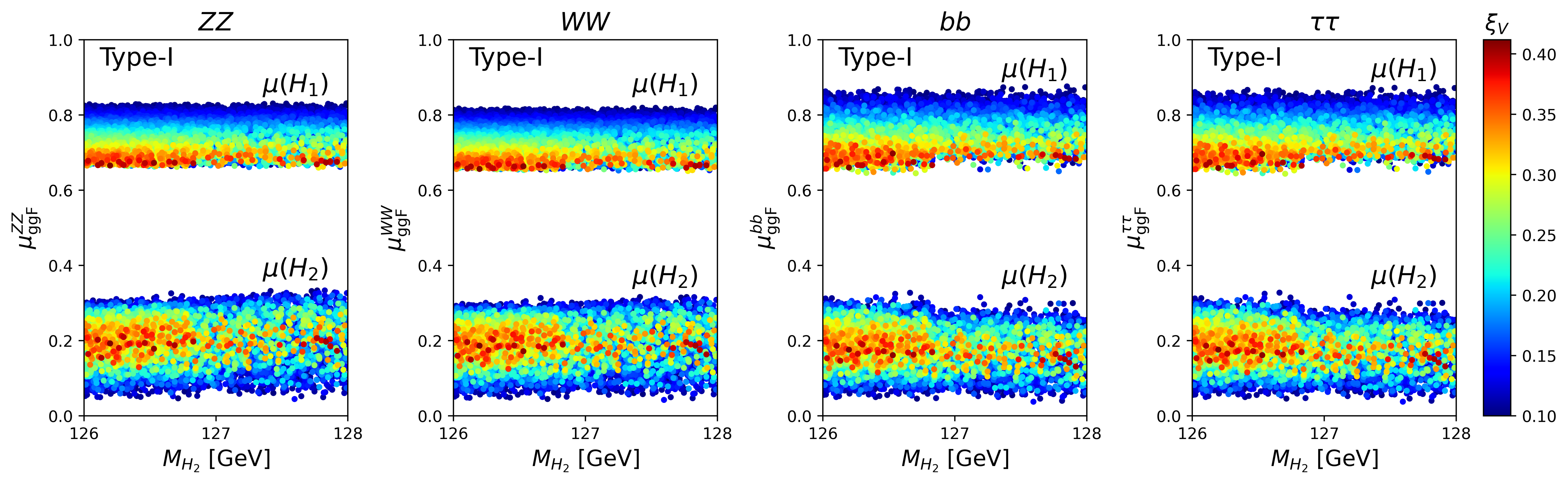}\\[7pt]
\includegraphics[width=\textwidth]{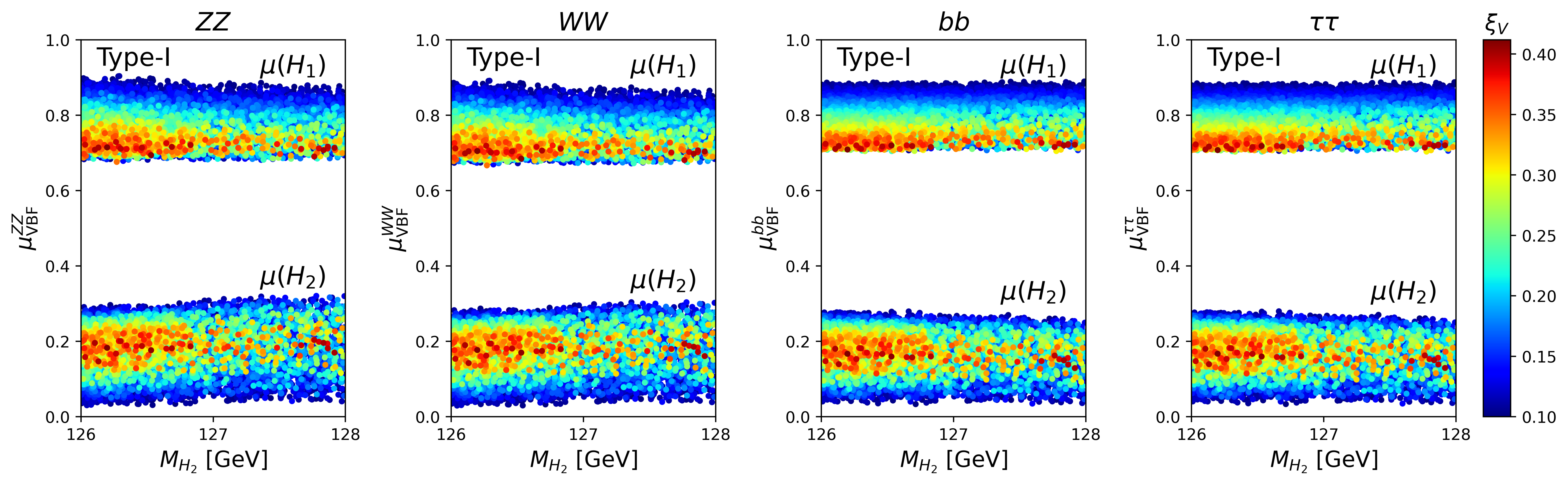}
\caption{%
Signal strengths $\mu^{\rm decay}_{\rm production}$ of $H_1$ and $H_2$ in the Near-Degenerate CPV Scenario:
$M_{H_2} \in [126,128]\,\text{GeV}$ and $\xiv>0.1$ in Type-I C2HDM.
Shown are decay modes into $ZZ$, $W^+W^-$, $\bb$, and $\ttau$ for gluon fusion (top) and VBF (bottom) productions. 
The color code denotes $\xiv$.
}
\label{fig-type1-mu}
\end{figure}
%------------------------

In \autoref{fig-type1-mu}, we display the resulting signal strengths of $H_1$ and $H_2$ for the main decay modes into vector bosons
and fermions ($ZZ$, $W^+W^-$, $\bb$, and $\ttau$), comparing gluon-fusion (upper panels) and VBF (lower panels) production.
It is evident that $H_1$ dominates the total signal, with $\mu(H_1)$ ranging roughly from 0.67 to 0.85, while $H_2$ contributes the remaining fraction, with $\mu(H_2)$ falling between 0.1 and 0.3.
This distribution directly reflects the coupling sum rule $c^2(H_1VV)+c^2(H_2VV)\!\simeq\!1$.
Hence, although each state individually yields a sub-SM rate, their combined signal remains close to unity (within $\sim 10\%$), effectively satisfying the current Higgs precision data.

A remarkable feature of \autoref{fig-type1-mu} is the strong correlation between the signal strengths and the CPV measure $\xiv$.
Larger values of $\xiv$ confine $\mu(H_1)$ and $\mu(H_2)$ to specific ranges; for instance, demanding $\xiv \gtrsim 0.4$ implies $\mu(H_1)\simeq 0.67$ and $\mu(H_2) \sim 0.2$.
In this high-CPV regime, the total signal strength---comprising the sum of the $H_1$ and $H_2$ contributions---falls distinctly below the SM expectation.
Consequently, improved precision in future Higgs measurements will effectively probe the CPV measure $\xiv$ within this scenario.

We now turn to the loop-induced final states $H_i\to \gamma\gamma$ and $H_i\to Z\gamma$.
As these decays are radiatively generated, they offer a unique window into BSM physics~\cite{CMS:2022ahq,ATLAS:2025aip}, being susceptible to both modifications of the SM-like couplings and new loop contributions from the charged Higgs boson.
Given the light $\mch$ preferred in the large-$\xiv$ regime (see \autoref{fig-type1-xiv-tb-mch}), these additional loops can induce observable deviations.
Moreover, the interference between $W$ and top loops differs between the $Z\gamma$ and $\gamma\gamma$ modes, making them complementary probes.

%------------------------
\begin{figure}[t]
\centering
\includegraphics[width=\textwidth]{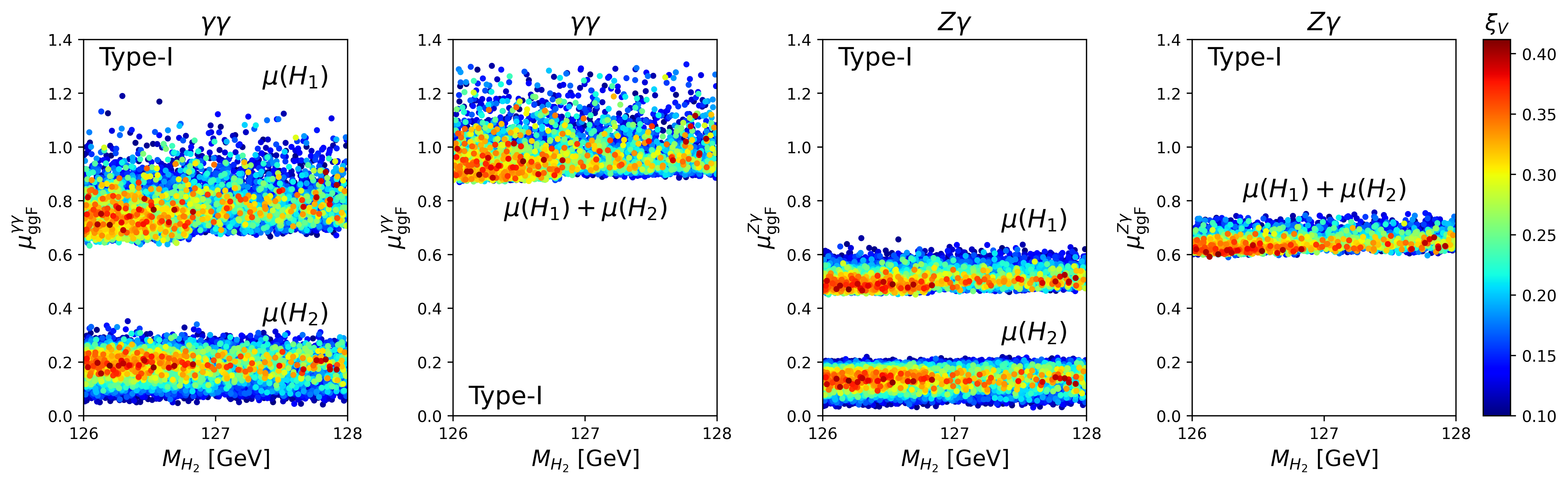}\\[7pt]
\includegraphics[width=\textwidth]{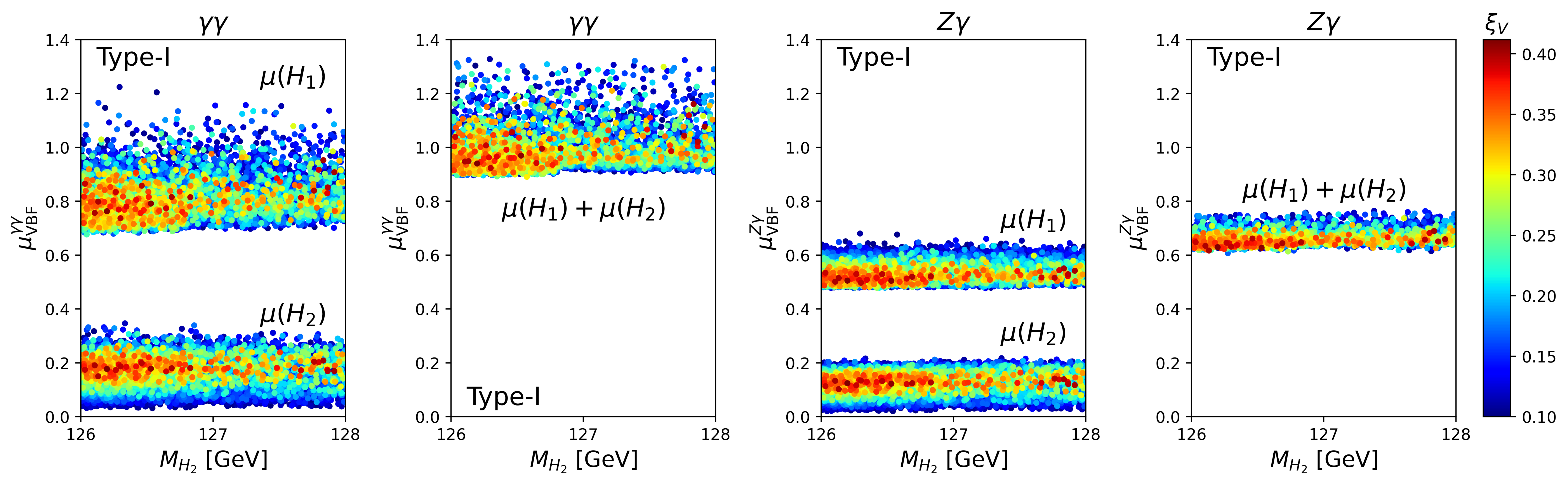}
\caption{%
Signal strengths of $H_1$ and $H_2$ for the decay modes into $\gamma\gamma$ and $Z\gamma$ 
in the Near-Degenerate CPV Scenario ($M_{H_2}\in[126,128]\,\text{GeV}$, $\xiv>0.1$).
Results for gluon fusion are shown in the top panels, and for VBF in the bottom.
The color code denotes $\xiv$.
}
\label{fig-type1-mu-rrZr}
\end{figure}
%------------------------

\autoref{fig-type1-mu-rrZr} presents the predicted $\mu^{\gamma\gamma}$ and $\mu^{Z\gamma}$ for both ggF (upper panels) and VBF (lower panels) production.
Since the individual values for $\mu^{\gamma\gamma}(H_1)$ and $\mu^{\gamma\gamma}(H_2)$ exhibit a broad spread, one might question whether the total rate is consistent with the current Higgs precision data.
To address this, the second column of \autoref{fig-type1-mu-rrZr} explicitly displays the combined rate $\mu^{\gamma\gamma}(H_1)+\mu^{\gamma\gamma}(H_2)$.
This demonstrates that the total diphoton signal remains well within the current ATLAS best-fit value of $\mu^{\gamma\gamma}=1.05\pm0.10$~\cite{ATLAS:2022tnm}.
In contrast, for the $Z\gamma$ channel, both the individual and combined signal strengths show a much narrower spread, with the combined rate consistently suppressed below $\sim 0.75$.
While this suppression is compatible with the current experimental value of $\mu^{Z\gamma}=1.3^{+0.6}_{-0.5}$~\cite{ATLAS:2025aip}, such a reduced rate is a robust prediction of the Near-Degenerate CPV Scenario.
Consequently, a future measurement of $\mu^{Z\gamma}$ significantly below unity would corroborate the robust suppression predicted by this Near-Degenerate CPV Scenario.

\noindent\textbf{Experimental outlook.}---The Near-Degenerate CPV Scenario predicts that the observed 125~GeV Higgs peak arises from a superposition of two states, $H_1$ and $H_2$. While their combined signal strengths mimic SM-like behavior, the individual states possess reduced couplings and narrower widths. Given the current LHC invariant-mass resolution of $\Delta m \sim 1$--$2$~GeV in the $H \to \gamma\gamma$ and $H \to ZZ \to 4\ell$ channels~\cite{CMS:2022ahq,ATLAS:2023fsi}, our predicted mass splitting of $\Delta M = M_{H_2} - M_{H_1} \lesssim 3$~GeV typically results in the two states appearing as a single, broadened resonance. Furthermore, mass-insensitive channels spanning the $122$--$130$~GeV window naturally accommodate this scenario~\cite{ATLAS:2023owm,ATLAS:2023tnc,CMS:2025zue}. As demonstrated in our analysis, the $H_2$ contribution remains subdominant ($\sim 10\%$ of the $H_1$ signal), preserving consistency with the observed resonance structure.

The theoretical foundation for analyzing such overlapping resonances was established by Pilaftsis~\cite{Pilaftsis:1997dr}, who systematically studied the lineshape effects of nearly degenerate, CP-violating neutral Higgs bosons using a propagator-level resonant-mixing formalism. This approach was further developed for the LHC in the context of the CP-violating MSSM~\cite{Ellis:2004fs}, employing a full propagator-matrix formalism to consistently treat the mixing, production, and decay of nearly degenerate states. 
Based on this framework, a high-statistics \textit{lineshape analysis} at the HL-LHC is expected to serve as a particularly powerful probe.
Specifically, a splitting of $\Delta M \sim 1$~GeV would produce a subtle non-Breit--Wigner distortion, while $\Delta M \gtrsim 2$~GeV could manifest as a visible shoulder or double-peak structure. Both ATLAS and CMS are pursuing such analyses, and the increased sensitivity of the HL-LHC will be decisive.

Ultimately, future $e^+e^-$ colliders (ILC, FCC-ee, CEPC) offer the definitive resolution to this ambiguity. Through precise threshold scans, these machines could not only disentangle the split resonance structure but also provide a model-independent measurement of the total width $\Gamma_{\text{tot}}$~\cite{Ruan:2014xxa,Jadach:2015cwa,deBlas:2019rxi,Azzurri:2021nmy}. This would allow for a critical test of the scenario's primary prediction: a reduced $H_1$ width in the range $\Gamma_{\rm tot}(H_1) \in [2.65, 3.49]$~MeV. Confirming such a deviation from the SM expectation of $\sim 4.1$~MeV would serve as a smoking-gun signature of the extended Higgs sector, even if the mass splitting remains near the resolution limit.

\subsection{The CP Structure of Top and Tau Couplings in Type-II}

While the Near-Degenerate CPV Scenario in Type-I highlights the possibility of large CP violation in the gauge sector ($\xiv$), the Type-II model exhibits a fundamentally different behavior.
As observed in our global scan, Type-II is severely constrained to the regime of negligible gauge-sector CP violation, effectively forcing $\xiv \lesssim 3\times 10^{-3}$.
Consequently, the primary manifestation of CP violation in Type-II shifts to the Yukawa sector.
Here, the interactions of the top quark and the $\tau$ lepton serve as the definitive probes of the Higgs CP parity.

To quantify the CP structure in this sector, we parameterize the Yukawa couplings between a neutral Higgs boson $H_i$ and fermions $f$ in terms of a coupling strength modifier $\kp_f^{H_i}$ and a CP-mixing angle $\al_f^{H_i}$:
\beq
\label{eq-Lagrangian-alphat}
\lg_{\rm Yukawa} \supset - \sum_{i=1,2,3} \sum_{f=t,\tau}\frac{m_f}{v} \kp_f^{H_i} 
\left[ \cos {\al_f^{H_i} } + i \gm_5 \sin {\al_f^{H_i} } \right]\bar{f}f H_i.
\eeq
In this notation, $\alpha_f = 0$ corresponds to a purely CP-even interaction, $\alpha_f = 90^\circ$ represents a purely CP-odd state, while intermediate values indicate a CP-mixed Higgs boson.

We begin with the CP properties of the observed 125~GeV Higgs boson, $H_1$.
Current measurements indicate that $H_1$ is dominantly CP-even~\cite{ATLAS:2013xga,CMS:2014nkk,ATLAS:2020ior,CMS:2020cga,CMS:2021sdq,ATLAS:2022akr}, though a small CP-odd admixture remains a viable possibility.
A powerful avenue for probing this possibility involves its direct Yukawa interactions.
At the LHC, the CP structure of the top--Higgs coupling is primarily accessed through the $t\bar t \, \hsm$ production mode, which offers two key advantages: (i) strong suppression of QCD backgrounds, and (ii) sensitivity to CP-odd effects through spin-correlation and angular observables involving the top quarks and the Higgs boson~\cite{Aoki:2018zgq,Bahl:2021dnc,Bahl:2022yrs,Barger:2023wbg,Esmail:2024gdc}.
Recent ATLAS and CMS analyses on $t\bar t \, \hsm$ production and decay measurements find no significant deviation from the SM CP-even hypothesis, constraining the CP-mixing angle  to
$
\left|\alpha_t^{H_1}\right| \lesssim 40^\circ ~\text{(95\% C.L.)}
$~\cite{ATLAS:2023cbt,CMS:2022dbt}.

Similarly, the CP-mixing angle $\alpha_\tau^{H_1}$ can be probed through the transverse spin correlations of $\tau$-leptons in $H_1 \to\tau\tau$ decays. These correlations are imprinted on the directions of the decay products, specifically in the acoplanarity angle $\Phi^\ast$ between the two $\tau$ decay planes. The differential decay rate is modulated by $\cos ( \Phi^* - 2 \alpha_\tau^{H_1} )$~\cite{Antusch:2020ngh,Alonso-Gonzalez:2021jsa,Mb:2022rxu,Cardini:2025svy}.
Since reconstructing $\Phi^\ast$ is experimentally challenging due to the missing energy from neutrinos,
the constraint on $\alpha_\tau^{H_1} $ is weaker than on $\alpha_t^{H_1}$.
Standard analyses by ATLAS and CMS, which utilize template-based fits to $\tau$-pair kinematics~\cite{CMS:2021sdq,ATLAS:2022akr}, currently yield loose constraints: $\alpha_\tau^{H_1} = 9^\circ \pm 34^\circ$ at 95\%~C.L.

%------------------------
\begin{figure}[t]
\centering
\includegraphics[width=\textwidth]{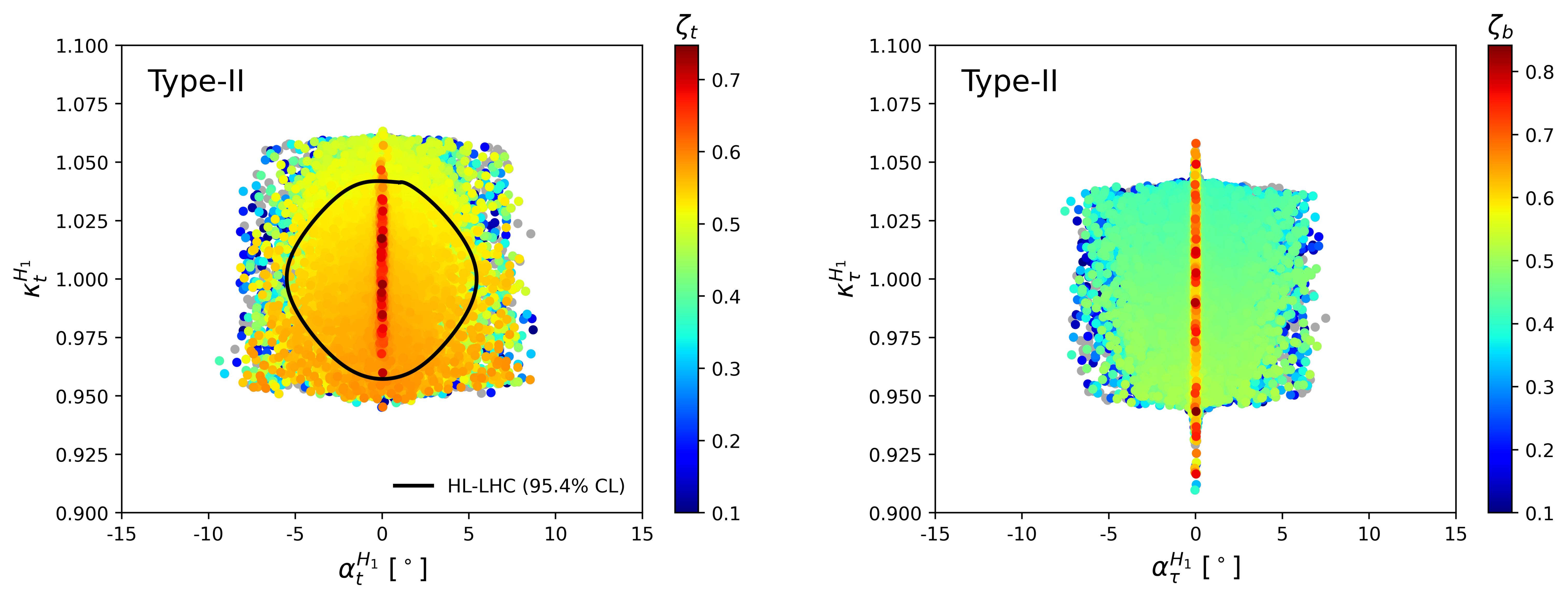}
\caption{%
Viable parameter points in the $(\al_t^{H_1}, \kappa_t^{H_1})$ plane (left) and $(\al_\tau^{H_1}, \kappa_\tau^{H_1})$ plane (right) for the Type-II C2HDM.
Gray points represent the full set of viable parameter points, while colored points indicate the subset satisfying the large CP violation condition: $\ztt>0.1$ (left) and $\ztb>0.1$ (right).
The color code denotes the magnitude of $\ztt$ and $\ztb$, respectively.
The solid black curve in the left panel displays the projected 95\%~C.L. exclusion contour for the HL-LHC ($14~\tev, 3~\iab$), obtained using the GNN-based method.
}
\label{fig-type2-zeta-alpha-tt-tautau}
\end{figure}

Given these experimental constraints, a key question is how large the CP violation in the top--$H_1$ and $\tau$--$H_1$ couplings can be within the Type-II C2HDM.
In \autoref{fig-type2-zeta-alpha-tt-tautau}, we present the viable parameter space, showing the $(\al_t^{H_1}, \kappa_t^{H_1})$ plane in the left panel and the $(\al_\tau^{H_1}, \kappa_\tau^{H_1})$ plane in the right panel.
The gray points denote the general viable parameter space, while the colored points represent the subset satisfying the large CP violation conditions ($\ztt>0.1$ for the left panel and $\ztb>0.1$ for the right).
As evident from the overlap, the accessible ranges for the Yukawa coupling parameters are nearly identical for both the general and large CP violation parameter sets.

The left panel of \autoref{fig-type2-zeta-alpha-tt-tautau} reveals that the CP violation in the top--$H_1$ coupling is quite limited.
The magnitude $\kappa_t^{H_1}$ deviates from the SM prediction by at most $\sim 6\%$, and the mixing angle $\alpha_t^{H_1}$ remains within approximately $\pm 8^\circ$.
Remarkably, we observe an inverse correlation: higher values of $\ztt$ tend to correspond to smaller values of $\alpha_t^{H_1}$. This indicates that the top--$H_1$ coupling is not the most efficient probe for the global CPV measure $\ztt$.	

To assess the HL-LHC's ability to probe this restricted parameter space, we show in the left panel of \autoref{fig-type2-zeta-alpha-tt-tautau} the projected 95\%~C.L. exclusion contour (solid black line) for $\sqrt{s}=14~\tev$ and $\mathcal{L}_{\rm tot}=3\iab$. 
This projection is obtained by extending the Machine Learning (ML) framework of Ref.~\cite{Esmail:2024gdc} to the general $\kappa_t^{H_1}$ case using the $t\bar t \, \hsm$ production channel. 
The conditional Graph Neural Network (GNN) approach of Ref.~\cite{Esmail:2024gdc} significantly improves sensitivity by directly learning event topology and angular correlations from reconstructed objects.
This offers a distinct advantage over traditional cut-and-count analyses, which suffer from suboptimal signal regions and limited sensitivity to CP-odd observables~\cite{Demartin:2014fia,Buckley:2015vsa,Azevedo:2017qiz,Goncalves:2018agy,ATLAS:2020ior,CMS:2020cga}.

However, even with these advanced techniques, the HL-LHC will lack the statistical power to conclusively test this scenario.
For instance, in the optimistic case where $\alpha_t^{H_1} \approx 8^\circ$, the HL-LHC would detect a non-zero CP phase with a significance of only $\sim 3.2\sigma$.
Conversely, as indicated by the 95\%~C.L. contour, a null result would not exclude the high $\ztt$ regime.
Crucially, because the parameter points exhibiting the largest global CP violation (e.g., $\ztt \gtrsim 0.5$) cluster near $\alpha_t^{H_1} \approx 0$, they mimic the purely CP-even $H_1$ state predicted by the SM.
Thus, merely increasing the statistical precision of the $t\bar{t}H_1$ measurement at future colliders will not suffice to uncover this Yukawa-sector CP violation.

The situation is equally challenging for the $\tau$--$H_1$ coupling.
The right panel of \autoref{fig-type2-zeta-alpha-tt-tautau} shows the viable parameter points in the $(\al_\tau^{H_1}, \kappa_\tau^{H_1})$ plane.
Deviations from the SM-like limit are again small: $\kappa_\tau^{H_1}$ is within $\sim 5\%$ of the SM value, and $\alpha_\tau^{H_1}$ is confined to $\pm 7^\circ$.
These values are well within current experimental limits.
While advanced deep learning techniques, such as the Graph Transformer Network proposed in Ref.~\cite{Esmail:2024jdg}, can significantly improve sensitivity (reaching $|\alpha_\tau| \sim 20^\circ$ at 95\% C.L. with $100~\mathrm{fb}^{-1}$), the predicted mixing angles in the Type-II C2HDM are substantially smaller than even these optimized projections. Thus, probing $\alpha_\tau^{H_1}$ at the HL-LHC remains a formidable challenge.

%------------------------
\begin{figure}[t]
\centering
\includegraphics[width=\textwidth]{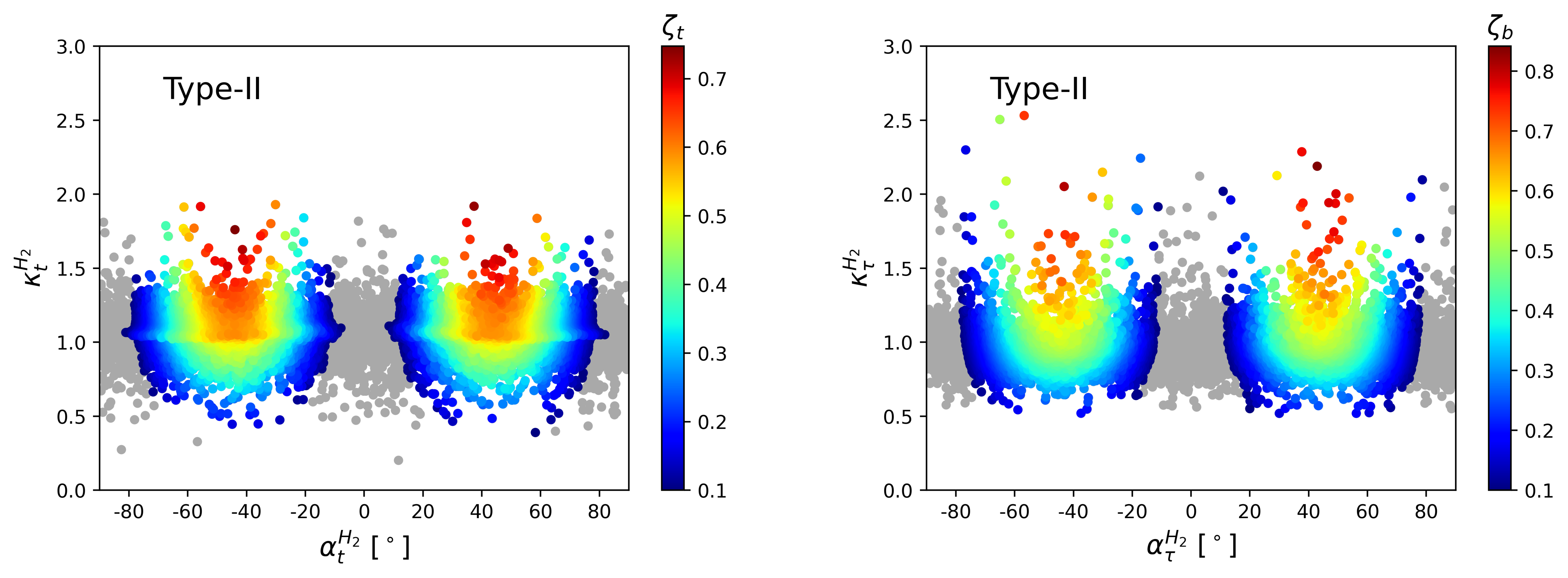}
\includegraphics[width=\textwidth]{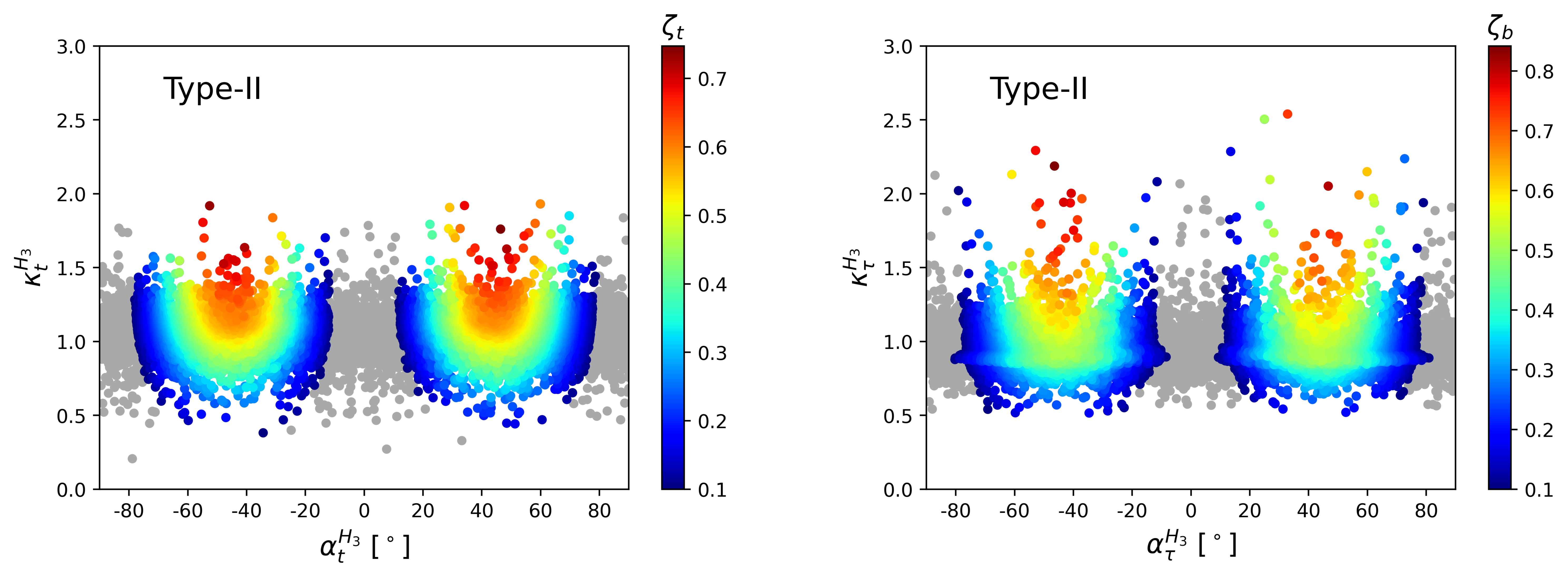}
\caption{%
Viable parameter points in the $(\al_t, \kappa_t)$ (left panels) and $(\al_\tau, \kappa_\tau)$ (right panels) planes for $H_2$ (upper panels) and $H_3$ (lower panels) in the Type-II C2HDM.
Gray points represent the full viable space; colored points satisfy the large CP violation conditions ($\ztt>0.1$ left, $\ztb>0.1$ right).
The color code indicates the magnitude of $\ztt$ and $\ztb$.
}
\label{fig-type2-CH23-tt-tautaug}
\end{figure}

Given that the CP-violating effects in the top and $\tau$ couplings of the $125\,\text{GeV}$ Higgs boson are effectively beyond the realistic reach of the HL-LHC, we turn to the heavier neutral Higgs states as the complementary probes.
\autoref{fig-type2-CH23-tt-tautaug} displays the viable points in the $(\alpha_t,\kappa_t)$ and $(\alpha_\tau,\kappa_\tau)$ planes for $H_2$ (upper panels) and $H_3$ (lower panels).
Gray points show the full viable parameter space, while the colored points highlight regions with large global CPV measures, $\ztt>0.1$ (left panels) or $\ztb>0.1$ (right panels).

The most striking feature to emerge from our analysis is that strong global CP violation forces the heavy-Higgs sector into a state of maximal mixing.
As $\ztt$ and $\ztb$ increase, the mixing angles $\alpha_{t}^{H_{2,3}}$ and $\alpha_{\tau}^{H_{2,3}}$ simultaneously converge to $\pi/4$, signifying that the heavy Higgs bosons manifest as maximally mixed CP-even/CP-odd states.
This distinct phenomenological pattern identifies the heavy neutral sector as the primary reservoir of CP-violating effects, offering a far more robust signature than the SM-like state.
We further observe that $H_2$ and $H_3$ exhibit very similar behavior in this regard, while their coupling magnitudes $\kappa_{t}^{H_{2,3}}$ and $\kappa_{\tau}^{H_{2,3}}$ remain broadly SM-like (typically within $1\pm 0.5$, though values up to $\sim 2.5$ are possible in rare high-CPV cases).
Consequently, extending CP-violation searches to these BSM Higgs bosons at the HL-LHC is essential, as they offer significantly greater sensitivity to the underlying CP structure than measurements restricted to the SM-like Higgs boson.

%------------

\subsection{Hidden CP violation in the near-alignment limit}
\label{subsec-hidden}

We conclude our analysis by examining the CP violation arising from the mixing of the heavy neutral states $H_2$ and $H_3$---a particularly compelling phenomenon that emerges in the near-alignment regime. We define this regime as
\beq
\text{Near-alignment limit: } \delta_\text{H-align}  < 0.1 \;.
\eeq
In this limit, the 125~GeV Higgs boson $H_1$ is approximately CP-even and largely decoupled from the heavy sector. Nevertheless, sizable CP violation can still be realized within the heavy sector through the mixing angle $\alpha_3$. Since this CP violation does not significantly manifest in the couplings of the observed 125~GeV state, we refer to it as \textit{hidden CP violation}.

Probing hidden CP violation via the gauge sector is experimentally challenging because the near-alignment condition inherently suppresses the couplings of the heavy neutral scalars to a weak gauge boson pair $VV$ ($V=W^\pm, Z$):
\beq
c(H_2VV) \simeq -c(H_3VV) \simeq (c_{\alpha_3} - s_{\alpha_3})\delta_\text{H-align} + \mathcal{O}(\delta_\text{H-align} ^3) \;.
\eeq
Furthermore, the off-diagonal neutral Higgs couplings to the $Z$ boson involving the SM-like state---specifically the $[H_1, H_2, Z^\mu]$ and $[H_1, H_3, Z^\mu]$ vertices---are similarly suppressed by factors of $\delta_\text{H-align} $.\footnote{The diagonal neutral Higgs couplings to the $Z$ boson, $[H_i,H_i,Z^\mu]$, vanish identically due to Bose symmetry.} In striking contrast, the off-diagonal $H_2$-$H_3$-$Z$ vertex remains robust~\cite{Fontes:2017zfn}:
\beq
\label{eq-H2H3Z}
[H_2, H_3, Z^\mu] \simeq \frac{g}{2 c_W} (p_{H_2} - p_{H_3})^\mu + \mathcal{O}(\delta_\text{H-align} ^2) \;.
\eeq
This indicates that the heavy states maintain a robust scalar-to-scalar transition through the $Z$ boson. Crucially, this vertex is determined by the gauge structure of the model and remains independent of the CP-violating mixing. Such a robust vertex suggests a potential production channel at the LHC via $pp \to Z^* \to H_2 H_3$. To directly probe the CP-violating structure of the heavy sector, we propose examining the Yukawa couplings, which retain a strong dependence on $\alpha_3$ even in the near alignment limit (see \autoref{eq-H23-modifiers-HAL}). This CP-violating structure remains fundamentally inaccessible through $H_1$ precision measurements alone.

\begin{figure}[t]
\centering
\includegraphics[width=\textwidth]{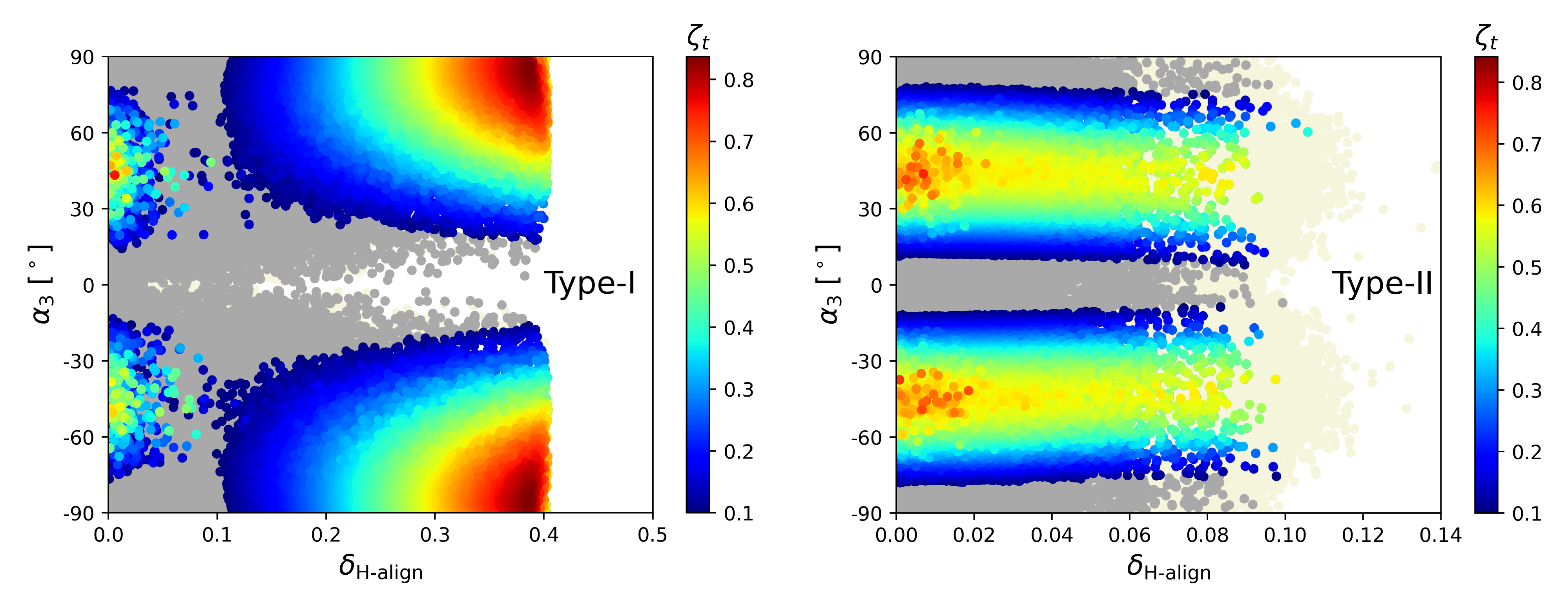}
\caption{%
Viable parameter points in the $(\delta_\text{H-align} , \alpha_3)$ plane for Type-I (left panel) and Type-II (right panel) C2HDM
within the near-alignment regime ($\delta_\text{H-align}  < 0.1$).
Beige, gray, and colored points satisfy the same conditions as in \autoref{fig-type1-xiv-mh3-mh2}.
}
\label{fig-type12-alpha3-dha}
\end{figure}

To quantify the scope of this hidden CP violation, \autoref{fig-type12-alpha3-dha} presents the distribution of viable parameter points in the $(\delta_\text{H-align} , \alpha_3)$ plane. Here, we focus specifically on the near-alignment regime by restricting the visualization to points satisfying $\delta_\text{H-align}  < 0.1$. This selective plotting highlights the behavior of the heavy-sector mixing in the limit where $H_1$ is most SM-like. The color coding of the points follows the conventions established in \autoref{fig-type1-xiv-mh3-mh2}, allowing for a consistent interpretation of the underlying physics across the different parameter scans.

Two significant features emerge from the analysis of \autoref{fig-type12-alpha3-dha}. First, our results demonstrate that sizable CP violation persists in the heavy neutral Higgs sector even as the 125~GeV state approaches the exact alignment limit. Specifically, for parameter points characterized by large CP violation (colored points), the mixing angle $\alpha_3$ remains significantly displaced from the CP-conserving values of $\{0, \pm\pi/2\}$. 
This persistence reflects the non-decoupling behavior identified by Pilaftsis and Wagner~\cite{Pilaftsis:1999qt}, whereby CP violation decouples from the SM-like Higgs sector at large mass scales but remains operative within the heavy neutral Higgs system. In this regime, the strong mixing between the heavy states can facilitate a resonant enhancement of CP-violating effects through scalar--pseudoscalar transitions, particularly when $H_2$ and $H_3$ are nearly degenerate. Our global scan confirms that the C2HDM provides a robust manifestation of this behavior, supporting sizable CP-violating observables while remaining consistent with the full suite of current theoretical and experimental constraints.

Second, the mixing angle $\alpha_3$ exhibits a strong correlation with the Yukawa-sector CPV measures. In the near-alignment regime ($\delta_\text{H-align}  \ll 1$), the leading contributions to these measures simplify to:
\begin{equation}
\zeta_{23} \approx s^2_{2\alpha_3} c^2_\beta, \quad \zeta_{13} \approx s^2_{2\alpha_3} s^2_\beta \;,
\end{equation}
where their specific relationships to $\ztt$ and $\ztb$ in Type-I and Type-II are detailed in \autoref{eq-zeta-tbtau}. From these expressions, it is evident that both $\ztt$ and $\ztb$ are maximized at $\alpha_3 = \pm\pi/4$, corresponding to maximal mixing between the heavy neutral states.

The near-alignment regime is therefore the physically relevant region for heavy-sector CP violation.
In this regime,  $\al_3$ retains physical meaning as the mixing angle of the heavy neutral mass eigenstates.
Although the neutral-gauge couplings of $[H_{2,3},V,V]$ and $[H_1, H_{2,3},Z]$ are strongly suppressed near alignment, interactions involving $[H_2, H_{3},Z]$ can still provide complementary information on the heavy-sector mixing structure.

\section{Conclusions}
\label{sec:conclusion}

In this work, we have performed a comprehensive global analysis of the complex two-Higgs-doublet model (C2HDM) with softly broken $Z_{2}$ symmetry, identifying the intrinsic features of the parameter space that support large CP violation while satisfying all current theoretical, collider, flavor, and EDM constraints. Our analysis reveals that Type-I and Type-II models follow sharply distinct pathways toward sizable CP violation and exhibit unique structural behaviors near the Higgs alignment limit.

In Type-I, both the gauge- and Yukawa-sector CPV measures can reach significant magnitudes, with $\xi_{V}$ rising to $\sim 0.4$ and $\zeta_{t}$ reaching values as high as $\sim 0.8$, but only within a narrow region where the second neutral Higgs boson is nearly degenerate with the 125~GeV state $H_1$. For the ensemble of parameter points satisfying the full set of theoretical and experimental constraints within our motivated mass range, the predicted electron EDM exhibits a characteristic scale of $|d_e| \gtrsim 10^{-31}\, e\cdot\mathrm{cm}$. This suggests that the physically viable Type-I parameter space largely lies within the projected sensitivity of next-generation eEDM experiments, irrespective of the size of the CP-violating measures.

By contrast, Type-II strongly suppresses gauge-sector CP violation while still allowing large Yukawa-sector CP violation. In addition, destructive interference among different contributions to the eEDM can suppress $|d_e|$ down to values of order $10^{-35}\,e\cdot\mathrm{cm}$. No phenomenologically relevant lower bound on $|d_e|$ emerges within the presently viable parameter space. 

Finally, we have identified a novel phenomenon in the near-alignment limit: hidden CP violation. Although the heavy neutral scalars lose their couplings to gauge-boson pairs in this limit, their mutual CP mixing—parameterized by the angle $\alpha_3$—remains unconstrained and can become nearly maximal as the Yukawa-sector CPV measures increase. We have shown that this hidden CP violation can nevertheless be probed through CP-violating Yukawa couplings, providing a novel and necessary pathway to experimentally expose this otherwise elusive source of CP violation at future colliders.

Taken together, our analysis demonstrates that large CP violation in the C2HDM is highly structured and tightly constrained. The results presented here provide a systematic map of the viable parameter space and outline clear strategies for testing these scenarios through future EDM measurements, Higgs precision studies, and dedicated searches for the heavy Higgs sector.

\section*{Acknowledgments}
We would like to thank the anonymous referee for their insightful comments and constructive suggestions, which have significantly improved the clarity and theoretical depth of this manuscript.
The work of SL was supported by the National Science and Technology Council (NSTC) of Taiwan under grant no. NSTC 113-2112-M-007-041-MY3.
AH is funded by grant number 22H05113, “Foundation of Machine Learning Physics”, Grant in Aid for Transformative Research Areas and 22K03626, Grant-in-Aid for Scientific Research (C). AH is partially supported by the Science, Technology and Innovation Funding Authority (STDF) under grant number 50806. 

%\bibliographystyle{JHEPMod}
%\bibliography{C2HDM-large-CPV}  

\begin{thebibliography}{100}

\bibitem{ATLAS:2012yve}
{\scshape ATLAS} collaboration, G.~Aad et~al., \textit{{Observation of a new
  particle in the search for the Standard Model Higgs boson with the ATLAS
  detector at the LHC}},
  \href{https://doi.org/10.1016/j.physletb.2012.08.020}{\textit{Phys. Lett. B}
  {\bfseries 716} (2012) 1--29},
  [\href{https://arxiv.org/abs/1207.7214}{{\ttfamily 1207.7214}}].

\bibitem{CMS:2012qbp}
{\scshape CMS} collaboration, S.~Chatrchyan et~al., \textit{{Observation of a
  New Boson at a Mass of 125 GeV with the CMS Experiment at the LHC}},
  \href{https://doi.org/10.1016/j.physletb.2012.08.021}{\textit{Phys. Lett. B}
  {\bfseries 716} (2012) 30--61},
  [\href{https://arxiv.org/abs/1207.7235}{{\ttfamily 1207.7235}}].

\bibitem{Planck:2015fie}
{\scshape Planck} collaboration, P.~A.~R. Ade et~al., \textit{{Planck 2015
  results. XIII. Cosmological parameters}},
  \href{https://doi.org/10.1051/0004-6361/201525830}{\textit{Astron.
  Astrophys.} {\bfseries 594} (2016) A13},
  [\href{https://arxiv.org/abs/1502.01589}{{\ttfamily 1502.01589}}].

\bibitem{Sakharov:1967dj}
A.~D. Sakharov, \textit{{Violation of CP Invariance, C asymmetry, and baryon
  asymmetry of the universe}},
  \href{https://doi.org/10.1070/PU1991v034n05ABEH002497}{\textit{Pisma Zh.
  Eksp. Teor. Fiz.} {\bfseries 5} (1967) 32--35}.

  
\bibitem{Huet:1994jb}
P.~Huet and E.~Sather, \textit{{Electroweak baryogenesis and standard model CP
  violation}}, \href{https://doi.org/10.1103/PhysRevD.51.379}{\textit{Phys.
  Rev. D} {\bfseries 51} (1995) 379--394},
  [\href{https://arxiv.org/abs/hep-ph/9404302}{{\ttfamily hep-ph/9404302}}].

\bibitem{Kajantie:1996mn}
K.~Kajantie, M.~Laine, K.~Rummukainen and M.~E. Shaposhnikov, \textit{{Is there
  a~ hot electroweak phase transition at $m_H \gtrsim m_W$?}},
  \href{https://doi.org/10.1103/PhysRevLett.77.2887}{\textit{Phys. Rev. Lett.}
  {\bfseries 77} (1996) 2887--2890},
  [\href{https://arxiv.org/abs/hep-ph/9605288}{{\ttfamily hep-ph/9605288}}].

\bibitem{Csikor:1998eu}
F.~Csikor, Z.~Fodor and J.~Heitger, \textit{{Endpoint of the hot electroweak
  phase transition}},
  \href{https://doi.org/10.1103/PhysRevLett.82.21}{\textit{Phys. Rev. Lett.}
  {\bfseries 82} (1999) 21--24},
  [\href{https://arxiv.org/abs/hep-ph/9809291}{{\ttfamily hep-ph/9809291}}].

\bibitem{Gavela:1993ts}
M.~B. Gavela, P.~Hernandez, J.~Orloff and O.~Pene, \textit{{Standard model CP
  violation and baryon asymmetry}},
  \href{https://doi.org/10.1142/S0217732394000629}{\textit{Mod. Phys. Lett. A}
  {\bfseries 9} (1994) 795--810},
  [\href{https://arxiv.org/abs/hep-ph/9312215}{{\ttfamily hep-ph/9312215}}].

\bibitem{Gavela:1994ds}
M.~B. Gavela, M.~Lozano, J.~Orloff and O.~Pene, \textit{{Standard model CP
  violation and baryon asymmetry. Part 1: Zero temperature}},
  \href{https://doi.org/10.1016/0550-3213(94)00409-9}{\textit{Nucl. Phys. B}
  {\bfseries 430} (1994) 345--381},
  [\href{https://arxiv.org/abs/hep-ph/9406288}{{\ttfamily hep-ph/9406288}}].

\bibitem{Gavela:1994dt}
M.~B. Gavela, P.~Hernandez, J.~Orloff, O.~Pene and C.~Quimbay,
  \textit{{Standard model CP violation and baryon asymmetry. Part 2: Finite
  temperature}},
  \href{https://doi.org/10.1016/0550-3213(94)00410-2}{\textit{Nucl. Phys. B}
  {\bfseries 430} (1994) 382--426},
  [\href{https://arxiv.org/abs/hep-ph/9406289}{{\ttfamily hep-ph/9406289}}].

\bibitem{Gunion:1989we}
J.~F. Gunion, H.~E. Haber, G.~L. Kane and S.~Dawson, \textit{{The Higgs
  Hunter's Guide}}, vol.~80.
Front. Phys. \textbf{80}, 1-404 (2000),
  \href{https://doi.org/10.1201/9780429496448}{10.1201/9780429496448}.

\bibitem{Branco:2011iw}
G.~C. Branco, P.~M. Ferreira, L.~Lavoura, M.~N. Rebelo, M.~Sher and J.~P.
  Silva, \textit{{Theory and phenomenology of two-Higgs-doublet models}},
  \href{https://doi.org/10.1016/j.physrep.2012.02.002}{\textit{Phys. Rept.}
  {\bfseries 516} (2012) 1--102},
  [\href{https://arxiv.org/abs/1106.0034}{{\ttfamily 1106.0034}}].

\bibitem{Basler:2021kgq}
P.~Basler, L.~Biermann, M.~M{\"u}hlleitner and J.~M{\"u}ller,
  \textit{{Electroweak baryogenesis in the CP-violating two-Higgs doublet
  model}}, \href{https://doi.org/10.1140/epjc/s10052-023-11192-9}{\textit{Eur.
  Phys. J. C} {\bfseries 83} (2023) 57},
  [\href{https://arxiv.org/abs/2108.03580}{{\ttfamily 2108.03580}}].

\bibitem{Goncalves:2023svb}
D.~Gon{\c{c}}alves, A.~Kaladharan and Y.~Wu, \textit{{Gravitational waves,
  bubble profile, and baryon asymmetry in the complex 2HDM}},
  \href{https://doi.org/10.1103/PhysRevD.108.075010}{\textit{Phys. Rev. D}
  {\bfseries 108} (2023) 075010},
  [\href{https://arxiv.org/abs/2307.03224}{{\ttfamily 2307.03224}}].

\bibitem{Basso:2012st}
L.~Basso, A.~Lipniacka, F.~Mahmoudi, S.~Moretti, P.~Osland, G.~M. Pruna et~al.,
  \textit{{Probing the charged Higgs boson at the LHC in the CP-violating
  type-II 2HDM}}, \href{https://doi.org/10.1007/JHEP11(2012)011}{\textit{JHEP}
  {\bfseries 11} (2012) 011},
  [\href{https://arxiv.org/abs/1205.6569}{{\ttfamily 1205.6569}}].

\bibitem{Keus:2015hva}
V.~Keus, S.~F. King, S.~Moretti and K.~Yagyu, \textit{{CP Violating
  Two-Higgs-Doublet Model: Constraints and LHC Predictions}},
  \href{https://doi.org/10.1007/JHEP04(2016)048}{\textit{JHEP} {\bfseries 04}
  (2016) 048}, [\href{https://arxiv.org/abs/1510.04028}{{\ttfamily
  1510.04028}}].

\bibitem{Aoki:2018zgq}
M.~Aoki, K.~Hashino, D.~Kaneko, S.~Kanemura and M.~Kubota, \textit{{Probing CP
  violating Higgs sectors via the precision measurement of coupling
  constants}}, \href{https://doi.org/10.1093/ptep/ptz038}{\textit{PTEP}
  {\bfseries 2019} (2019) 053B02},
  [\href{https://arxiv.org/abs/1808.08770}{{\ttfamily 1808.08770}}].

\bibitem{Arhrib:2018qmw}
A.~Arhrib, R.~Benbrik, M.~El~Kacimi, L.~Rahili and S.~Semlali,
  \textit{{Extended Higgs sector of 2HDM with real singlet facing LHC data}},
  \href{https://doi.org/10.1140/epjc/s10052-019-7472-2}{\textit{Eur. Phys. J.
  C} {\bfseries 80} (2020) 13},
  [\href{https://arxiv.org/abs/1811.12431}{{\ttfamily 1811.12431}}].

\bibitem{deBlas:2018mhx}
{\scshape CLIC} collaboration, J.~de~Blas et~al., \textit{{The CLIC Potential
  for New Physics}}, \href{https://doi.org/10.23731/CYRM-2018-003}{\textit{CERN
  Yellow Rep. Monogr.} {\bfseries 3} (2018) 1--282},
  [\href{https://arxiv.org/abs/1812.02093}{{\ttfamily 1812.02093}}].

\bibitem{Wang:2019pet}
X.~Wang, F.~P. Huang and X.~Zhang, \textit{{Gravitational wave and collider
  signals in complex two-Higgs doublet model with dynamical CP-violation at
  finite temperature}},
  \href{https://doi.org/10.1103/PhysRevD.101.015015}{\textit{Phys. Rev. D}
  {\bfseries 101} (2020) 015015},
  [\href{https://arxiv.org/abs/1909.02978}{{\ttfamily 1909.02978}}].

\bibitem{Biekotter:2019kde}
T.~Biek{\"o}tter, M.~Chakraborti and S.~Heinemeyer, \textit{{A 96 GeV Higgs
  boson in the N2HDM}},
  \href{https://doi.org/10.1140/epjc/s10052-019-7561-2}{\textit{Eur. Phys. J.
  C} {\bfseries 80} (2020) 2},
  [\href{https://arxiv.org/abs/1903.11661}{{\ttfamily 1903.11661}}].

\bibitem{Basler:2019iuu}
P.~Basler, M.~M{\"u}hlleitner and J.~M{\"u}ller, \textit{{Electroweak Phase
  Transition in Non-Minimal Higgs Sectors}},
  \href{https://doi.org/10.1007/JHEP05(2020)016}{\textit{JHEP} {\bfseries 05}
  (2020) 016}, [\href{https://arxiv.org/abs/1912.10477}{{\ttfamily
  1912.10477}}].

\bibitem{Chen:2020soj}
N.~Chen, T.~Li, Z.~Teng and Y.~Wu, \textit{{Collapsing domain walls in the
  two-Higgs-doublet model and deep insights from the EDM}},
  \href{https://doi.org/10.1007/JHEP10(2020)081}{\textit{JHEP} {\bfseries 10}
  (2020) 081}, [\href{https://arxiv.org/abs/2006.06913}{{\ttfamily
  2006.06913}}].

\bibitem{Han:2020lta}
T.~Han, S.~Li, S.~Su, W.~Su and Y.~Wu, \textit{{Comparative Studies of 2HDMs
  under the Higgs Boson Precision Measurements}},
  \href{https://doi.org/10.1007/JHEP01(2021)045}{\textit{JHEP} {\bfseries 01}
  (2021) 045}, [\href{https://arxiv.org/abs/2008.05492}{{\ttfamily
  2008.05492}}].

\bibitem{Abouabid:2021yvw}
H.~Abouabid, A.~Arhrib, D.~Azevedo, J.~E. Falaki, P.~M. Ferreira,
  M.~M{\"u}hlleitner et~al., \textit{{Benchmarking di-Higgs production in
  various extended Higgs sector models}},
  \href{https://doi.org/10.1007/JHEP09(2022)011}{\textit{JHEP} {\bfseries 09}
  (2022) 011}, [\href{https://arxiv.org/abs/2112.12515}{{\ttfamily
  2112.12515}}].

\bibitem{Biekotter:2022jyr}
T.~Biek{\"o}tter, S.~Heinemeyer and G.~Weiglein, \textit{{Mounting evidence for
  a 95 GeV Higgs boson}},
  \href{https://doi.org/10.1007/JHEP08(2022)201}{\textit{JHEP} {\bfseries 08}
  (2022) 201}, [\href{https://arxiv.org/abs/2203.13180}{{\ttfamily
  2203.13180}}].

\bibitem{Azevedo:2023zkg}
D.~Azevedo, T.~Biek{\"o}tter and P.~M. Ferreira, \textit{{2HDM interpretations
  of the CMS diphoton excess at 95 GeV}},
  \href{https://doi.org/10.1007/JHEP11(2023)017}{\textit{JHEP} {\bfseries 11}
  (2023) 017}, [\href{https://arxiv.org/abs/2305.19716}{{\ttfamily
  2305.19716}}].

\bibitem{Biekotter:2024ykp}
T.~Biek{\"o}tter, D.~Fontes, M.~M{\"u}hlleitner, J.~C. Rom{\~a}o, R.~Santos and
  J.~P. Silva, \textit{{Impact of new experimental data on the C2HDM: the
  strong interdependence between LHC Higgs data and the electron EDM}},
  \href{https://doi.org/10.1007/JHEP05(2024)127}{\textit{JHEP} {\bfseries 05}
  (2024) 127}, [\href{https://arxiv.org/abs/2403.02425}{{\ttfamily
  2403.02425}}].

\bibitem{Biekotter:2025fjx}
T.~Biek{\"o}tter and M.~O. Olea-Romacho, \textit{{Benchmarking a fading window:
  electroweak baryogenesis in the C2HDM, LHC constraints after Run 2 and
  prospects for LISA}},  \href{https://arxiv.org/abs/2505.09670}{{\ttfamily
  2505.09670}}.
  
  %\cite{Davila:2025goc}
\bibitem{Davila:2025goc}
J.~M.~D{\'a}vila, A.~Karan, E.~Passemar, A.~Pich and L.~Vale Silva,
``The Electric Dipole Moment of the electron in the decoupling limit of the aligned two-Higgs doublet model,''
JHEP \textbf{10} (2025), 053
doi:10.1007/JHEP10(2025)053
[arXiv:2504.16700 [hep-ph]].
%9 citations counted in INSPIRE as of 30 Dec 2025


\bibitem{Coy:2019rfr}
R.~Coy, M.~Frigerio, F.~Mescia and O.~Sumensari, \textit{{New physics in $b\to
  s\ell\ell$ transitions at one loop}},
  \href{https://doi.org/10.1140/epjc/s10052-019-7581-y}{\textit{Eur. Phys. J.
  C} {\bfseries 80} (2020) 52},
  [\href{https://arxiv.org/abs/1909.08567}{{\ttfamily 1909.08567}}].

\bibitem{Frank:2021pkc}
M.~Frank, E.~G. Fuakye and M.~Toharia, \textit{{Restricting the parameter space
  of type-II two-Higgs-doublet models with CP violation}},
  \href{https://doi.org/10.1103/PhysRevD.106.035010}{\textit{Phys. Rev. D}
  {\bfseries 106} (2022) 035010},
  [\href{https://arxiv.org/abs/2112.14295}{{\ttfamily 2112.14295}}].

\bibitem{Jung:2013hka}
M.~Jung and A.~Pich, \textit{{Electric Dipole Moments in Two-Higgs-Doublet
  Models}}, \href{https://doi.org/10.1007/JHEP04(2014)076}{\textit{JHEP}
  {\bfseries 04} (2014) 076},
  [\href{https://arxiv.org/abs/1308.6283}{{\ttfamily 1308.6283}}].

\bibitem{Yamanaka:2018uud}
{\scshape JLQCD} collaboration, N.~Yamanaka, S.~Hashimoto, T.~Kaneko and
  H.~Ohki, \textit{{Nucleon charges with dynamical overlap fermions}},
  \href{https://doi.org/10.1103/PhysRevD.98.054516}{\textit{Phys. Rev. D}
  {\bfseries 98} (2018) 054516},
  [\href{https://arxiv.org/abs/1805.10507}{{\ttfamily 1805.10507}}].

\bibitem{Fuyuto:2019svr}
K.~Fuyuto, W.-S. Hou and E.~Senaha, \textit{{Cancellation mechanism for the
  electron electric dipole moment connected with the baryon asymmetry of the
  Universe}}, \href{https://doi.org/10.1103/PhysRevD.101.011901}{\textit{Phys.
  Rev. D} {\bfseries 101} (2020) 011901},
  [\href{https://arxiv.org/abs/1910.12404}{{\ttfamily 1910.12404}}].

\bibitem{Altmannshofer:2020shb}
W.~Altmannshofer, S.~Gori, N.~Hamer and H.~H. Patel, \textit{{Electron EDM in
  the complex two-Higgs doublet model}},
  \href{https://doi.org/10.1103/PhysRevD.102.115042}{\textit{Phys. Rev. D}
  {\bfseries 102} (2020) 115042},
  [\href{https://arxiv.org/abs/2009.01258}{{\ttfamily 2009.01258}}].

\bibitem{Cheung:2020ugr}
K.~Cheung, A.~Jueid, Y.-N. Mao and S.~Moretti, \textit{{Two-Higgs-doublet model
  with soft $CP$ violation confronting electric dipole moments and colliders}},
  \href{https://doi.org/10.1103/PhysRevD.102.075029}{\textit{Phys. Rev. D}
  {\bfseries 102} (2020) 075029},
  [\href{https://arxiv.org/abs/2003.04178}{{\ttfamily 2003.04178}}].

\bibitem{Inoue:2014nva}
S.~Inoue, M.~J. Ramsey-Musolf and Y.~Zhang, \textit{{CP-violating phenomenology
  of flavor conserving two Higgs doublet models}},
  \href{https://doi.org/10.1103/PhysRevD.89.115023}{\textit{Phys. Rev. D}
  {\bfseries 89} (2014) 115023},
  [\href{https://arxiv.org/abs/1403.4257}{{\ttfamily 1403.4257}}].

\bibitem{Cheung:2014oaa}
K.~Cheung, J.~S. Lee, E.~Senaha and P.-Y. Tseng, \textit{{Confronting
  Higgcision with Electric Dipole Moments}},
  \href{https://doi.org/10.1007/JHEP06(2014)149}{\textit{JHEP} {\bfseries 06}
  (2014) 149}, [\href{https://arxiv.org/abs/1403.4775}{{\ttfamily 1403.4775}}].

\bibitem{Chen:2015gaa}
C.-Y. Chen, S.~Dawson and Y.~Zhang, \textit{{Complementarity of LHC and EDMs
  for Exploring Higgs CP Violation}},
  \href{https://doi.org/10.1007/JHEP06(2015)056}{\textit{JHEP} {\bfseries 06}
  (2015) 056}, [\href{https://arxiv.org/abs/1503.01114}{{\ttfamily
  1503.01114}}].

\bibitem{Roussy:2022cmp}
T.~S. Roussy et~al., \textit{{An improved bound on the
  electron{\textquoteright}s electric dipole moment}},
  \href{https://doi.org/10.1126/science.adg4084}{\textit{Science} {\bfseries
  381} (2023) adg4084}, [\href{https://arxiv.org/abs/2212.11841}{{\ttfamily
  2212.11841}}].

\bibitem{ACME:2018yjb}
{\scshape ACME} collaboration, V.~Andreev et~al., \textit{{Improved limit on
  the electric dipole moment of the electron}},
  \href{https://doi.org/10.1038/s41586-018-0599-8}{\textit{Nature} {\bfseries
  562} (2018) 355--360}.

\bibitem{BhupalDev:2014bir}
P.~S.~Bhupal Dev and A.~Pilaftsis, \textit{{Maximally Symmetric Two Higgs
  Doublet Model with Natural Standard Model Alignment}},
  \href{https://doi.org/10.1007/JHEP12(2014)024}{\textit{JHEP} {\bfseries 12}
  (2014) 024}, [erratum:
  \href{https://doi.org/10.1007/JHEP11(2015)147}{\textit{JHEP} {\bfseries 11}
  (2015) 147}], [\href{https://arxiv.org/abs/1408.3405}{{\ttfamily 1408.3405}}].

\bibitem{Darvishi:2023fjh}
N.~Darvishi, A.~Pilaftsis and J.~H.~Yu, \textit{{Maximising CP Violation in
  naturally aligned Two-Higgs Doublet Models}},
  \href{https://doi.org/10.1007/JHEP05(2024)233}{\textit{JHEP} {\bfseries 05}
  (2024) 233}, [\href{https://arxiv.org/abs/2312.00882}{{\ttfamily 2312.00882}}].

\bibitem{Pilaftsis:2025wpz}
A.~Pilaftsis and N.~Darvishi, \textit{{Mixed CP Violation and Natural Alignment
  in 2HDMs}}, \href{https://doi.org/10.22323/1.490.0119}{\textit{PoS}
  {\bfseries CORFU2024} (2025) 119},
  [\href{https://arxiv.org/abs/2503.18588}{{\ttfamily 2503.18588}}].
  
  \bibitem{Khater:2003ym}
W.~Khater and P.~Osland, \textit{{Maximal CP nonconservation in the two Higgs
  doublet model}}, {\textit{Acta Phys. Polon. B} {\bfseries 34} (2003)
  4531--4548}, [\href{https://arxiv.org/abs/hep-ph/0305308}{{\ttfamily
  hep-ph/0305308}}].

\bibitem{Glashow:1976nt}
S.~L. Glashow and S.~Weinberg, \textit{{Natural Conservation Laws for Neutral
  Currents}}, \href{https://doi.org/10.1103/PhysRevD.15.1958}{\textit{Phys.
  Rev. D} {\bfseries 15} (1977) 1958}.


\bibitem{Grzadkowski:2014ada}
B.~Grzadkowski, O.~M.~Ogreid and P.~Osland, \textit{{Measuring CP violation in
  Two-Higgs-Doublet models in light of the LHC Higgs data}},
  \href{https://doi.org/10.1007/JHEP11(2014)084}{\textit{JHEP} {\bfseries 11}
  (2014) 084}, [\href{https://arxiv.org/abs/1409.7265}{{\ttfamily 1409.7265}}].

\bibitem{Grzadkowski:2015zma}
B.~Grzadkowski, O.~M.~Ogreid and P.~Osland, \textit{{Testing the presence of CP
  violation in the 2HDM}},
  \href{https://doi.org/10.22323/1.231.0086}{\textit{PoS} {\bfseries CORFU2014}
  (2015) 086}, [\href{https://arxiv.org/abs/1504.06076}{{\ttfamily 1504.06076}}].

\bibitem{Fontes:2017zfn}
D.~Fontes, M.~M{\"u}hlleitner, J.~C. Rom{\~a}o, R.~Santos, J.~P. Silva and
  J.~Wittbrodt, \textit{{The C2HDM revisited}},
  \href{https://doi.org/10.1007/JHEP02(2018)073}{\textit{JHEP} {\bfseries 02}
  (2018) 073}, [\href{https://arxiv.org/abs/1711.09419}{{\ttfamily
  1711.09419}}].

\bibitem{Fontes:2022izp}
D.~Fontes and J.~C. Rom{\~a}o, \textit{{The one-loop impact of a dependent
  mass: the role of m$_{3}$ in the C2HDM}},
  \href{https://doi.org/10.1007/JHEP03(2022)144}{\textit{JHEP} {\bfseries 03}
  (2022) 144}, [\href{https://arxiv.org/abs/2201.02479}{{\ttfamily
  2201.02479}}].

\bibitem{Capucha:2025qgr}
R.~Capucha, {\'A}.~Lozano-Onrubia, L.~Merlo, J.~M. No and R.~Santos,
  \textit{{V-associated production and vector boson fusion as an LHC signature
  of CP violation}}, \href{https://doi.org/10.1103/3s5k-h11w}{\textit{Phys.
  Rev. D} {\bfseries 112} (2025) 095005},
  [\href{https://arxiv.org/abs/2507.05942}{{\ttfamily 2507.05942}}].

\bibitem{Mendez:1991gp}
A.~Mendez and A.~Pomarol, \textit{{Signals of CP violation in the Higgs
  sector}}, \href{https://doi.org/10.1016/0370-2693(91)91836-K}{\textit{Phys.
  Lett. B} {\bfseries 272} (1991) 313--318}.


\bibitem{Paschos:1976ay}
E.~A. Paschos, \textit{{Diagonal Neutral Currents}},
  \href{https://doi.org/10.1103/PhysRevD.15.1966}{\textit{Phys. Rev. D}
  {\bfseries 15} (1977) 1966}.

\bibitem{Ginzburg:2002wt}
I.~F. Ginzburg, M.~Krawczyk and P.~Osland, \textit{{Two Higgs doublet models
  with CP violation}},  in \textit{{International Workshop on Linear Colliders
  (LCWS 2002)}}, pp.~703--706, 11, 2002,
  \href{https://arxiv.org/abs/hep-ph/0211371}{{\ttfamily hep-ph/0211371}}.

\bibitem{Lavoura:1994fv}
L.~Lavoura and J.~P. Silva, \textit{{Fundamental CP violating quantities in a
  SU(2) x U(1) model with many Higgs doublets}},
  \href{https://doi.org/10.1103/PhysRevD.50.4619}{\textit{Phys. Rev. D}
  {\bfseries 50} (1994) 4619--4624},
  [\href{https://arxiv.org/abs/hep-ph/9404276}{{\ttfamily hep-ph/9404276}}].

\bibitem{Botella:1994cs}
F.~J. Botella and J.~P. Silva, \textit{{Jarlskog - like invariants for theories
  with scalars and fermions}},
  \href{https://doi.org/10.1103/PhysRevD.51.3870}{\textit{Phys. Rev. D}
  {\bfseries 51} (1995) 3870--3875},
  [\href{https://arxiv.org/abs/hep-ph/9411288}{{\ttfamily hep-ph/9411288}}].

\bibitem{Davidson:2005cw}
S.~Davidson and H.~E. Haber, \textit{{Basis-independent methods for the
  two-Higgs-doublet model}},
  \href{https://doi.org/10.1103/PhysRevD.72.099902}{\textit{Phys. Rev. D}
  {\bfseries 72} (2005) 035004},
  [\href{https://arxiv.org/abs/hep-ph/0504050}{{\ttfamily hep-ph/0504050}}].
  [Erratum: Phys.Rev.D 72, 099902 (2005)].

\bibitem{Haber:2010bw}
H.~E. Haber and D.~O'Neil, \textit{{Basis-independent methods for the
  two-Higgs-doublet model III: The CP-conserving limit, custodial symmetry, and
  the oblique parameters S, T, U}},
  \href{https://doi.org/10.1103/PhysRevD.83.055017}{\textit{Phys. Rev. D}
  {\bfseries 83} (2011) 055017},
  [\href{https://arxiv.org/abs/1011.6188}{{\ttfamily 1011.6188}}].

\bibitem{ElKaffas:2007rq}
A.~W. El~Kaffas, P.~Osland and O.~M. Ogreid, \textit{{CP violation, stability
  and unitarity of the two Higgs doublet model}}, {\textit{Nonlin. Phenom.
  Complex Syst.} {\bfseries 10} (2007) 347--357},
  [\href{https://arxiv.org/abs/hep-ph/0702097}{{\ttfamily hep-ph/0702097}}].

\bibitem{ATLAS:2022vkf}
{\scshape ATLAS} collaboration, G.~Aad et~al., \textit{{A detailed map of Higgs
  boson interactions by the ATLAS experiment ten years after the discovery}},
  \href{https://doi.org/10.1038/s41586-022-04893-w}{\textit{Nature} {\bfseries
  607} (2022) 52--59}, [\href{https://arxiv.org/abs/2207.00092}{{\ttfamily
  2207.00092}}]. [Erratum: Nature 612, E24 (2022)].

\bibitem{CMS:2022dwd}
{\scshape CMS} collaboration, A.~Tumasyan et~al., \textit{{A portrait of the
  Higgs boson by the CMS experiment ten years after the discovery.}},
  \href{https://doi.org/10.1038/s41586-022-04892-x}{\textit{Nature} {\bfseries
  607} (2022) 60--68}, [\href{https://arxiv.org/abs/2207.00043}{{\ttfamily
  2207.00043}}]. [Erratum: Nature 623, (2023)].

\bibitem{Sopczak:2025ptc}
{\scshape ATLAS, CMS} collaboration, A.~Sopczak, \textit{{Prospects for Higgs
  Boson Research at the LHC}},
  \href{https://doi.org/10.5506/APhysPolB.56.9-A4}{\textit{Acta Phys. Polon. B}
  {\bfseries 56} (2025) 9--A4},
  [\href{https://arxiv.org/abs/2509.22455}{{\ttfamily 2509.22455}}].

\bibitem{Branco:1999fs}
G.~C. Branco, L.~Lavoura and J.~P. Silva, \textit{{CP Violation}}, Published in: Int.Ser.Monogr.Phys. 103 (1999) 1-536,  \href{https://doi.org/10.1093/oso/9780198503996.001.0001}{10.1093/oso/9780198503996.001.0001}.

\bibitem{Fontes:2015xva}
D.~Fontes, J.~C. Rom{\~a}o, R.~Santos and J.~P. Silva, \textit{{Undoubtable
  signs of $CP$-violation in Higgs boson decays at the LHC run 2}},
  \href{https://doi.org/10.1103/PhysRevD.92.055014}{\textit{Phys. Rev. D}
  {\bfseries 92} (2015) 055014},
  [\href{https://arxiv.org/abs/1506.06755}{{\ttfamily 1506.06755}}].

\bibitem{Coimbra:2013qq}
R.~Coimbra, M.~O.~P. Sampaio and R.~Santos, \textit{{ScannerS: Constraining the
  phase diagram of a complex scalar singlet at the LHC}},
  \href{https://doi.org/10.1140/epjc/s10052-013-2428-4}{\textit{Eur. Phys. J.
  C} {\bfseries 73} (2013) 2428},
  [\href{https://arxiv.org/abs/1301.2599}{{\ttfamily 1301.2599}}].

\bibitem{Muhlleitner:2020wwk}
M.~M{\"u}hlleitner, M.~O.~P. Sampaio, R.~Santos and J.~Wittbrodt,
  \textit{{ScannerS: parameter scans in extended scalar sectors}},
  \href{https://doi.org/10.1140/epjc/s10052-022-10139-w}{\textit{Eur. Phys. J.
  C} {\bfseries 82} (2022) 198},
  [\href{https://arxiv.org/abs/2007.02985}{{\ttfamily 2007.02985}}].

\bibitem{Ivanov:2006yq}
I.~P. Ivanov, \textit{{Minkowski space structure of the Higgs potential in
  2HDM}}, \href{https://doi.org/10.1103/PhysRevD.75.035001}{\textit{Phys. Rev.
  D} {\bfseries 75} (2007) 035001},
  [\href{https://arxiv.org/abs/hep-ph/0609018}{{\ttfamily hep-ph/0609018}}].
  [Erratum: Phys.Rev.D 76, 039902 (2007)].

\bibitem{Ivanov:2008cxa}
I.~P. Ivanov, \textit{{General two-order-parameter Ginzburg-Landau model with
  quadratic and quartic interactions}},
  \href{https://doi.org/10.1103/PhysRevE.79.021116}{\textit{Phys. Rev. E}
  {\bfseries 79} (2009) 021116},
  [\href{https://arxiv.org/abs/0802.2107}{{\ttfamily 0802.2107}}].

\bibitem{Barroso:2012mj}
A.~Barroso, P.~M. Ferreira, I.~P. Ivanov, R.~Santos and J.~P. Silva,
  \textit{{Evading death by vacuum}},
  \href{https://doi.org/10.1140/epjc/s10052-013-2537-0}{\textit{Eur. Phys. J.
  C} {\bfseries 73} (2013) 2537},
  [\href{https://arxiv.org/abs/1211.6119}{{\ttfamily 1211.6119}}].

\bibitem{Barroso:2013awa}
A.~Barroso, P.~M. Ferreira, I.~P. Ivanov and R.~Santos, \textit{{Metastability
  bounds on the two Higgs doublet model}},
  \href{https://doi.org/10.1007/JHEP06(2013)045}{\textit{JHEP} {\bfseries 06}
  (2013) 045}, [\href{https://arxiv.org/abs/1303.5098}{{\ttfamily 1303.5098}}].

\bibitem{Peskin:1991sw}
M.~E. Peskin and T.~Takeuchi, \textit{{Estimation of oblique electroweak
  corrections}}, \href{https://doi.org/10.1103/PhysRevD.46.381}{\textit{Phys.
  Rev. D} {\bfseries 46} (1992) 381--409}.

\bibitem{ParticleDataGroup:2024cfk}
{\scshape Particle Data Group} collaboration, S.~Navas et~al., \textit{{Review
  of particle physics}},
  \href{https://doi.org/10.1103/PhysRevD.110.030001}{\textit{Phys. Rev. D}
  {\bfseries 110} (2024) 030001}.

\bibitem{Haller:2018nnx}
J.~Haller, A.~Hoecker, R.~Kogler, K.~M{\"o}nig, T.~Peiffer and J.~Stelzer,
  \textit{{Update of the global electroweak fit and constraints on
  two-Higgs-doublet models}},
  \href{https://doi.org/10.1140/epjc/s10052-018-6131-3}{\textit{Eur. Phys. J.
  C} {\bfseries 78} (2018) 675},
  [\href{https://arxiv.org/abs/1803.01853}{{\ttfamily 1803.01853}}].

\bibitem{Arbey:2017gmh}
A.~Arbey, F.~Mahmoudi, O.~Stal and T.~Stefaniak, \textit{{Status of the Charged
  Higgs Boson in Two Higgs Doublet Models}},
  \href{https://doi.org/10.1140/epjc/s10052-018-5651-1}{\textit{Eur. Phys. J.
  C} {\bfseries 78} (2018) 182},
  [\href{https://arxiv.org/abs/1706.07414}{{\ttfamily 1706.07414}}].

\bibitem{Sanyal:2019xcp}
P.~Sanyal, \textit{{Limits on the Charged Higgs Parameters in the Two Higgs
  Doublet Model using CMS $\sqrt{s}=13$ TeV Results}},
  \href{https://doi.org/10.1140/epjc/s10052-019-7431-y}{\textit{Eur. Phys. J.
  C} {\bfseries 79} (2019) 913},
  [\href{https://arxiv.org/abs/1906.02520}{{\ttfamily 1906.02520}}].

\bibitem{Misiak:2017bgg}
M.~Misiak and M.~Steinhauser, \textit{{Weak radiative decays of the B meson and
  bounds on $M_{H^\pm }$ in the Two-Higgs-Doublet Model}},
  \href{https://doi.org/10.1140/epjc/s10052-017-4776-y}{\textit{Eur. Phys. J.
  C} {\bfseries 77} (2017) 201},
  [\href{https://arxiv.org/abs/1702.04571}{{\ttfamily 1702.04571}}].

\bibitem{Belle:2017hum}
{\scshape Belle} collaboration, T.~Horiguchi et~al., \textit{{Evidence for
  Isospin Violation and Measurement of $CP$ Asymmetries in $B \to K^{\ast}(892)
  \gamma$}},
  \href{https://doi.org/10.1103/PhysRevLett.119.191802}{\textit{Phys. Rev.
  Lett.} {\bfseries 119} (2017) 191802},
  [\href{https://arxiv.org/abs/1707.00394}{{\ttfamily 1707.00394}}].

\bibitem{Belle:2014sac}
{\scshape Belle} collaboration, D.~Dutta et~al., \textit{{Search for
  $B_{s}^{0}\rightarrow\gamma\gamma$ and a measurement of the branching
  fraction for $B_{s}^{0}\rightarrow\phi\gamma$}},
  \href{https://doi.org/10.1103/PhysRevD.91.011101}{\textit{Phys. Rev. D}
  {\bfseries 91} (2015) 011101},
  [\href{https://arxiv.org/abs/1411.7771}{{\ttfamily 1411.7771}}].

\bibitem{Bahl:2022igd}
H.~Bahl, T.~Biek{\"o}tter, S.~Heinemeyer, C.~Li, S.~Paasch, G.~Weiglein et~al.,
  \textit{{HiggsTools: BSM scalar phenomenology with new versions of
  HiggsBounds and HiggsSignals}},
  \href{https://doi.org/10.1016/j.cpc.2023.108803}{\textit{Comput. Phys.
  Commun.} {\bfseries 291} (2023) 108803},
  [\href{https://arxiv.org/abs/2210.09332}{{\ttfamily 2210.09332}}].

\bibitem{Bechtle:2020uwn}
P.~Bechtle, S.~Heinemeyer, T.~Klingl, T.~Stefaniak, G.~Weiglein and
  J.~Wittbrodt, \textit{{HiggsSignals-2: Probing new physics with precision
  Higgs measurements in the LHC 13 TeV era}},
  \href{https://doi.org/10.1140/epjc/s10052-021-08942-y}{\textit{Eur. Phys. J.
  C} {\bfseries 81} (2021) 145},
  [\href{https://arxiv.org/abs/2012.09197}{{\ttfamily 2012.09197}}].

\bibitem{Bechtle:2020pkv}
P.~Bechtle, D.~Dercks, S.~Heinemeyer, T.~Klingl, T.~Stefaniak, G.~Weiglein
  et~al., \textit{{HiggsBounds-5: Testing Higgs Sectors in the LHC 13 TeV
  Era}}, \href{https://doi.org/10.1140/epjc/s10052-020-08557-9}{\textit{Eur.
  Phys. J. C} {\bfseries 80} (2020) 1211},
  [\href{https://arxiv.org/abs/2006.06007}{{\ttfamily 2006.06007}}].

\bibitem{ATLAS:2022akr}
{\scshape ATLAS} collaboration, G.~Aad et~al., \textit{{Measurement of the CP
  properties of Higgs boson interactions with $\tau $-leptons with the ATLAS
  detector}},
  \href{https://doi.org/10.1140/epjc/s10052-023-11583-y}{\textit{Eur. Phys. J.
  C} {\bfseries 83} (2023) 563},
  [\href{https://arxiv.org/abs/2212.05833}{{\ttfamily 2212.05833}}].

\bibitem{Barr:1990vd}
S.~M.~Barr and A.~Zee, \textit{{Electric Dipole Moment of the Electron and of
  the Neutron}}, \href{https://doi.org/10.1103/PhysRevLett.65.21}{\textit{Phys.
  Rev. Lett.} {\bfseries 65} (1990) 21--24}, [erratum:
  \href{https://doi.org/10.1103/PhysRevLett.65.2920}{\textit{Phys. Rev. Lett.}
  {\bfseries 65} (1990) 2920}].

\bibitem{Pospelov:2005pr}
M.~Pospelov and A.~Ritz, \textit{{Electric dipole moments as probes of new
  physics}}, \href{https://doi.org/10.1016/j.aop.2005.04.002}{\textit{Annals
  Phys.} {\bfseries 318} (2005) 119--169},
  [\href{https://arxiv.org/abs/hep-ph/0504231}{{\ttfamily hep-ph/0504231}}].

\bibitem{Abe:2013qla}
T.~Abe, J.~Hisano, T.~Kitahara and K.~Tobioka, \textit{{Gauge invariant
  Barr-Zee type contributions to fermionic EDMs in the two-Higgs doublet
  models}}, \href{https://doi.org/10.1007/JHEP01(2014)106}{\textit{JHEP}
  {\bfseries 01} (2014) 106},
  [\href{https://arxiv.org/abs/1311.4704}{{\ttfamily 1311.4704}}]. [Erratum:
  JHEP 04, 161 (2016)].

\bibitem{Bechtle:2013xfa}
P.~Bechtle, S.~Heinemeyer, O.~St{\r{a}}l, T.~Stefaniak and G.~Weiglein,
  \textit{{$HiggsSignals$: Confronting arbitrary Higgs sectors with
  measurements at the Tevatron and the LHC}},
  \href{https://doi.org/10.1140/epjc/s10052-013-2711-4}{\textit{Eur. Phys. J.
  C} {\bfseries 74} (2014) 2711},
  [\href{https://arxiv.org/abs/1305.1933}{{\ttfamily 1305.1933}}].

\bibitem{ParticleDataGroup:2020ssz}
{\scshape Particle Data Group} collaboration, P.~A. Zyla et~al.,
  \textit{{Review of Particle Physics}},
  \href{https://doi.org/10.1093/ptep/ptaa104}{\textit{PTEP} {\bfseries 2020}
  (2020) 083C01}.

\bibitem{ATLAS:2016neq}
{\scshape ATLAS, CMS} collaboration, G.~Aad et~al., \textit{{Measurements of
  the Higgs boson production and decay rates and constraints on its couplings
  from a combined ATLAS and CMS analysis of the LHC pp collision data at $
  \sqrt{s}=7 $ and 8 TeV}},
  \href{https://doi.org/10.1007/JHEP08(2016)045}{\textit{JHEP} {\bfseries 08}
  (2016) 045}, [\href{https://arxiv.org/abs/1606.02266}{{\ttfamily
  1606.02266}}].

\bibitem{Misiak:2020vlo}
M.~Misiak, A.~Rehman and M.~Steinhauser, \textit{{Towards $ \overline{B}\to
  {X}_s\gamma $ at the NNLO in QCD without interpolation in m$_{c}$}},
  \href{https://doi.org/10.1007/JHEP06(2020)175}{\textit{JHEP} {\bfseries 06}
  (2020) 175}, [\href{https://arxiv.org/abs/2002.01548}{{\ttfamily
  2002.01548}}].

\bibitem{ACME:2013pal}
{\scshape ACME} collaboration, J.~Baron et~al., \textit{{Order of Magnitude
  Smaller Limit on the Electric Dipole Moment of the Electron}},
  \href{https://doi.org/10.1126/science.1248213}{\textit{Science} {\bfseries
  343} (2014) 269--272}, [\href{https://arxiv.org/abs/1310.7534}{{\ttfamily
  1310.7534}}].

\bibitem{ATLAS:2020hpj}
{\scshape ATLAS} collaboration, G.~Aad et~al., \textit{{Evidence for
  $t\bar{t}t\bar{t}$ production in the multilepton final state in
  proton{\textendash}proton collisions at $\sqrt{s}=13$ $\text {TeV}$ with the
  ATLAS detector}},
  \href{https://doi.org/10.1140/epjc/s10052-020-08509-3}{\textit{Eur. Phys. J.
  C} {\bfseries 80} (2020) 1085},
  [\href{https://arxiv.org/abs/2007.14858}{{\ttfamily 2007.14858}}].

\bibitem{Aoude:2022deh}
R.~Aoude, H.~El~Faham, F.~Maltoni and E.~Vryonidou, \textit{{Complete SMEFT
  predictions for four top quark production at hadron colliders}},
  \href{https://doi.org/10.1007/JHEP10(2022)163}{\textit{JHEP} {\bfseries 10}
  (2022) 163}, [\href{https://arxiv.org/abs/2208.04962}{{\ttfamily
  2208.04962}}].

\bibitem{vanBeekveld:2022hty}
M.~van Beekveld, A.~Kulesza and L.~M. Valero, \textit{{Threshold Resummation
  for the Production of Four Top Quarks at the LHC}},
  \href{https://doi.org/10.1103/PhysRevLett.131.211901}{\textit{Phys. Rev.
  Lett.} {\bfseries 131} (2023) 211901},
  [\href{https://arxiv.org/abs/2212.03259}{{\ttfamily 2212.03259}}].

\bibitem{CMS:2023ftu}
{\scshape CMS} collaboration, A.~Hayrapetyan et~al., \textit{{Observation of
  four top quark production in proton-proton collisions at s=13TeV}},
  \href{https://doi.org/10.1016/j.physletb.2023.138290}{\textit{Phys. Lett. B}
  {\bfseries 847} (2023) 138290},
  [\href{https://arxiv.org/abs/2305.13439}{{\ttfamily 2305.13439}}].

\bibitem{ATLAS:2023ajo}
{\scshape ATLAS} collaboration, G.~Aad et~al., \textit{{Observation of
  four-top-quark production in the multilepton final state with the ATLAS
  detector}},
  \href{https://doi.org/10.1140/epjc/s10052-023-11573-0}{\textit{Eur. Phys. J.
  C} {\bfseries 83} (2023) 496},
  [\href{https://arxiv.org/abs/2303.15061}{{\ttfamily 2303.15061}}]. [Erratum:
  Eur.Phys.J.C 84, 156 (2024)].

\bibitem{NL-eEDM:2018lno}
{\scshape NL-eEDM} collaboration, P.~Aggarwal et~al., \textit{{Measuring the
  electric dipole moment of the electron in BaF}},
  \href{https://doi.org/10.1140/epjd/e2018-90192-9}{\textit{Eur. Phys. J. D}
  {\bfseries 72} (2018) 197},
  [\href{https://arxiv.org/abs/1804.10012}{{\ttfamily 1804.10012}}].

\bibitem{Chupp:2014gka}
T.~Chupp and M.~Ramsey-Musolf, \textit{{Electric Dipole Moments: A Global
  Analysis}}, \href{https://doi.org/10.1103/PhysRevC.91.035502}{\textit{Phys.
  Rev. C} {\bfseries 91} (2015) 035502},
  [\href{https://arxiv.org/abs/1407.1064}{{\ttfamily 1407.1064}}].

\bibitem{Chupp:2017rkp}
T.~Chupp, P.~Fierlinger, M.~Ramsey-Musolf and J.~Singh, \textit{{Electric
  dipole moments of atoms, molecules, nuclei, and particles}},
  \href{https://doi.org/10.1103/RevModPhys.91.015001}{\textit{Rev. Mod. Phys.}
  {\bfseries 91} (2019) 015001},
  [\href{https://arxiv.org/abs/1710.02504}{{\ttfamily 1710.02504}}].

\bibitem{EuropwanEDMprojects:2025okn}
{\scshape Europwan EDM projects} collaboration, M.~Athanasakis-Kaklamanakis
  et~al., \textit{{Community input to the European Strategy on particle
  physics: Searches for Permanent Electric Dipole Moments}},
  \href{https://arxiv.org/abs/2505.22281}{{\ttfamily 2505.22281}}.

\bibitem{Pospelov:2025vzj}
M.~Pospelov and A.~Ritz, \textit{{Electric Dipole Moments and New Physics}},
  \href{https://arxiv.org/abs/2509.23531}{{\ttfamily 2509.23531}}.

\bibitem{Djouadi:1997yw}
A.~Djouadi, J.~Kalinowski and M.~Spira, \textit{{HDECAY: A Program for Higgs
  boson decays in the standard model and its supersymmetric extension}},
  \href{https://doi.org/10.1016/S0010-4655(97)00123-9}{\textit{Comput. Phys.
  Commun.} {\bfseries 108} (1998) 56--74},
  [\href{https://arxiv.org/abs/hep-ph/9704448}{{\ttfamily hep-ph/9704448}}].

\bibitem{Djouadi:2018xqq}
{\scshape HDECAY} collaboration, A.~Djouadi, J.~Kalinowski, M.~Muehlleitner and
  M.~Spira, \textit{{HDECAY: Twenty$_{++}$ years after}},
  \href{https://doi.org/10.1016/j.cpc.2018.12.010}{\textit{Comput. Phys.
  Commun.} {\bfseries 238} (2019) 214--231},
  [\href{https://arxiv.org/abs/1801.09506}{{\ttfamily 1801.09506}}].

\bibitem{Harlander:2012pb}
R.~V. Harlander, S.~Liebler and H.~Mantler, \textit{{SusHi: A program for the
  calculation of Higgs production in gluon fusion and bottom-quark annihilation
  in the Standard Model and the MSSM}},
  \href{https://doi.org/10.1016/j.cpc.2013.02.006}{\textit{Comput. Phys.
  Commun.} {\bfseries 184} (2013) 1605--1617},
  [\href{https://arxiv.org/abs/1212.3249}{{\ttfamily 1212.3249}}].

\bibitem{Harlander:2016vzb}
R.~V. Harlander and T.~Neumann, \textit{{The perturbative QCD gradient flow to
  three loops}}, \href{https://doi.org/10.1007/JHEP06(2016)161}{\textit{JHEP}
  {\bfseries 06} (2016) 161},
  [\href{https://arxiv.org/abs/1606.03756}{{\ttfamily 1606.03756}}].

\bibitem{Brein:2012ne}
O.~Brein, R.~V. Harlander and T.~J.~E. Zirke, \textit{{vh@nnlo - Higgs
  Strahlung at hadron colliders}},
  \href{https://doi.org/10.1016/j.cpc.2012.11.002}{\textit{Comput. Phys.
  Commun.} {\bfseries 184} (2013) 998--1003},
  [\href{https://arxiv.org/abs/1210.5347}{{\ttfamily 1210.5347}}].

\bibitem{Harlander:2018zpi}
R.~V. Harlander, Y.~Kluth and F.~Lange, \textit{{The two-loop
  energy{\textendash}momentum tensor within the gradient-flow formalism}},
  \href{https://doi.org/10.1140/epjc/s10052-018-6415-7}{\textit{Eur. Phys. J.
  C} {\bfseries 78} (2018) 944},
  [\href{https://arxiv.org/abs/1808.09837}{{\ttfamily 1808.09837}}]. [Erratum:
  Eur.Phys.J.C 79, 858 (2019)].

\bibitem{CMS:2022ahq}
{\scshape CMS} collaboration, A.~Tumasyan et~al., \textit{{Search for Higgs
  boson decays to a Z boson and a photon in proton-proton collisions at $
  \sqrt{s} $ = 13 TeV}},
  \href{https://doi.org/10.1007/JHEP05(2023)233}{\textit{JHEP} {\bfseries 05}
  (2023) 233}, [\href{https://arxiv.org/abs/2204.12945}{{\ttfamily
  2204.12945}}].

\bibitem{ATLAS:2025aip}
{\scshape ATLAS} collaboration, G.~Aad et~al., \textit{{Search for the Higgs
  boson decay to a $Z$ boson and a photon in $pp$ collisions at $\sqrt{s}=13$
  TeV and $13.6$ TeV with the ATLAS detector}},
  \href{https://arxiv.org/abs/2507.12598}{{\ttfamily 2507.12598}}.

\bibitem{ATLAS:2022tnm}
{\scshape ATLAS} collaboration, G.~Aad et~al., \textit{{Measurement of the
  properties of Higgs boson production at $\sqrt{s} = 13$ TeV in the
  $H\to\gamma\gamma$ channel using $139$ fb$^{-1}$ of $pp$ collision data with
  the ATLAS experiment}},
  \href{https://doi.org/10.1007/JHEP07(2023)088}{\textit{JHEP} {\bfseries 07}
  (2023) 088}, [\href{https://arxiv.org/abs/2207.00348}{{\ttfamily
  2207.00348}}].

\bibitem{ATLAS:2023fsi}
{\scshape ATLAS} collaboration, \textit{{Improved W boson Mass Measurement
  using 7 TeV Proton-Proton Collisions with the ATLAS Detector}}, 
  [\href{https://cds.cern.ch/record/2853290}{{\ttfamily [ATLAS-CONF-2023-004]}}].

\bibitem{ATLAS:2023owm}
{\scshape ATLAS} collaboration, G.~Aad et~al., \textit{{Measurement of the Higgs boson mass with $H\to \gamma\gamma$ decays in 140 fb$^{-1}$ of s=13 TeV pp collisions with the ATLAS detector
}},
  \href{https://doi.org/10.1016/j.physletb.2023.138315}{\textit{Phys. Lett. B}
  {\bfseries 847} (2023) 138315},
  [\href{https://arxiv.org/abs/2308.07216}{{\ttfamily 2308.07216}}].

\bibitem{ATLAS:2023tnc}
{\scshape ATLAS} collaboration, G.~Aad et~al., \textit{{Measurement of the $H
  \rightarrow \gamma \gamma $ and $H \rightarrow ZZ^* \rightarrow 4 \ell $
  cross-sections in pp collisions at $\sqrt{s}=13.6$ TeV with the ATLAS
  detector}},
  \href{https://doi.org/10.1140/epjc/s10052-023-12130-5}{\textit{Eur. Phys. J.
  C} {\bfseries 84} (2024) 78},
  [\href{https://arxiv.org/abs/2306.11379}{{\ttfamily 2306.11379}}].

\bibitem{CMS:2025zue}
{\scshape CMS} collaboration, \textit{{Constraints on the Higgs boson total
  decay width using signal-background interference in the diphoton final state
  with proton-proton collisions at $\sqrt s$ = 13 TeV}}, .
    [\href{https://cds.cern.ch/record/2941115}{{\ttfamily [CMS-PAS-HIG-25-004]}}].

\bibitem{Pilaftsis:1997dr}
A.~Pilaftsis, \textit{{Resonant CP violation induced by particle mixing in
  transition amplitudes}},
  \href{https://doi.org/10.1016/S0550-3213(97)00469-0}{\textit{Nucl. Phys. B}
  {\bfseries 504} (1997) 61},
  [\href{https://arxiv.org/abs/hep-ph/9702393}{{\ttfamily hep-ph/9702393}}].
  
\bibitem{Ellis:2004fs}
J.~R.~Ellis, J.~S.~Lee and A.~Pilaftsis, \textit{{CERN LHC signatures of resonant CP violation in a minimal supersymmetric Higgs sector}},
\href{https://doi.org/10.1103/PhysRevD.70.075010}{\textit{Phys. Rev. D} {\bfseries 70} (2004) 075010},
[\href{https://arxiv.org/abs/hep-ph/0404167}{{\ttfamily hep-ph/0404167}}].

\bibitem{Ruan:2014xxa}
M.~Ruan, \textit{{Higgs Measurement at $e^+e^-$ Circular Colliders}},
  \href{https://doi.org/10.1016/j.nuclphysbps.2015.09.132}{\textit{Nucl. Part.
  Phys. Proc.} {\bfseries 273-275} (2016) 857--862},
  [\href{https://arxiv.org/abs/1411.5606}{{\ttfamily 1411.5606}}].

\bibitem{Jadach:2015cwa}
S.~Jadach and R.~A. Kycia, \textit{{Lineshape of the Higgs boson in future
  lepton colliders}},
  \href{https://doi.org/10.1016/j.physletb.2016.01.065}{\textit{Phys. Lett. B}
  {\bfseries 755} (2016) 58--63},
  [\href{https://arxiv.org/abs/1509.02406}{{\ttfamily 1509.02406}}].

\bibitem{deBlas:2019rxi}
J.~de~Blas et~al., \textit{{Higgs Boson Studies at Future Particle Colliders}},
  \href{https://doi.org/10.1007/JHEP01(2020)139}{\textit{JHEP} {\bfseries 01}
  (2020) 139}, [\href{https://arxiv.org/abs/1905.03764}{{\ttfamily
  1905.03764}}].

\bibitem{Azzurri:2021nmy}
P.~Azzurri, G.~Bernardi, S.~Braibant, D.~d'Enterria, J.~Eysermans, P.~Janot
  et~al., \textit{{A special Higgs challenge: measuring the mass and production
  cross section with ultimate precision at FCC-ee}},
  \href{https://doi.org/10.1140/epjp/s13360-021-02202-4}{\textit{Eur. Phys. J.
  Plus} {\bfseries 137} (2022) 23},
  [\href{https://arxiv.org/abs/2106.15438}{{\ttfamily 2106.15438}}].

\bibitem{ATLAS:2013xga}
{\scshape ATLAS} collaboration, G.~Aad et~al., \textit{{Evidence for the spin-0
  nature of the Higgs boson using ATLAS data}},
  \href{https://doi.org/10.1016/j.physletb.2013.08.026}{\textit{Phys. Lett. B}
  {\bfseries 726} (2013) 120--144},
  [\href{https://arxiv.org/abs/1307.1432}{{\ttfamily 1307.1432}}].

\bibitem{CMS:2014nkk}
{\scshape CMS} collaboration, V.~Khachatryan et~al., \textit{{Constraints on
  the spin-parity and anomalous HVV couplings of the Higgs boson in proton
  collisions at 7 and 8 TeV}},
  \href{https://doi.org/10.1103/PhysRevD.92.012004}{\textit{Phys. Rev. D}
  {\bfseries 92} (2015) 012004},
  [\href{https://arxiv.org/abs/1411.3441}{{\ttfamily 1411.3441}}].

\bibitem{ATLAS:2020ior}
{\scshape ATLAS} collaboration, G.~Aad et~al., \textit{{$CP$ Properties of
  Higgs Boson Interactions with Top Quarks in the $t\bar{t}H$ and $tH$
  Processes Using $H \rightarrow \gamma\gamma$ with the ATLAS Detector}},
  \href{https://doi.org/10.1103/PhysRevLett.125.061802}{\textit{Phys. Rev.
  Lett.} {\bfseries 125} (2020) 061802},
  [\href{https://arxiv.org/abs/2004.04545}{{\ttfamily 2004.04545}}].

\bibitem{CMS:2020cga}
{\scshape CMS} collaboration, A.~M. Sirunyan et~al., \textit{{Measurements of
  $\mathrm{t\bar{t}}H$ Production and the CP Structure of the Yukawa
  Interaction between the Higgs Boson and Top Quark in the Diphoton Decay
  Channel}},
  \href{https://doi.org/10.1103/PhysRevLett.125.061801}{\textit{Phys. Rev.
  Lett.} {\bfseries 125} (2020) 061801},
  [\href{https://arxiv.org/abs/2003.10866}{{\ttfamily 2003.10866}}].

\bibitem{CMS:2021sdq}
{\scshape CMS} collaboration, A.~Tumasyan et~al., \textit{{Analysis of the $CP$
  structure of the Yukawa coupling between the Higgs boson and $\tau$ leptons
  in proton-proton collisions at $ \sqrt{s} $ = 13 TeV}},
  \href{https://doi.org/10.1007/JHEP06(2022)012}{\textit{JHEP} {\bfseries 06}
  (2022) 012}, [\href{https://arxiv.org/abs/2110.04836}{{\ttfamily
  2110.04836}}].

\bibitem{Bahl:2021dnc}
H.~Bahl and S.~Brass, \textit{{Constraining $ \mathcal{CP} $-violation in the
  Higgs-top-quark interaction using machine-learning-based inference}},
  \href{https://doi.org/10.1007/JHEP03(2022)017}{\textit{JHEP} {\bfseries 03}
  (2022) 017}, [\href{https://arxiv.org/abs/2110.10177}{{\ttfamily
  2110.10177}}].

\bibitem{Bahl:2022yrs}
H.~Bahl, E.~Fuchs, S.~Heinemeyer, J.~Katzy, M.~Menen, K.~Peters et~al.,
  \textit{{Constraining the ${\mathcal {C}}{\mathcal {P}}$ structure of
  Higgs-fermion couplings with a global LHC fit, the electron EDM and
  baryogenesis}},
  \href{https://doi.org/10.1140/epjc/s10052-022-10528-1}{\textit{Eur. Phys. J.
  C} {\bfseries 82} (2022) 604},
  [\href{https://arxiv.org/abs/2202.11753}{{\ttfamily 2202.11753}}].

\bibitem{Barger:2023wbg}
V.~Barger, K.~Hagiwara and Y.-J. Zheng, \textit{{CP-violating top-Higgs
  coupling in SMEFT}},
  \href{https://doi.org/10.1016/j.physletb.2024.138547}{\textit{Phys. Lett. B}
  {\bfseries 850} (2024) 138547},
  [\href{https://arxiv.org/abs/2310.10852}{{\ttfamily 2310.10852}}].

\bibitem{Esmail:2024gdc}
W.~Esmail, A.~Hammad, A.~Jueid and S.~Moretti, \textit{{Boosting probes of $
  \mathcal{CP} $ violation in the top Yukawa coupling with Deep Learning}},
  \href{https://doi.org/10.1007/JHEP12(2025)050}{\textit{JHEP} {\bfseries 12}
  (2025) 050}, [\href{https://arxiv.org/abs/2405.16499}{{\ttfamily
  2405.16499}}].

\bibitem{ATLAS:2023cbt}
{\scshape ATLAS} collaboration, G.~Aad et~al., \textit{{Probing the CP nature
  of the top{\textendash}Higgs Yukawa coupling in tt{\textasciimacron}H and tH
  events with H{\textrightarrow}bb{\textasciimacron} decays using the ATLAS
  detector at the LHC}},
  \href{https://doi.org/10.1016/j.physletb.2024.138469}{\textit{Phys. Lett. B}
  {\bfseries 849} (2024) 138469},
  [\href{https://arxiv.org/abs/2303.05974}{{\ttfamily 2303.05974}}].

\bibitem{CMS:2022dbt}
{\scshape CMS} collaboration, A.~Tumasyan et~al., \textit{{Search for $CP$
  violation in ttH and tH production in multilepton channels in proton-proton
  collisions at $\sqrt{s}$ = 13 TeV}},
  \href{https://doi.org/10.1007/JHEP07(2023)092}{\textit{JHEP} {\bfseries 07}
  (2023) 092}, [\href{https://arxiv.org/abs/2208.02686}{{\ttfamily
  2208.02686}}].

\bibitem{Antusch:2020ngh}
S.~Antusch, O.~Fischer, A.~Hammad and C.~Scherb, \textit{{Testing CP Properties
  of Extra Higgs States at the HL-LHC}},
  \href{https://doi.org/10.1007/JHEP03(2021)200}{\textit{JHEP} {\bfseries 03}
  (2021) 200}, [\href{https://arxiv.org/abs/2011.10388}{{\ttfamily
  2011.10388}}].

\bibitem{Alonso-Gonzalez:2021jsa}
J.~Alonso-Gonz{\'a}lez, L.~Merlo and S.~Pokorski, \textit{{A new bound on CP
  violation in the {\ensuremath{\tau}} lepton Yukawa coupling and electroweak
  baryogenesis}}, \href{https://doi.org/10.1007/JHEP06(2021)166}{\textit{JHEP}
  {\bfseries 06} (2021) 166},
  [\href{https://arxiv.org/abs/2103.16569}{{\ttfamily 2103.16569}}].

\bibitem{Mb:2022rxu}
{\scshape CMS} collaboration, V.~K. Mb, \textit{{Measurement of the
  $\textit{CP}$ structure of the Higgs-tau Yukawa coupling}},
  \href{https://doi.org/10.21468/SciPostPhysProc.8.008}{\textit{SciPost Phys.
  Proc.} {\bfseries 8} (2022) 008},
  [\href{https://arxiv.org/abs/2203.08191}{{\ttfamily 2203.08191}}].

\bibitem{Cardini:2025svy}
A.~Cardini, \textit{{Tau leptons as a tool to investigate the CP properties of
  the Higgs boson at CMS}},
  \href{https://doi.org/10.21468/SciPostPhysProc.16.019}{\textit{SciPost Phys.
  Proc.} {\bfseries 16} (2025) 019}.

\bibitem{Demartin:2014fia}
F.~Demartin, F.~Maltoni, K.~Mawatari, B.~Page and M.~Zaro, \textit{{Higgs
  characterisation at NLO in QCD: CP properties of the top-quark Yukawa
  interaction}},
  \href{https://doi.org/10.1140/epjc/s10052-014-3065-2}{\textit{Eur. Phys. J.
  C} {\bfseries 74} (2014) 3065},
  [\href{https://arxiv.org/abs/1407.5089}{{\ttfamily 1407.5089}}].

\bibitem{Buckley:2015vsa}
M.~R. Buckley and D.~Goncalves, \textit{{Boosting the Direct CP Measurement of
  the Higgs-Top Coupling}},
  \href{https://doi.org/10.1103/PhysRevLett.116.091801}{\textit{Phys. Rev.
  Lett.} {\bfseries 116} (2016) 091801},
  [\href{https://arxiv.org/abs/1507.07926}{{\ttfamily 1507.07926}}].

\bibitem{Azevedo:2017qiz}
D.~Azevedo, A.~Onofre, F.~Filthaut and R.~Gon{\c{c}}alo, \textit{{CP tests of
  Higgs couplings in $t\bar{t}h$ semileptonic events at the LHC}},
  \href{https://doi.org/10.1103/PhysRevD.98.033004}{\textit{Phys. Rev. D}
  {\bfseries 98} (2018) 033004},
  [\href{https://arxiv.org/abs/1711.05292}{{\ttfamily 1711.05292}}].

\bibitem{Goncalves:2018agy}
D.~Gon{\c{c}}alves, K.~Kong and J.~H. Kim, \textit{{Probing the top-Higgs
  Yukawa CP structure in dileptonic $ t\overline{t}h $ with M$_{2}$-assisted
  reconstruction}},
  \href{https://doi.org/10.1007/JHEP06(2018)079}{\textit{JHEP} {\bfseries 06}
  (2018) 079}, [\href{https://arxiv.org/abs/1804.05874}{{\ttfamily
  1804.05874}}].

\bibitem{Esmail:2024jdg}
W.~Esmail, A.~Hammad, M.~Nojiri and C.~Scherb, \textit{{Testing CP properties
  of the Higgs boson coupling to {\ensuremath{\tau}} leptons with heterogeneous
  graphs}}, \href{https://doi.org/10.1007/JHEP04(2025)083}{\textit{JHEP}
  {\bfseries 04} (2025) 083},
  [\href{https://arxiv.org/abs/2409.06132}{{\ttfamily 2409.06132}}].

\bibitem{Pilaftsis:1999qt}
A.~Pilaftsis and C.~E.~M.~Wagner, \textit{{Higgs bosons in the minimal
  supersymmetric standard model with explicit CP violation}},
  \href{https://doi.org/10.1016/S0550-3213(99)00261-8}{\textit{Nucl. Phys. B}
  {\bfseries 553} (1999) 3--42},
  [\href{https://arxiv.org/abs/hep-ph/9902371}{{\ttfamily hep-ph/9902371}}].

\end{thebibliography}

\end{document}